\documentclass[11 pt, a4paper]{article}
\usepackage{array}
\usepackage{epsfig}
\usepackage{amsthm}
\usepackage{amsmath}
\usepackage{color}
\usepackage{enumitem}
\usepackage{booktabs}
\usepackage{sidecap}
\usepackage{setspace}
\usepackage{comment}
\usepackage{url}
\usepackage{float}
\usepackage{apacite}
\usepackage{parskip}
\usepackage{standalone}
\usepackage{threeparttable}
\usepackage{tabu}
\usepackage{adjustbox}
\usepackage{amssymb}
\usepackage{dsfont}
\usepackage[font=footnotesize,labelfont=bf]{caption}
\usepackage{bm, bbm, mdframed, mathtools, mathabx, multirow, subfigure, tikz, multirow, tcolorbox, soul}
\usepackage[left=2.5cm, right=2.5cm, top=3cm, bottom=3cm]{geometry}
\usepackage[english]{babel}
\usepackage[utf8]{inputenc}
\usepackage[mathscr]{euscript}
\newcommand{\E}{\mathrm{E}}

\numberwithin{equation}{section} 
\tcbuselibrary{breakable}
\usepackage{mathdots}
\usepackage{lscape}
\usepackage{pdflscape}

\usepackage[round, longnamesfirst, authoryear, sort]{natbib}
\usepackage[autostyle]{csquotes}  
\usepackage[citecolor=black, linkcolor = black, colorlinks=true, allcolors=black]{hyperref}
\usepackage{glossaries}
\usepackage{appendix}
\usepackage{verbatim}
\usepackage{tcolorbox}
\usepackage{xcolor}
\usepackage{listings}
\usepackage{rotating}
\usepackage{comment}
\usepackage{graphicx}
\usepackage{dsfont}
\usepackage{appendix}
\usepackage{algorithmicx}
\usepackage{algorithm}
\usepackage{algpseudocode}
\usepackage{mathrsfs}
\numberwithin{equation}{section} 
\tcbuselibrary{breakable}
\usepackage{tikz}
\usetikzlibrary{shapes.geometric, arrows, positioning, arrows.meta, patterns}
\tikzstyle{c_solid} = [circle, minimum width=1cm, minimum height=1cm,text centered,draw=black, fill=white]
\tikzstyle{c_dashed} = [circle, minimum width=1cm, minimum height=1cm,text centered,draw=black, dashed]
\tikzstyle{arrow} = [thick,->,>={Stealth[scale=1.3]}]



\title{\textbf{Beyond Baby Blues: The Child Penalty in Mental Health in Switzerland}}
\author{Nora Bearth\thanks{Swiss Institute for Empirical Economic Research of the University of St. Gallen (SEW-HSG), 
Rosenbergstrasse 22, 9000 St. Gallen, CH, E-mail: \texttt{nora.bearth@unisg.ch} \\
Financial support from the crowdfunding campaign ``Parenting together'' is gratefully acknowledged. I thank Ana Paula Armendariz Pacheco, Daniele Ballinari, Gianna Bearth, Beatrix Eugster, Astrid Kunze, Michael Lechner, Dominik Sachs for comments and discussions on this and related topics. Moreover, I thank the health insurance CSS, especially, Lukas Kauer, Christian Schmid and Nicolas Schreiner, and the health insurance Swica, especially, Aurélien Sallin for discussions and providing me the anonymised prescription data for this project. This paper was presented at the University of St. Gallen and University of Lucern. I thank the participants for their helpful comments. GPT-4 and Grammarly were used to edit the manuscript.}}
\date{}

\begin{document}
\begingroup
\let\newpage\relax
\maketitle
\endgroup
\begin{center}
 \vspace{1.2cm}
  \textbf{Abstract}   \\ \vspace{0.5cm}
\end{center}

\begin{minipage}{\textwidth}
    \small

This paper investigates the mental health penalty for women after childbirth in Switzerland. Leveraging insurance data, we employ a staggered difference-in-difference research design. The findings reveal a substantial mental health penalty for women following the birth of their first child. Approximately four years after childbirth, there is a one percentage point (p.p.) increase in antidepressant prescriptions, representing a 50\% increase compared to pre-birth levels. This increase rises to 1.7 p.p. (a 75\% increase) six years postpartum. The mental health penalty is likely not only a direct consequence of giving birth but also a consequence of the changed life circumstances and time constraints that accompany it, as the penalty is rising over time and is higher for women who are employed. \\[0.5cm]
\end{minipage}


\thispagestyle{empty}
\newpage
\setcounter{page}{1}


\section{Introduction}

Mental health awareness has increased in recent years and is key to welfare in a society \citep{Votruba:2016}. Approximately 13\% of women and 8\% of men in Switzerland between 15 and 44 years experienced a mental health crisis that needed care in 2023, and the costs for mental illness in Switzerland are estimated to be around 7 billion Swiss francs per year \citep{SRF:2023a}.\footnote{This is around 2\% of the Swiss yearly government spending.} In general, the risk of suffering from mental health problems varies between individuals due to environmental, social and economic risk factors and individual attributes and behaviours \citep{WHO:2012}.\footnote{Environmental factors include, for example, cultural beliefs, economic policies or gender inequality. Social and economic factors include, for example, the engagement with family members or friends or earnings. Individual attributes and behaviours include dealing with feelings, managing oneself in life, and participating in social activities. Furthermore, they can be influenced by biological factors, such as genes.}

The birth of a child is a significant life event that is often associated with happiness. However, it also increases some risk factors for mental health problems. Women carry the physical burden during pregnancy and childbirth, and biological factors change. Afterwards, a child leads to more stress related to sleep deprivation, a change in the family structure and the roles and responsibilities of the parents, career uncertainties and maybe also financial pressure \citep{Heshmati:2023}. These circumstances do not only change for women but also for men. Hence, the risk of mental health problems increases. The prevalence of postpartum depression in industrialised countries for women lies between 10\% and 15\% \citep{WHO:2020} and for men around 5\% \citep*{Bradley:2011}. This higher risk of psychiatric disorders does also not decrease if the pregnancy was planned \citep{Munk:2015}. Hence, even if a couple wants a child, it does not prevent parents from suffering from a mental health crisis. This paper examines the impact of childbirth on maternal mental health in Switzerland, using anonymized prescription data and a staggered difference-in-differences approach.

Several studies summarised by \cite*{Blanchflower:2009}, \cite*{Clark:2008} and \cite*{Dolan:2008} already looked at the association between parenthood and self-reported well-being and find a negative effect between having a child and happiness in the short-run. However, self-reported outcome variables might be prone to systematic under-reporting or selection bias due to lower response rates from individuals with health problems \citep*{Baetschmann:2016}. Recent research by \cite{Ahammer:2023} has shed light on the different effects of having a child on mental health among men and women using drug prescription data. Women in Austria and Denmark face enduring mental health challenges after giving birth. The consumption of antidepressants significantly increases post-childbirth. Similarly, the percentage of men taking antidepressants also increases after childbirth. However, the increase is higher for women, with no statistically significant reduction in the mental health gap between men and women over the years. This gap appears to be less related to the act of childbirth but more to the evolving roles and responsibilities that follow \citep{Ahammer:2023}. Similar effects of a decline in mental health after childbirth for women are found in Germany \citep{Dehos:2024}. They further suggest that key factors contributing to this decline may include reduced sleep, less physical activity, and limited leisure time, alongside the extensive involvement in childcare.

Women generally invest more time in childcare and household duties than men \citep*{Borra:2021, Guryan:2008}. Consequently, they are more likely to reduce their working hours outside the home \citep*[e.g.][]{Kleven:2020}. Even when women outearn their husbands, they spend more time doing household chores \citep*{Bertrand:2015}. Hence, the evolving roles and responsibilities that mothers face both at home and in the workplace can significantly impact their mental health as time pressure mediates the relationship between parenthood and mental health \citep*{Ruppanner:2019}. The combination of these dual roles, along with insufficient support at home, often leads to stress and sleep deprivation, thereby increasing the risk of mental health issues for women more than for men, resulting in an increased mental health gender gap.\footnote{Another strand of the literature shows that intimate partner violance could also be associated with an increase in mental health problems after childbirth as being pregnant and having a child makes women more vulnerable and dependent on their partner \citep{Woolhouse:2012, Chan:2022}. As we are not able to add any evidence on this, we do not discuss this further.}

Simple measures such as household and childcare assistance, extended time off work, or access to childcare facilities can significantly reduce the increasing mental health burden after childbirth. Therefore, family policies play a crucial role in reducing the risk of mental health issues such as depression, posttraumatic stress disorder, anxiety, and postpartum psychosis \citep*[e.g.,][]{Chatterji:2008, Avendano:2015, Aitken:2015, Hewitt:2017}. A systematic review study by \cite*{Heshmati:2023} suggest that an introduction of maternity leave improves mental health outcomes after childbirth as found in the majority of studies \citep*[e.g.][]{Chatterji:2008, Avendano:2015} but the length of the leave seems to be crucial. Studies in Austria find that an increased maternity leave exacerbates the gender gap in mental health through further enforcement of traditional gender roles \citep[e.g.,][]{Chuard:2023, Ahammer:2023}.\footnote{The parental leave reform extended parental leave from 24 months up to 36 months. The maximum leave for one parent was increased from 18 months to 30 months.} The evidence on the effect of paternity leave on maternal mental health is scarce. Some research suggests that womens' mental health improves when partners have access to paternity leave \citep*{Bilgrami:2020} or to flexible paternity leave policies \citep*{Persson:2019}.

This study primarily contributes to the literature on the effect of a child on the mental health of women using anonymized prescription data and looking at medium-term outcomes \citep{Ahammer:2023, Kravdal:2017}.\footnote{\cite{Ahammer:2023} use data from Austria and Denmark and \cite{Kravdal:2017} use data from Norway.} We investigate the mental health penalty for women after the birth of their first child in Switzerland by using insurance data from two of the biggest Swiss insurance companies and employing a staggered difference-in-difference estimation strategy \citep{Callaway:2021}. Mental health is proxied by antidepressant prescriptions as the primary outcome. Additionally, the number of visits to psychiatrists and the number of general practitioners (GP) are used as secondary measures, as these healthcare providers issue most of these prescriptions \citep{NZZ:2019}. Switzerland has one of the highest female employment rates \citep{BfS:2023c} in Europe, but many mothers only work part-time. This situation may be influenced by the country's relatively short maternity leave (14 weeks), the absence of paternity leave until 2021 (extended to two weeks in 2021), and the high cost of childcare, as public kindergarten starts when a child is between three and five years depending on the region in Switzerland. Furthermore, Switzerland has one of the highest gender pay gaps in Europe \citep{BfS:2023c}, which may also contribute to the tendency for women to reduce their working hours more than men. As a result, many women in Switzerland face a double workload without the same level of support found in more family-friendly countries like Denmark or Norway, which may lead to different outcomes for working mothers.

The findings reveal that four years after childbirth, the percentage of women having an antidepressant prescription in a specific month rises from 2\% to 3\%, representing a 50\% increase.\footnote{Women who are about to have a child, especially in the years leading up to childbirth, are less likely to have an antidepressant prescription compared to women who will never have a child or those who already have children.}  This percentage further climbs to 3.3\% six years post-childbirth, marking a 75\% increase. The results align with the findings of \cite{Ahammer:2023} for Austria and Denmark. The effect found is substantial and economically relevant as mental illnesses impose high costs on the healthcare system \citep{Swissinfo:2008}. Additionally, considering that mental health is a prerequisite for labour market participation, there is a notable loss of productivity in an economy already experiencing a shortage of skilled workers \citep*[e.g.][]{Biasi:2021}. While we do not find statistically significant effects for visits to psychiatrists or GPs in our main specification, we observe an increase in psychiatrist visits in certain alternative specifications, suggesting some sensitivity of the results to sample choice.

Additionally, this paper sheds light on some potential mechanisms by looking at heterogeneous treatment effects. There is a higher penalty for women employed before childbirth than those not employed.\footnote{In Switzerland, accident insurance is paid by the employer if an individual is employed. Accident coverage included in the health insurance is only needed for individuals who are unemployed, out of the labour market or self-employed. Therefore, we can identify the employment status.} The double workload at home and work outside the home might exacerbate the mental health penalty. Moreover, despite some evidence from the literature which finds that cesarean sections are associated with a higher risk of postpartum depression \citep*{Moameri:2019, Xu:2017}, there is no indication of a difference in antidepressant prescriptions by the mode of delivery. Hence, the surgery itself does not seem to be the reason for the mental health penalty. Similarly, despite some literature showing that individuals with low socioeconomic status are more likely to suffer from mental health problems after childbirth \citep*[e.g.][]{Goyal:2010,Rich:2006}, we find that the mental health penalty is lower for individuals with a low income. This can partly be explained by the fact that those women are more likely to not be employed, as income correlates with employment. We also find that the number of first-time antidepressant prescriptions is stable in the first six years after childbirth, which suggests that the increase in antidepressant prescriptions does not only stem from women having postpartum depression and then getting addicted to the drugs. Concluding, these different findings suggest that the increase in antidepressant prescriptions only partly comes from postpartum depression. The other part of the increase probably stems from changing circumstances, leading to more stress. 

Last, this paper contributes to the literature on the effect of paternity leave on maternal mental health. We investigate the effect of the introduction of paternity leave in 2021 in Switzerland on the mental health of the mother. The literature on this highly relevant question is scarce. Mental health is crucial for women to enter the labour market after childbirth \citep[e.g.,][]{Biasi:2021}, and as mentioned, mental illness is costly for an economy. Therefore, economists should also consider the impact of family policies on mental health outcomes and not only on labour market outcomes \citep[e.g.,][]{Kleven:2019, Olivetti:2017, Havnes:2011}. Since the introduction was in 2021, only short-run effects can be analysed. We do not find any statistically significant effect on mental health. Due to the low power of the analysis, only a few women have an antidepressant prescription in the first two years after childbirth; therefore, the outcome is rare, the null result should be interpreted with caution.

The remainder of the paper is structured as follows: Section \ref{1_institutional_background} explains the healthcare system and provides some facts about womens' general mental health, the labour force participation and the different family policies in Switzerland. Section \ref{1_data} describes the data used, how the sample has been selected for the analysis and shows some descriptives. Section \ref{stagg_diff_diff: empirical_strategy} explains the empirical strategy, and Section \ref{results_main} presents the different results, including robustness checks, results for men and heterogeneities. The results are further discussed in Section \ref{discussion}, and potential mechanisms are explained. Last, Section \ref{1_conclusion} concludes. Additional descriptives and results are presented in Appendix \ref{appendix: staggered_diff_diff}. 
Appendix \ref{appendix: effect_on_men} presents the effect of childbirth on mens' mental health, while Appendix \ref{appendix: paternal_leave} covers the impact of paternal leave on mothers' mental health.

\section{Institutional Setting}
\label{1_institutional_background}

\subsection{Healthcare System} \label{1_health_insurance}

Having basic health insurance is mandatory in Switzerland. Residents of Switzerland can choose among 50 different health insurers, but over 90\% of individuals are insured at the ten biggest health insurances \citep{SRF:2024}. The contracts are made for one year, meaning individuals can change insurers yearly. Between 2010 and 2022, around 10\% of individuals regularly changed health insurers \citep{versicherung_schweiz:2021}. In the event of illness, maternity and accident, the health insurance protects the insured people, and the insurer is not allowed to treat insured individuals differently, for example, by differentiating between healthy and non-healthy individuals. This also applies to insurance admission. The insurance is financed by the monthly contributions from the individuals (premium), with the individuals' cost-sharing (deductible, excess and hospital contributions) and partly with contributions from the cantons and the Swiss Confederation. How much an individual pays monthly depends on the place of residence and the chosen form of insurance, for example, how high the deductible is and what the first point of contact is in case of medical need (e.g. a GP, a telemedicine service or directly a specialist). The average monthly premium for an adult is slightly over CHF 300. The most often used yearly deductible is CHF 2,500 \citep{BAG:2023}. Until the deductible is met, an individual pays for the medical treatments by himself or herself. Afterwards, the insurance covers the costs. Despite paying the costs individually, all services are billed to the insurance company, which reclaims the costs from the individuals. Hence, the insurance company knows how much of the deductible has already been used and what services and treatments an individual has used in a year.

Home visits by midwives for postnatal care and monitoring the health of both mother and newborn are covered by basic health insurance within the first 56 days postpartum. In cases of premature birth, multiple births, first-time mothers, or cesarean sections, up to 16 visits are permitted. Otherwise, the limit is 10 visits. During the first 10 days, up to five additional second daily visits are allowed, and any further visits beyond these limits require a physician's prescription. In conclusion, the midwife visits once to twice weekly in the first two months after childbirth \citep{BAG:2024}.

Additionally, individuals in Switzerland can sign up for supplementary insurance, which covers additional services, such as alternative medicine, glasses, or a private room in the hospital. It is not mandatory to have supplementary insurance, the insurance companies can refuse individuals, and it is possible to have basic and supplementary insurance at different insurance companies. Therefore, supplementary insurance is not included in the main analysis for consistency reasons.\footnote{Data on the private hospital insurance is used in the identification section to explore if individuals are signing up for it at a certain point before having a child, which could be a sign of planning a child.}

\subsection{Mental Health} \label{1_mental_health}
Around 9\% of women suffer from a major depression in Switzerland \citep{BfS:2023a} and around 5\% to 7\% of the Swiss population take antidepressants at least once a year \citep{BfS:2019}. The prevalence is lower for younger women, as the risk of depression increases with age. Furthermore, it is noteworthy that many antidepressants are prescribed by GPs rather than psychiatrists, highlighting the crucial role GPs play in the mental health care of the Swiss population \citep{Obsan:2022}. The basic insurance covers the costs of antidepressants and psychiatrist visits (after using up the deductible). From 2022 on, it also covers visits to a psychotherapist who is not working for a psychiatrist \citep{BAG:2022}. For consistency reasons, those visits are not included in the analysis.

Postpartum depression refers to depressive episodes that arise during pregnancy or within the first year after childbirth. Common symptoms include persistent sadness, loss of interest or pleasure in previously enjoyed activities, sleep disturbances, feelings of worthlessness or being a bad mother, frequent crying, and thoughts of death or suicide \citep{APA:2020}. These symptoms largely mirror those of non-postpartum depression. For mild cases of depression, psychotherapy is typically the first line of treatment. In moderate cases, psychotherapy is often combined with antidepressant medication, while in severe cases, antidepressant medication is considered essential. Depending on individual needs, additional medications such as sleep aids or anxiolytics may also be prescribed \citep{Swissmom:2024}.

\subsection{Family Policies and Labour Force} \label{family_policies}

As pointed out in the literature, part of the mental health problems after childbirth are likely mediated by an increased level of stress, sleep deprivation and workload due to household responsibilities, childcare duties, and work outside the home. Therefore, family policies are crucial in supporting maternal well-being, primarily if the mother works outside the home.

The employment rate for women in Switzerland lies around 90\% before childbirth \citep{BfS:2022}. This is a high participation rate compared to other European countries. The percentage of women working after a first child decreases to 80\% and after a second to 69\%. As their children grow older, more women return to work, with about 87.5\% of mothers with children aged 12 to 14 working again. Most mothers work only part-time, namely 78.1 \%, while only 18\% of Swiss fathers do \citep{BfS:2022}. Part of the reason for part-time work is that childcare is expensive in Switzerland \citep{BSV:2024a}. In 2003, the Swiss government decided to incentivise the building of new childcare places, which was relatively successful in increasing the availability but not in reducing childcare costs.

Maternity and paternity leave can help to reduce the mental health burden of mothers after childbirth. In 2005, a 14-weeks maternity leave has been introduced in Switzerland. During this period, 80\% of the salary is paid, and job protection extends for up to 16 weeks \citep{BSV:2024}. Compared to other European countries, this is a relatively short period. A two-week paternity leave has only been introduced in 2021 \citep{BSV:2024}. Furthermore, Switzerland has no flexible parental leave system that allows parents to choose who stays home with the children.  

\section{Data} \label{1_data}

\subsection{Data Sources}

This paper uses data from two of the biggest Swiss insurance companies (CSS and Swica), including all incurred insurance costs for each individual over time. It is a selected sample of women insured at CSS and Swica, which corresponds to roughly 25\% of the Swiss population. The dataset from CSS contains information on insured women and men born between 1971 and 2003 from the years 2010 to 2022, with information on the drugs that a person buys and the visits to GPs and psychiatrists over the years. The data from Swica contains the same information but only for women and starts in 2012. Hence, all costs related to mental health and childbirth that occur for an individual are observed. Additionally, the dataset provides each insured person's basic sociodemographic characteristics, such as age, civil status, and region of residence.

The advantage of using insurance data is that there are no self-reported health outcomes, which could easily lead to systematic under-reporting \citep{Bound:2001}. Furthermore, survey data could suffer from selection bias due to lower response rates from individuals with health problems.\footnote{Theoretically, some individuals might prefer to pay for antidepressants out of pocket to prevent their insurance from knowing about their mental health issues, potentially leading to an under-reporting bias. However, this scenario is highly improbable for several reasons. First, personal health data is protected by law, ensuring that insurance providers cannot disclose an individual's use of antidepressants. Second, individuals can switch basic insurance plans annually, with other insurers prohibited from refusing coverage. Third, when applying for supplementary insurance, a mandatory health assessment requires disclosing all relevant health information. Fourth, if a psychiatrist or GP directly distributes the antidepressants, the cost is included in the overall bill for the visit. Even if dispensed by a pharmacy, individuals must explicitly request to pay out of pocket.} A disadvantage is that mental health changes are only observed if the person uses some medical treatment, i.e. either some drugs or a visit to a doctor. Hence, small changes in the mental health conditions of individuals are not detectable. Furthermore, due to the sensitivity of the data, it cannot be linked to other administrative sources. Since individuals can easily switch insurance companies yearly, the dataset only provides a partial panel for some women.

Additionally, this paper uses administrative data from the Swiss government, i.e. income compensation data, population and household statistics, vital statistics and social security data. The data is used to check if the insurance data is representative of the Swiss population. For more information on the representativeness of the data, see Appendix \ref{appendix: data_descriptives}.

\subsection{Selected Sample} \label{section: analysis_sample}

The analysis focuses on women born between 1985 and 1995. We exclude older women from the sample because they are more likely to have had their first child before 2010. For example, women born in 1985 would have been 25 years old in 2010, and while many may have delayed childbearing, some are likely to have become mothers before 2010. Hence, second births would be misclassified as first births.\footnote{Some births may still be misclassified, with a slightly higher potential for misclassification among individuals insured with Swica, as the data only begins in 2012 rather than 2010.} Conversely, women born after 1995 are less likely to have had their first child by 2022, as those born in 1995 would only be 27 years old. Consequently, including women born after 1995 would shift the distribution of the age at the first birth to the left. Therefore, they are excluded from the analysis.\footnote{These thresholds have been defined with the median age of women giving birth to their first child in Switzerland in mind, which is 31 years \citep{BfS:2024}.} The sample is further refined to include only women who gave birth to their first child between 20 and 40.\footnote{By construction, the oldest women in the sample are 37 years old when they have their first child - women born 1985 that have their first child in 2022. Hence, only the lower age is binding.} This age range allows us to focus on womens' reproductive years.

We further constrain the samples to the years between 2012 and 2022, a choice influenced by the introduction of the DRG on the 1st of January 2012 \citep{DRG:2024}.\footnote{A Diagnostic-Related Group (DRG) is a system used by health insurance companies to categorise hospitalisation costs and determine payment amounts for hospital stays. Instead of billing for each individual service received, a fixed payment amount is predetermined based on the DRG assigned to your stay.} This significant event in the healthcare system could affect data reliability prior to 2012. After applying these criteria, the sample of women comprises $N = 83,841$ individuals.

To further mitigate issues regarding misclassifying first births, we use a balanced sample of women who have remained with the same insurance provider between 2012 and 2022. While this approach helps reduce the risk of bias caused by compositional changes over time, as it captures the same group of women throughout the study period, it may introduce a limitation by focusing on a specific population, as the individuals have not changed insurance for over a decade.\footnote{Because women have a child at different points in time, and we only contemplate the dataset from 2012 to 2022, we still have a compositional change if using a staggered difference-in-difference approach \citep{Callaway:2021} but a smaller one than by using an unbalanced sample. Assuming a woman has her first child in 2018, we contemplate six years before and five years after childbirth. Hence, if we look at the effect six years after childbirth, this woman is not in the sample anymore. To eliminate all compositional changes, we would have to restrict the sample to women for whom we contemplate the whole event-time of interest. This would reduce the sample size even more.} Specifically, individuals with poorer mental health may have less energy or motivation to change their insurance provider, leading to a sample that overrepresents those with mental health challenges. Table \ref{Table:samples_staggered_diff_diff} in Section \ref{section: covariates} shows that the balanced sample only differs from the unbalanced sample in terms of nationality, language, and number of children. Thus, the balanced sample appears mostly representative of the unbalanced sample and, hence, of the broader Swiss population. 

This decision to use a balanced sample comes at the cost of reducing the sample size of women to $N =  25,120$. Hence, the power of the analysis is reduced. An unbalanced sample is used to check for robustness due to the reasons mentioned in the previous paragraph. Furthermore, a balanced sample only going until 2019 ($N = 18,714$) is used to avoid the potential bias of the COVID-19 pandemic on mothers' mental health. If mothers with specific mental health choose to have a child during the pandemic and, at the same time, the pandemic has an impact on antidepressant prescriptions, this could bias the results in both directions.\footnote{We do not find any change in the pattern of antidepressant prescriptions due to the Corona pandemic (see Figure \ref{fig:descriptives_antidepri_over_years} in Appendix \ref{1_data_paternity_leave}).} 

\subsection{Descriptive Statistics} \label{stagg_diff_diff: descriptive_statistics}

\subsubsection{Outcome Variables} \label{section: outcone_variables}
The primary outcome variable is the mental health status of a woman, proxied by whether a woman has an antidepressant prescription (ATC Code: N06A) in a particular month. Since antidepressant prescriptions can last several months, the outcome variable is smoothed over the months the individual likely consumed the antidepressant. This smoothing is based on the number of prescribed packages and pills per package. Typically, patients consume one pill per day, so a package containing 30 pills is assumed to last one month. Additionally, if a woman lacks an antidepressant prescription for one or two months between two months with prescriptions, continuous consumption during these months is assumed. This smoothing approach provides a less volatile outcome variable without altering the results.\footnote{A robustness check with the non-smoothed outcome can be found in Appendix \ref{appendix_non_smoothed}.} Nevertheless, antidepressant prescriptions are potentially still a noisy measure of mental health, and they most likely underestimate mental health problems, as not everyone with mental health issues receives medication. Antidepressants are typically prescribed for moderate to severe cases, but many individuals with milder symptoms rely on therapy, other treatments, or do not seek treatment at all \citep{Obsan:2022}.

In addition to the primary outcome, we consider other mental health proxies, such as the number of visits to a GP or psychiatrist. These measures are relevant since GPs or psychiatrists prescribe most antidepressants. It is essential to note that there are limitations to these outcomes. Increased GP visits may result from various factors unrelated to mental health, such as childcare-related illnesses, and time constraints, particularly in the first years after childbirth, might influence the number of visits to a psychiatrist and GP more than the antidepressant prescriptions. Furthermore, many GPs are located in a Health Maintenance Organisation (HMO) with other physicians, such as gynaecologists or specialists. We combine GP and HMO visits. Therefore, there is a slight measurement error in this outcome variable. Moreover, the number of visits to the GP or HMO can only be identified consistently from 2012 to 2022 for the CSS data, which leads to a smaller sample size for this outcome. The outcome variables are measured relative to the event-time of birth $t$, with $t+1, t+2, \dots$ representing the months post-childbirth, and $t-1, t-2, \dots$ the months pre-childbirth. 

\begin{figure}[h!]
    \centering
    \begin{minipage}{15cm}
    \caption{Descriptives: Antidepressants prescriptions}
    \label{fig:antidepressants_prescription_descriptives}
    \includegraphics[width=15cm]{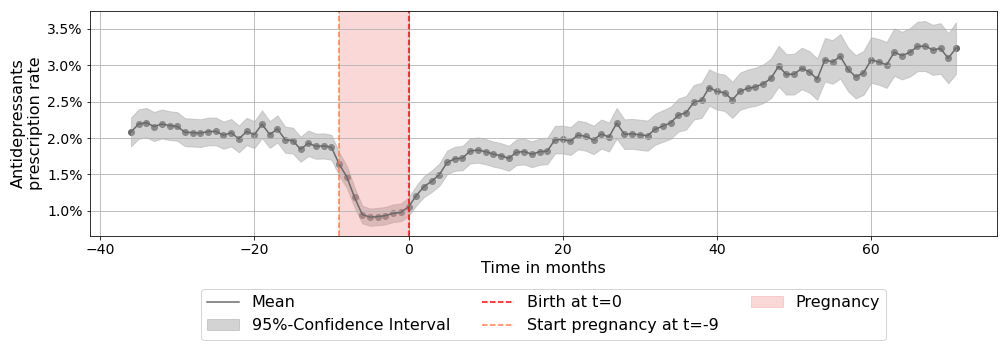}
    \footnotesize \textit{Note:} The figure shows the percentage of women taking antidepressants around the time of birth ($t = 0$: time of first birth). The red-shaded area illustrates the pregnancy, and the grey-shaded area represents the 95\% confidence interval. The slow decrease in antidepressants at the start of the pregnancy and also the slow increase after giving birth are partly coming from the smoothing of the antidepressant prescriptions. Compared to Denmark or Austria, the antidepressant prescription rate is lower before childbirth. A reason, therefore, could be that even after monthly health insurance payments, Swiss individuals have to pay for medical visits and drugs up to a yearly deductible (often CHF 2,500). ($N = 25,120$) 
    \end{minipage}
\end{figure}

Figure \ref{fig:antidepressants_prescription_descriptives} illustrates the number of women with an antidepressant prescription in the months before and after childbirth. The red-shaded area represents the pregnancy period, culminating in the first child's birth at $t=0$. Approximately 2\% of women have an antidepressant prescription in any given month before childbirth.\footnote{If we look at yearly prescriptions, approximately 4\% of women have an antidepressant prescription before childbirth. This number is slightly lower than the average in the Swiss population as antidepressant consumption increases with age.} This number increases over the years following childbirth, reaching 2.5\% four and 3\% six years after childbirth. The observed decrease in antidepressant prescriptions during pregnancy and the first year postpartum is likely due to the recommendation that it is better for the child's health if the mother avoids consuming antidepressants.

\begin{figure}[h!]
    \centering
    \begin{minipage}{15cm}
    \caption{Descriptives: First-time antidepressants prescriptions}
    \label{fig:antidepressants_prescription_descriptives_first_time}
    \includegraphics[width=15cm]{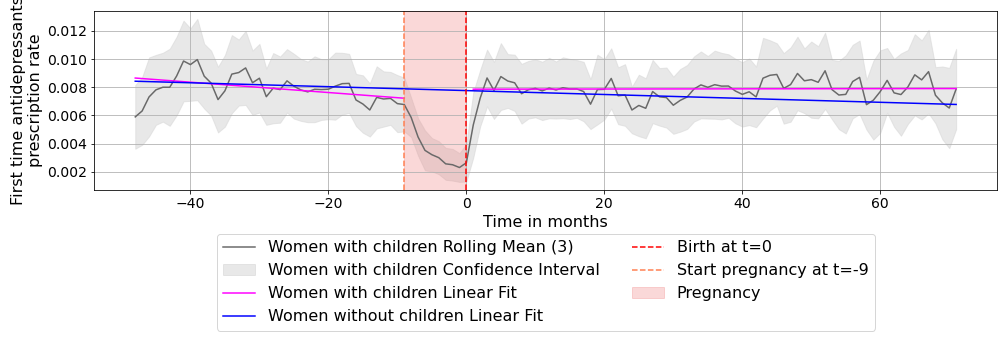}
    \footnotesize \textit{Note:} The figure shows the percentage of women taking for the first time antidepressants around the time of first birth ($t = 0$). In the sample are only women who have at least one antidepressant prescription between 2012 and 2022 and who have had at least one year without antidepressant prescriptions before their first recorded prescription. The pink line shows the linear trend before and after childbirth for women giving birth in $t = 0$. The blue line shows the linear trend for women who never have a child but get assigned random birthdates. The factual distribution of age at first birth for women without children has been approximated by a birth-cohort specific log-normal distribution $LN(\mu, \sigma^2)$. The month is randomly drawn from a uniform distribution. The red-shaded area illustrates the pregnancy, and the grey-shaded area represents the 95\% confidence interval.   ($N = 7,489$)
    \end{minipage}
\end{figure}

To assess whether the increase in antidepressant use for women is due to an immediate rise after childbirth leading to drug dependence, we examine first-time antidepressant prescriptions. Since the prescription data only begins in 2012, identifying actual first-time prescriptions is challenging. As we move further in time, the likelihood increases that the observed prescriptions are genuinely first-time prescriptions, explaining the negative trend in antidepressant use among women without children (blue line) in Figure \ref{fig:antidepressants_prescription_descriptives_first_time}.\footnote{For the women without children placebo dates for the first birth are randomly drawn. We follow \cite{Ahammer:2023} and approximate the factual distribution of age at first birth by a birth-cohort specific log-normal distribution $LN(\mu, \sigma^2)$. The month is randomly drawn from a uniform distribution.} To minimise this issue, we select women who had at least one year without antidepressant prescriptions before their first recorded prescription. We find that the number of women receiving their first antidepressant prescription stabilizes at a consistent level in the years following childbirth. This suggests that the observed increase in antidepressant use is not solely driven by postpartum depression but persists even two to six years after childbirth, indicating a longer-term effect.

\begin{figure}[h!]
    \centering
    \begin{minipage}{15cm}
    \caption{Descriptives: Visits to psychiatrist}
    \label{fig:psychiatrist_descriptives} 
    \includegraphics[width=15cm]{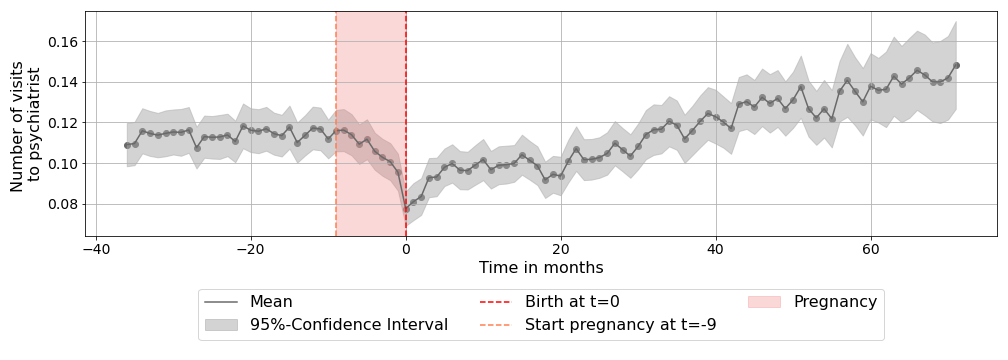}
    \footnotesize \textit{Note:} The figure shows the number of visits to a psychiatrist around the time of birth ($t = 0$: time of first birth). The red-shaded area illustrates the pregnancy, and the grey-shaded area represents the 95\% confidence interval.  ($N = 25,120$)
    \end{minipage}
\end{figure}

Figure \ref{fig:psychiatrist_descriptives} shows that a woman visits on average 0.12 times per month a psychiatrist.\footnote{Slightly more than 3.5\% of women go to a psychiatrist before childbirth.} The number of visits decreases during pregnancy, childbirth and the first years after birth, likely due to time constraints. Two years after childbirth, the number of visits increases, reaching approximately 0.14 times per month six years post-childbirth.

\begin{figure}[h!]
    \centering
    \begin{minipage}{15cm}
    \caption{Descriptives: Visits to GP or HMO}
    \label{fig:GP_descriptives} 
    \includegraphics[width=15cm]{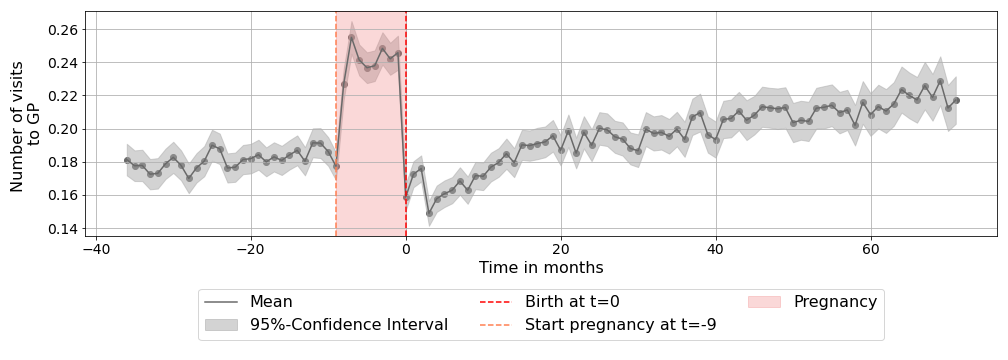}
    \footnotesize \textit{Note:} The figure shows the number of visits to a GP or HMO around the time of birth ($t = 0$: time of first birth). The red-shaded area illustrates the pregnancy, and the grey-shaded area represents the 95\% confidence interval. ($N = 16,368$) 
    \end{minipage}
\end{figure}

Figure \ref{fig:GP_descriptives} depicts that, on average, women visit a GP or HMO around 0.18 times per month (over two visits per year). The frequency of visits increases during pregnancy, likely due to gynaecological check-ups at HMO centres, then declines after childbirth. However, the rate of visits begins to rise again, reaching 0.22 visits per month by the time children are six years old.

\subsubsection{Covariates} \label{section: covariates}

\begin{figure}[h!]
    \centering
    \begin{minipage}{12cm}
    \caption{Descriptives: Age at first birth}
    \label{fig:age_descriptives}
    \includegraphics[width=12cm]{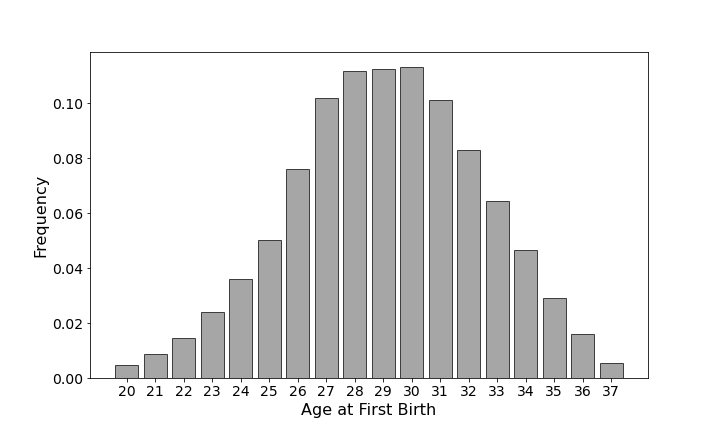}
    \footnotesize \textit{Note:} The figure shows the age distribution of women having a child between 2012 and 2022. ($N = 25,120$)
    \end{minipage}
\end{figure}

We control for the individual's age at their first birth in the analysis and use additional covariates to analyse heterogeneities. Controlling for age is important, as the risk of depression increases with age and because women who are mentally unhealthy might postpone having a child. Consequently, women giving birth at different ages may exhibit different underlying trends in mental health, which could bias estimates if not accounted for. The average age of having a first child is around 29 years, and the age distribution is shown in Figure \ref{fig:age_descriptives}.\footnote{The average age of Swiss women when they give birth for the first time is 30.  However, the mean in our sample is lower due to selecting only women born between 1985 and 1995 and having the years from 2010 to 2022.} By construction, the highest age at the birth of a first child is 37 years, and the lowest is 20 due to some sample restrictions.

\begin{figure}[h!]
    \centering
    \begin{minipage}{15cm}
    \caption{Descriptives: Employment}
    \label{fig:employed_descriptives} 
    \includegraphics[width=15cm]{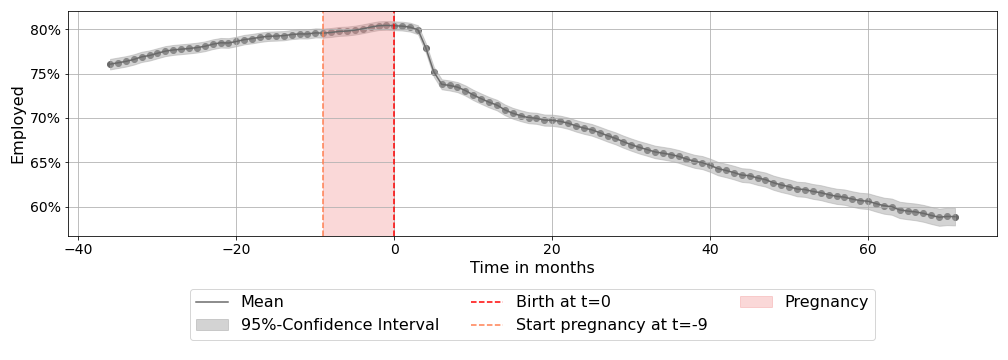}
    \footnotesize \textit{Note:} The figure shows the proportion of women employed around the time of birth ($t = 0$: time of first birth). The red shaded area illustrates the pregnancy, and the shaded area represents the 95\% confidence interval. ($N = 25,120$)
    \end{minipage}
\end{figure}

Employment status can be identified by looking at the accident coverage of the health insurance. Employers generally provide it for their employees, hence, employed individuals do not have accident coverage in the health insurance. On the other hand, it is needed by those unemployed, out of the labour force or self-employed.
Figure \ref{fig:employed_descriptives} shows a notable decrease in employement following childbirth. Around 80\% of women are employed before childbirth.\footnote{The increase in employment pre-birth is partly due to young women and men not yet started to work as they still are, for example, studying or doing an apprenticeship.}
This contrasts with the 90\% pre-childbirth employment rate for women in Switzerland as reported by official statistics, indicating a discrepancy due to self-employment  \citep{BfS:2022}. 8.1\% of women without and 11.8\% of women with children in Switzerland are self-employed. As depicted by Figure \ref{fig:employed_descriptives}, employment rates drop to about 70\% post-childbirth, and further reduce to 60\%. Official figures show a decrease in employment to 80\% after the first child and 69\% after the second \citep{BfS:2022}. Thus, the proxy continuously underestimates employment by approximately ten percentage points (p.p.). This ten p.p. discrepancy is likely due to the exclusion of self-employed individuals.\footnote{There is a concern that individuals might forget to cancel their accident insurance when they find a job, as it is inexpensive. However, the more common scenario is that people stop working after childbirth and need to sign up for accident coverage. Since accident insurance is mandatory and reminders are sent if it is not in place, this issue is less pronounced.}

The left part of Table \ref{Table:samples_staggered_diff_diff} presents the sociodemographic characteristics of women in the month they give birth, across the different samples used in the analysis. This helps to evaluate whether the sample selection might introduce bias and whether robustness checks are warranted. The right part of the table compares women with and without an antidepressant prescription prior to childbirth, highlighting potentially meaningful differences between these groups that motivate further investigations of heterogeneous treatment effects.

In the balanced sample, 19\% of women receive subsidies for health care, 63\% of women have more than one child, and 30\% have a cesarean section. Furthermore, 78\% of the women have German as a contact language, 19\% have French, and 3\% have Italian. The balanced and the unbalanced samples differ in some sociodemographic characteristics. For example, the balanced sample has a higher number of Swiss women and fewer women who speak English. A reason could be that some foreign women entered Switzerland later than 2012 and, hence, are not insured at the same insurance for the whole period. Similarly, some sociodemographic characteristics differ between the sample going until 2022 and the sample only going until 2019. On average, the latter has a lower age at first birth. By construction, the oldest woman giving birth can only be 34 years old instead of 37, which pushes the average age downwards. The unbalanced sample has more women with only one child by construction. We do not see all the years for the women in the unbalanced sample, hence, we miss some second births. As there are some differences between the two samples, the unbalanced sample is used as a robustness check (see Section \ref{robustness:unbalanced_sample}).

\begin{table}[h!]\centering
    \caption{Descriptives: Comparison of different samples for staggered difference-in-difference analysis}
    \label{Table:samples_staggered_diff_diff}
    \begin{adjustbox}{max width=0.93\textwidth}
    \begin{threeparttable}
    \begin{tabular}{l|rrrrr|rrr} \toprule
    Sample & \multicolumn{1}{c}{Balanced} & \multicolumn{2}{c}{Unbalanced} & \multicolumn{2}{c}{w/o Corona years} & \multicolumn{1}{c}{Antidepressants} & \multicolumn{2}{c}{No antidepressants} \vspace{0.1cm}\\ 
        &\textit{Mean} & \textit{Mean} & \textit{Std. diff.} & \textit{Mean}  & \textit{Std. diff.}& \textit{Mean} & \textit{Mean} & \textit{Std. diff.} \\\midrule
       & \multicolumn{8}{c}{\textbf{Discrete variables}} \\ \midrule
    \textbf{Nationality Regions}\\
    East Europe & 0.05 & 0.08 & 15.23 & 0.06 & 6.18 & 0.05 & 0.04 & 2.28\\
South Europe & 0.02 & 0.05 & 14.17 & 0.03 & 3.31 & 0.03 & 0.02 & 10.48\\
Switzerland & 0.55 & 0.43 & 25.10 & 0.53 & 4.34 & 0.52 & 0.56 & 7.71\\
West Europe & 0.01 & 0.04 & 17.69 & 0.01 & 2.69 & 0.01 & 0.01 & 1.59\\
not available & 0.36 & 0.37 & 1.93 & 0.35 & 0.63 & 0.37 & 0.36 & 2.17\\
other & 0.01 & 0.03 & 15.36 & 0.01 & 2.35 & 0.02 & 0.01 & 4.16\\
 \\
    \textbf{Civil status} \\
    divorced & 0.01 & 0.01 & 0.09 & 0.01 & 1.07 & 0.02 & 0.01 & 9.14\\
single & 0.35 & 0.30 & 11.97 & 0.34 & 2.75 & 0.37 & 0.35 & 5.18\\
not available & 0.20 & 0.24 & 11.00 & 0.20 & 0.51 & 0.21 & 0.19 & 4.21\\
married & 0.44 & 0.45 & 2.11 & 0.45 & 2.02 & 0.40 & 0.45 & 10.39\\
 \\
    \textbf{Contact language} \\
    English & 0.00 & 0.04 & 25.22 & 0.00 & 0.85 & 0.00 & 0.00 & 1.10\\
French & 0.19 & 0.19 & 1.49 & 0.20 & 4.18 & 0.23 & 0.18 & 12.99\\
German & 0.78 & 0.71 & 15.50 & 0.76 & 5.20 & 0.74 & 0.79 & 11.58\\
Italian & 0.03 & 0.06 & 12.82 & 0.04 & 2.72 & 0.03 & 0.03 & 1.45\\
 \\
    \textbf{Region} \\
    Central Switzerland & 0.16 & 0.13 & 8.19 & 0.15 & 2.90 & 0.12 & 0.17 & 14.18\\
Eastern Switzerland & 0.21 & 0.18 & 6.61 & 0.21 & 0.29 & 0.19 & 0.21 & 7.01\\
Espace Mittelland & 0.16 & 0.17 & 1.73 & 0.16 & 0.97 & 0.18 & 0.15 & 7.27\\
Northwestern Switzerland & 0.13 & 0.14 & 2.66 & 0.13 & 0.52 & 0.13 & 0.13 & 0.21\\
Region lemanique & 0.12 & 0.13 & 3.02 & 0.13 & 3.26 & 0.15 & 0.11 & 11.08\\
Ticino & 0.03 & 0.05 & 9.31 & 0.03 & 2.48 & 0.03 & 0.03 & 2.40\\
Zurich & 0.19 & 0.20 & 2.87 & 0.19 & 1.37 & 0.20 & 0.19 & 4.06\\
 \midrule
    & \multicolumn{8}{c}{\textbf{Binary variables}} \\ \midrule
    
Employed & 0.80 & 0.76 & 11.61 & 0.77 & 9.31 & 0.72 & 0.83 & 24.86\\

Subsidy & 0.19 & 0.22 & 5.93 & 0.23 & 9.74 & 0.27 & 0.17 & 23.80\\

Cesarean section & 0.30 & 0.29 & 1.23 & 0.28 & 2.44 & 0.32 & 0.29 & 7.95\\

One child & 0.37 & 0.46 & 17.43 & 0.34 & 5.82 & 0.42 & 0.36 & 12.70\\
\midrule
    & \multicolumn{8}{c}{\textbf{Continuous variables}}  \\ \midrule
    income index & -0.52 & -0.54 & 7.18 & -0.54 & 8.05 & -0.52 & -0.52 & 2.12\\
age at first birth & 29.10 & 29.49 & 11.37 & 27.96 & 36.19 & 29.10 & 29.25 & 4.39\\
 \midrule
    N &   25,120      &  83,841 &   & 18,714 & &  5,634      &  19,486 &    \\ \midrule
    \end{tabular}
    \begin{tablenotes}
        \small \item \textit{Note:} The table shows the covariates of women in the month they have a child. Column (2) shows the mean for the balanced sample. Column (3) shows it for the unbalanced sample, and column (4) shows the standardised difference comparing the unbalanced with the balanced sample. Columns (5) and (6) do the same for a balanced sample, only containing the years until 2019. Column (7) shows descriptives for women taking antidepressants and column (8) for women not taking antidepressants and one point between 2012 and 2022. The income index is between -1 and 1 and is estimated by the insurance company and only available for the data from the CSS insurance.
    \end{tablenotes}
    \end{threeparttable}
    \end{adjustbox}
\end{table}
Comparing women who have an antidepressant prescription with those who do not, we see that slightly more women speaking French and fewer women speaking German have an antidepressant prescription. This pattern can also be seen in the regions of residence, as more women living in the French-speaking part of Switzerland have an antidepressant prescription compared to, for example, central Switzerland. As expected, women with an antidepressant prescription are more likely to not be employed. Furthermore, they are more likely to have a subsidy for health care, have a cesarean section and only have one child. Hence, those are potentially interesting dimensions for a heterogeneity analysis.

\section{Empirical Strategy} \label{stagg_diff_diff: empirical_strategy}

A staggered difference-in-difference approach by \cite{Callaway:2021} is used as the preferred empirical strategy.\footnote{We also conducted the analysis using a two-way fixed effects (TWFE) regression, which is a common approach in the literature \citep{Ahammer:2023, Kleven:2019}. The results were largely consistent with those from the staggered difference-in-difference approach. Due to several limitations identified in recent research regarding the TWFE model in settings with heterogeneous treatment effects and staggered adoption \citep{SantAnna:2020, Callaway:2021, Sun:2021, Borusyak:2024}, we opt to present the results based on the staggered difference-in-difference approach.} The data is i.i.d. $\{Y_{i,1}, \dots, Y_{i,\tau}, X_i, D_{i,1}, \dots, D_{i, \tau}\}_{i = 1}^N$. A time period (month) is denoted as $\tau \in \{1, \dots, \mathcal{T}\}$ with a total of $\mathcal{T}$ time periods and an individual is denoted with $n \in \{1, \dots, N\}$ with a total number of individuals $N$. The outcome variable is defined as $Y_{i, \tau}$ and $X_i$ are some pre-treatment covariates that do not change over time. The treatment variable $D_{i,\tau }$ takes the value one if a woman $i$ has a child in time $\tau $ and zero otherwise. In the first period, no one is treated, and once a person is treated, this person is treated for all subsequent periods. This assumption is reasonable since once an individual has a child, it is impossible to return the child. The first period an individual is treated is denoted as $G_i$ and let $\bar g = \max_{i = 1, \dots, n}G_i$. $G_{i,g}$ is a binary variable that takes the value one if a woman has her first child in period $g$ and zero otherwise.

The parameter of interest is the average treatment effect on the treated (ATT):
\begin{equation}\label{eq:ATT}
    ATT(g,\tau ) = \E[Y_{i,\tau} (g) - Y_{i,\tau} (0) | G_{i,g} = 1]
\end{equation}

The ATT is particularly insightful because it explicitly measures the effect of having a child on mental health for women who have children rather than for the entire population. This distinction is crucial because the group of individuals with children is not a random sample of the entire population. Understanding the ATT is important for evaluating and designing family policies, as it provides a clear picture of the impact on the specific target group, ensuring that interventions are tailored to those who experience the treatment. The interest lies in the dynamics of the ATT over time elapsed since childbirth. Therefore, $t$ is the event-time, i.e., $t = \tau -g$, which denotes the time elapsed since childbirth, and the effects of interest are:
\begin{align} \label{eq:eventstudy_effect}
    \theta_{es}(t) = \sum_{g \in \mathcal{G}} \mathds{1}\{g + t \leq \mathcal{T}\}P(G_i = g | G + t \leq \mathcal{T})ATT(g, g+t)
\end{align}

In order to be able to identify the ATT, the following identifying assumptions are needed.

\textbf{Assumption 1: Limited Treatment Anticipation}
\begin{align*}
    \E[Y_{i,\tau} (g) | X_i, G_{i,g} = 1] = \E[Y_{i,\tau} (0) | X_i, G_{i,g} = 1] \quad \forall g \in \mathcal{G}, \tau  \in \{1, \dots, \mathcal{T}\} \text{ such that }  \tau  < g - \delta
\end{align*}
Assumption 1 implies that the anticipation period $\delta$ is known and set to nine in this analysis as the anticipation period of having a child is around nine months. This assumption might be violated if women select to have a child at a certain point. However, planning the exact date of having a child is not that easy, and there is a certain randomness in the exact birth month. On average, a waiting time of around one year until pregnancy is seen as standard from a medical point of view \citep{BGA:2024}. Hence, the exact birth month can only be anticipated around nine months in advance.

\begin{figure}[h!]
    \centering
    \begin{minipage}{15cm}
    \caption{Descriptives: Hospital insurance}
    \label{fig:hospital_descriptives} 
    \includegraphics[width=15cm]{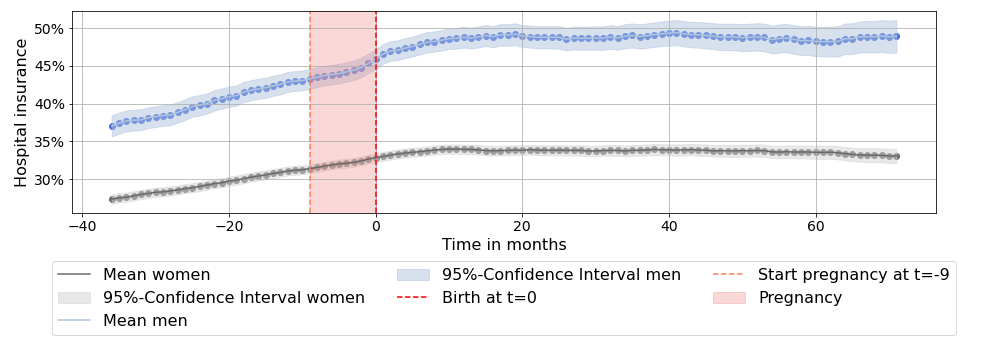}
    \footnotesize \textit{Note:} The figure shows the number of women having an additional hospital insurance around the time of birth ($t = 0$: time of first birth). The red-shaded area illustrates the pregnancy, and the grey-shaded area represents the 95\% confidence interval. ($N = 25,120$)
    \end{minipage}
\end{figure}

An indication of childbirth anticipation would be an increase in individuals signing up for private hospital insurance at a certain point before childbirth. By only having basic insurance, it could happen that a woman has to stay in a room with up to three other people after childbirth. Hospital insurances have a waiting period of at least nine months. Therefore, women must sign up before pregnancy to profit from the insurance. Figure \ref{fig:hospital_descriptives} shows an increase until a year post-childbirth without a break in the trend at a certain point before childbirth.\footnote{An explanation for the increase until after childbirth could be that women sign up to profit from the insurance at the birth of the second child.} This supports the assumption that the exact time of birth cannot be entirely anticipated. Additionally, the registration rate for hospital insurance for men follows a similar pattern despite them not giving birth. This further indicates that the increase in hospital insurance is not due to anticipation of childbirth but due to age.

\textbf{Assumption 2: Conditional Parallel Trends Based on a "Not-Yet-Treated" Group} 
\begin{align*}
    \E[Y_{i,\tau} (0) - Y_{i,\tau -1}(0) | X_i, G_{i,g} = 1] = \E[Y_{i,\tau} (0) - Y_{i,\tau -1}(0) | X_i, D_{i,s} = 0, G_{i,g} = 0] 
\end{align*}
for all $g \in \mathcal{G}$ and each $(s,\tau ) \in \{2, \dots, \mathcal{T}\} \times \{2, \dots, \mathcal{T}\}$  such that $\tau  \geq g - \delta$ and $\tau  + \delta \leq s < \bar g$.
Assumption 2 implies a parallel trend without treatment conditional on some covariates. In this analysis, the only covariate used is the individual's age at first birth. Hence, the assumption states that the trend in the outcome variable (e.g. antidepressant prescriptions) would be the same for women who have a child in a particular month and year and women who do not have a child yet at this specific time, conditional on their age.

This assumption could be violated if the trend in the outcome variable is different for women born at different points in time. For example, if a woman has a child in 2012 and is 30 years old at that time, we compare her in year two after childbirth to a woman who also has a child when she is 30 years old but does not have a child yet (she has no child yet in 2014). Hence, this woman was born later than the one with a child in 2012. The cohorts that are the furthest apart are the ones born in 1985 and 1995. One reason for different trends in mental health could be the different exposure to social media, as it is known that social media leads to worse mental health \citep{Braghieri:2022}. Social media gained popularity in the mid-2000s. Hence, individuals born in 1995 were exposed to social media while being teenagers, whereas individuals born in 1985 were already young adults when it became popular. Similarly, individuals born in 1985 entered the labour market during the financial crisis, whereas those born in 1995 entered the labour market during economic growth. Several papers have shown that the economic conditions when the labour market have long-lasting effects on wages and labour supply, but also on social outcomes such as fertility, marriage and divorce \citep{Wachter:2020, Raaum:2006}. Even though economic conditions and social media might affect the level of mental health, it is less probable that they change the trend over time. Certain life events increasing the risk of mental health problems, such as advancing in a career, starting a family or dealing with ageing, still happen at the same or similar time in life. As long as only the level between the cohorts is different but the trend is the same the assumption still holds.

\begin{figure}[h!]
    \centering
    \begin{minipage}{15cm}
    \caption{Antidepressant prescriptions over the years by birthyear of mother}
    \label{fig:antidepressant_outcome_years_by_group}
    \includegraphics[width=15cm]{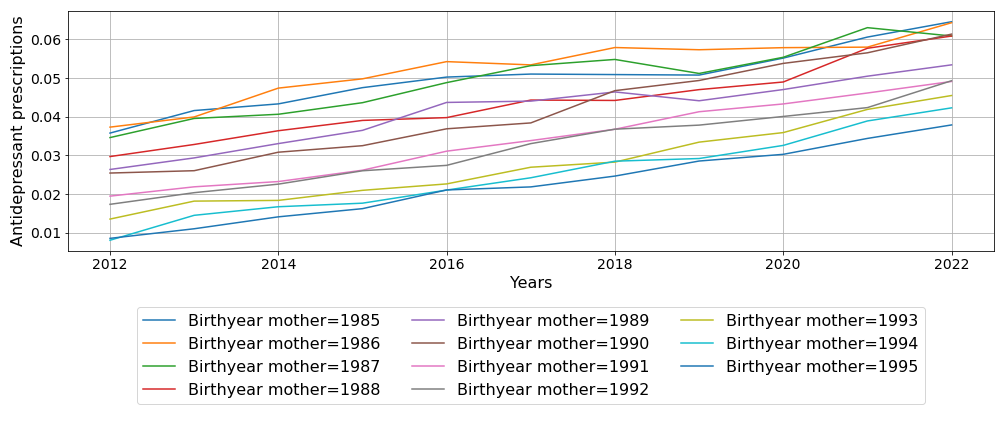}
    \footnotesize \textit{Note:}  This figure shows the percentage of women taking antidepressants over the years (2012 to 2022) grouped by the different birthyears of the women. Only women without children have been used. ($N = 34,252$)
    \end{minipage}
\end{figure}

It is impossible to test this in the data of women who have a child because the child changes the trend. Under the assumption that women who never have a child have comparable trends in the outcome variable, we can look at this group to check if there are different trends between birth cohorts. Figure \ref{fig:antidepressant_outcome_years_by_group} depicts no different trends for different cohorts. Antidepressant prescriptions are increasing over time for all cohorts. For the number of visits to the psychiatrist, the trends between birth cohorts are also similar, however, for the number of visits to a GP or HMO, they are not as clear. The figures for psychiatrist and GP or HMO visits are shown in Appendix \ref{appendix: identification_assumptions}.

Placebo analysis is often used to check the assumption of parallel trends, but this approach is challenging in this study due to the extended time frame we analyse. If we were to conduct a placebo analysis two years before childbirth, the analysis period would already overlap with pregnancy by the second year after childbirth. As a result, we would start detecting effects just one year into the placebo period, since some women would already be pregnant. To avoid this issue, we would need to set the placebo childbirth approximately four years, or even earlier, before the actual childbirth. However, this is not feasible due to the limited years of available data.

 \textbf{Assumption 3: Overlap}
 \begin{align*}
    P(G_{i,g} = 1) > \epsilon,  \quad p_{g,\tau }(X_i) < 1 - \epsilon \quad \text{for some } \epsilon > 0, 
 \end{align*}
with $p_{g,\tau }(X_i) = P(G_{i,g} = 1 | X_i, G_{i,g} + (1 - D_{i,\tau} )(1-G_{i,g}) = 1)$. Assumption 3 states that a positive number of individuals have a child in period g and that the propensity score is bounded away from one for all $g$ and $\tau $. The propensity score indicates the probability of having a child in period g, conditional on some pre-treatment covariates and not already being treated in a period before. So, only individuals treated in period $g$ or not-yet-treated individuals are used to calculate the propensity score. Both conditions are met. If all assumptions hold, we can identify the ATT (Equation \ref{eq:ATT}) as follows by using a double robust score function:

\begin{align*}
    ATT(g,\tau ; \delta) = \E\left[\left(\frac{G_g}{\E[G_{i,g}]} - \frac{\frac{p_{g, \tau + \delta}(X_i)(1-D_{i,\tau +\delta})(1-G_{i,g})}{1-p_{g, \tau + \delta}(X_i)}}{\E\left[\frac{p_{g, \tau + \delta}(X_i)(1-D_{i,\tau +\delta})(1-G_{i,g})}{1-p_{g, \tau +\delta}(X_i)}\right]} \right)(Y_{i,\tau}  - Y_{i,g-\delta-1} - m_{g, \tau , \delta}(X_i))\right]
\end{align*}
with $m_{g,\tau ,\delta}(X_i) = \E[Y_{i,\tau}  - Y_{i,g - \delta - 1} | X_i, D_{i,\tau  + \delta} = 0, G_{i,g} = 0]$. 
This can again easily be averaged to $\theta_{es}(t)$ (Equation \ref{eq:eventstudy_effect}).

The advantage of using a double robust score function is that it only requires one of the two models (outcome regression or propensity score) to be correctly specified \citep{Callaway:2021}. The estimation uses a linear regression to estimate the outcome regression and a logistic model for the propensity score. Uniform confidence bands proposed by \cite{Callaway:2021} are used for inference. In contrast to commonly used pointwise confidence bands, the simultaneous confidence bands offer asymptotic coverage of the entire path of group-time average treatment effects with a fixed probability. These also consider the interdependence between different group-time average treatment effect estimators. The analysis is done with the R package \textit{did} by Callaway and Sant'Anna.\footnote{https://cran.r-project.org/web/packages/did/index.html}

\section{Results} \label{results_main}

\subsection{Average Effects of Childbirth on Mental Health}
\begin{figure}[h!]
    \centering
    \begin{minipage}{15cm}
    \caption{Average effect: Antidepressants prescriptions}
    \label{fig:antidepressants_stagg_diff_diff_balanced}
    \includegraphics[width=15cm]{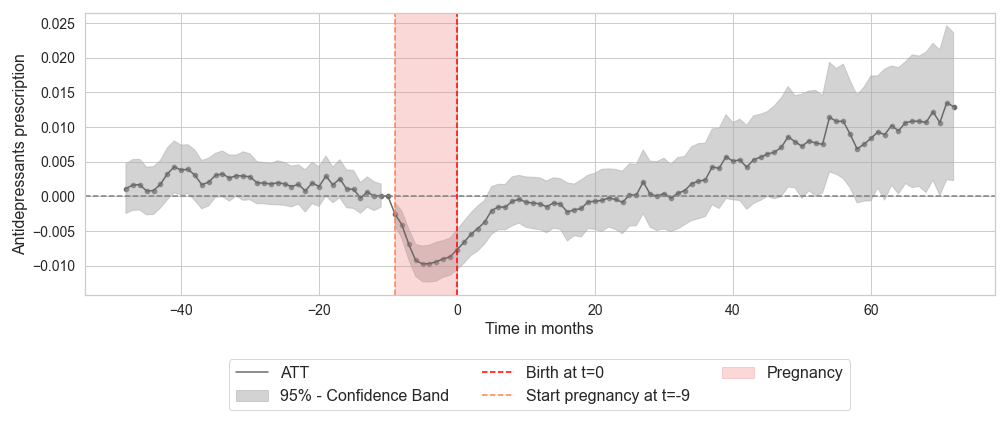}
    \footnotesize \textit{Note:} The figure shows the percentage point increase for women having an antidepressant prescription in comparison to the women who are ``not yet treated''. On average, 1.9\% of women in the sample have an antidepressant prescription before childbirth. The red-shaded area illustrates the pregnancy, and the grey-shaded area represents the 95\% confidence band. The slow decrease in antidepressants at the start of the pregnancy and also the slow increase after giving birth are partly coming from the smoothing of the antidepressant prescriptions. ($N= 25,120$)
    \end{minipage}
\end{figure}

Figure \ref{fig:antidepressants_stagg_diff_diff_balanced} illustrates the estimated effects of childbirth on the probability of having an antidepressant prescription over time. The red-shaded area represents the pregnancy period, culminating in the birth of the first child at $t=0$, while the grey-shaded area shows the 95\% confidence band.\footnote{The critical value for a 95\% uniform confidence band is 2.78 instead of 1.96.} There is a notable increase in antidepressant prescriptions for women, rising by one p.p four years after childbirth. Given that the average number of women with an antidepressant prescription one year before giving birth is around two percent, this one p.p. increase represents a substantial 50\% rise. This upward trend continues, reaching 3.3\% of women having an antidepressant prescription six years post-childbirth, which accounts for a 75\% increase. There is no observed effect in the years before birth, suggesting that the parallel trend assumption before the first child's birth is not violated. Additionally, a decrease in antidepressant prescriptions is visible during pregnancy and breastfeeding, likely due to the potential transmission of antidepressants through blood or milk to the child. Therefore, taking antidepressants during pregnancy and breastfeeding is not medically recommended. The flat trend in antidepressant prescriptions from year one to year three after childbirth can be attributed to a substantial number of women having a second child during this period. Similar to the first pregnancy, the second is associated with a reduction in antidepressant use during pregnancy and breastfeeding. When examining the effect of a second child, we observe a direct increase in antidepressant prescriptions following the end of the second breastfeeding period (see Figure \ref{fig:antidepressants_stagg_diff_diff_second_birth} in Appendix \ref{appendix: heterogeneity_analysis}).

\begin{figure}[h!]
    \centering
    \begin{minipage}{15cm}
    \caption{Average effect: Number of visits to psychiatrist}
    \label{fig:Psychiater_stagg_diff_diff_balanced}
    \includegraphics[width=15cm]{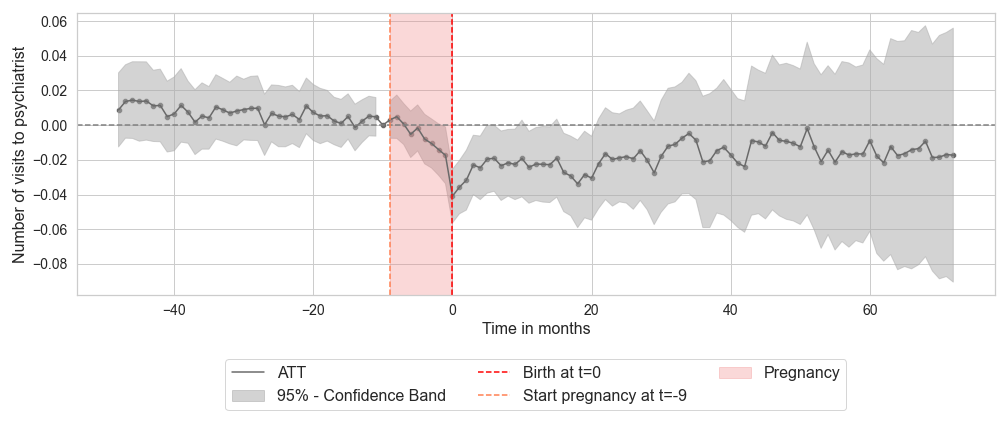}
    \footnotesize \textit{Note:} The figure shows the increase in the number of visits to the psychiatrist compared to the women and men who are ``not yet treated''. On average, women go to a psychiatrist 0.12 times per month before childbirth. The red-shaded area illustrates the pregnancy, and the grey-shaded area represents the 95\% confidence band.  ($N= 25,120$)
    \end{minipage}
\end{figure}

Figure \ref{fig:Psychiater_stagg_diff_diff_balanced} shows no increase in the number of visits to a psychiatrist post-childbirth. There is a decrease during pregnancy, especially during the first months after childbirth, which might be explained by a lack of time. Hence, visiting a psychiatrist might not be a priority. Moreover, there are no signs of a violation of the parallel trend assumption.

\begin{figure}[h!]
    \centering
    \begin{minipage}{15cm}
    \caption{Average effect: Number of visits to GP or HMO}
    \label{fig:Hausarzt_stagg_diff_diff_balanced}
    \includegraphics[width=15cm]{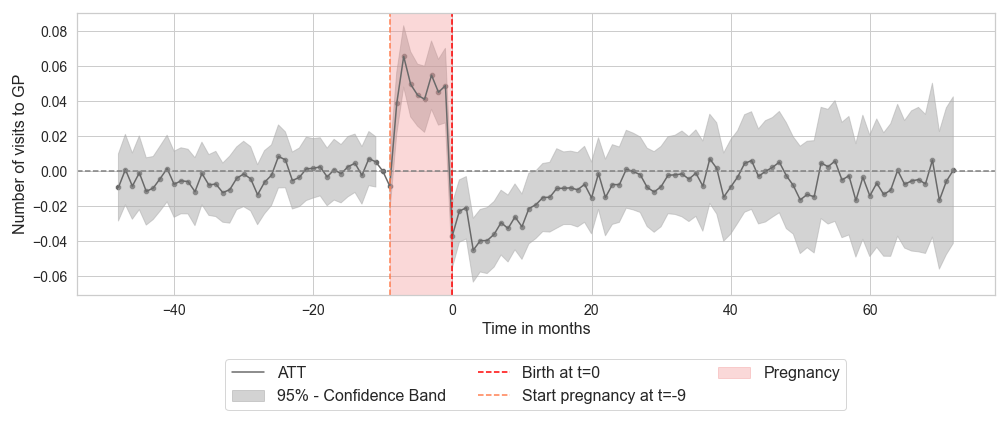}
    \footnotesize \textit{Note:} The figure shows the increase in the number of visits to the GP or HMO compared to the women who are ``not yet treated''. On average, women go to a GP or HMO centre approximately 0.19 times per month before childbirth. The red-shaded area illustrates the pregnancy, and the grey-shaded area represents the 95\% confidence band. ($N= 16,368$)
    \end{minipage}
\end{figure}

Figure \ref{fig:Hausarzt_stagg_diff_diff_balanced} shows the number of visits to a GP or HMO center. As noted in Section \ref{section: outcone_variables}, the number of visits to a GP can only be estimated for individuals insured at CSS, leading to fewer observations for this analysis. The increase during pregnancy can be explained by misclassified visits to a gynaecologist located at an HMO centre, as we combine GP and HMO centres due to many GPs working in HMO centres.\footnote{Another possible explanation is that women may visit their GP for less severe cases, as general medical and nursing services are provided free of charge from the 13th week of pregnancy until eight weeks postpartum \citep{BAG:2024}. However, since we do not observe an increase in visits after childbirth, despite the continued free access, this explanation seems less plausible.} There is a decrease in the number of visits in the first months after childbirth, which can probably again be explained by some time constraints shortly after childbirth. Similarly to the number of visits to a psychiatrist, there is no increase in the number of visits to a GP or HMO center after four or six years. Again, no signs of a violation of the parallel trend assumption are visible.

\subsection{Robustness Checks}

We check if the results are robust to different sample definitions. As we construct a balanced sample of women we can follow from 2012 to 2022, we lose many observations, which might lead to a particular subpopulation of individuals who do not change health insurance. Hence, the first robustness check is to see if we find similar results for an unbalanced sample. Second, there was a worldwide pandemic in 2020 and 2021, which might have affected the results. Therefore, we exclude these years from the analysis.

\subsubsection{Unbalanced Sample} \label{robustness:unbalanced_sample}

Using an unbalanced sample increases the number of observations from $N = 25,120$ to $N = 83,841$, which leads to more power at the cost of having a more considerable change in the sample over time and misclassifying more second births as first births. 

Antidepressant prescriptions show a 0.5 p.p. increase two years after childbirth, reaching a 1.5 p.p. increase after four years and rising by 1.7 p.p. six years post-childbirth. This represents more than a 100\% increase compared to pre-childbirth levels.\footnote{In the unbalanced sample, only 1.4\% of women had an antidepressant prescription before childbirth.} Due to the larger number of observations, the effect estimates become statistically significant one year post-childbirth. Interestingly, also the number of visits to psychiatrists increases approximately two years after birth. After four years, there is an increase of 0.07 visits per month, and after six years, this rises to 0.1 visits more per month — slightly more than a 100\% increase compared to before childbirth.\footnote{In the unbalanced sample, women visited psychiatrists 0.09 times per month before childbirth.} This increase is different from the finding in the balanced sample. Consistent with the balanced sample, the unbalanced sample shows no increase in GP or HMO visits. There is an increase during pregnancy, a decrease in the first months after childbirth and a gradual return to pre-childbirth levels.

The figures illustrating these results are in Appendix \ref{staggered_diff_diff_robustness_checks}. In conclusion, the results are robust for the antidepressant prescriptions and the number of visits to the GP or HMO when using a balanced or an unbalanced sample. The results for the visits to the psychiatrist do change when using an unbalanced sample. 

\subsubsection{Excluding Years in Corona Pandemic}

The worldwide pandemic in 2020 and 2021 affected Switzerland, making visiting a psychiatrist, GP or HMO more challenging due to contact restrictions. To account for potential biases caused by the pandemic, we exclude 2020 and 2021 from the analysis. Due to the shorter period, we can only look at four years after childbirth.

Antidepressant prescriptions increase by one p.p. four years after childbirth, similar to the results that include the pandemic years. However, the number of visits to the psychiatrist increases by 0.05 visits four years after childbirth, representing a 50\% increase compared to before childbirth. This finding differs from the analysis that extends until 2022, which shows no effect on the number of visits to the psychiatrist when the pandemic years are included. Despite this increase not being statistically significantly different from zero, it might point to an influence of the pandemic on the number of visits to a psychiatrist. We do not find any statistically significant effects in the number of visits to a GP or HMO in either sample, including or excluding the pandemic years. 

All relevant graphs are shown in Appendix \ref{appendix: longer_time_period}. Summing up, we again find that the results for antidepressant prescriptions and visits to GPs or HMOs are robust to excluding the pandemic years, while the results for visits to psychiatrists are not.

\subsection{Effect of Having a Child on Mental Health of Men} \label{effect_men}

Not only women but also men can suffer from postpartum depression \citep[e.g.][]{Bradley:2011}. \cite{Ahammer:2023} find that a child also leads to an increase in antidepressant prescriptions for men. As we can only identify fathers being insured at the same insurance company as the mother, the sample size for men is small, and the sample might not be representative of the Swiss population.\footnote{We have 5,749 fathers that we can identify and that are insured at the same company from 2012 to 2022.\\
See Appendix \ref{appendix: data_descriptives} for a check on the representativeness of the sample.} Descriptive statistics and the results of the staggered difference-in-difference analysis are shown in Appendix \ref{appendix: effect_on_men}.

In general, men have fewer antidepressant prescriptions and are less often going to a psychiatrist than women, but the antidepressant prescriptions and visits to a psychiatrist also increase over time. Looking at the effect of the first child, we do not find any effect of a first child on antidepressant prescriptions, visits to psychiatrists, GPs or HMOs for men if using the balanced sample constructed in the same way as for women. This is different from the findings of \cite{Ahammer:2023}, who also find an increase in antidepressant prescriptions for men.

\subsection{Heterogeneous Effects}\label{staggered_diff_diff_heterogeneities}

This section examines heterogeneous treatment effects, as the average effect might mask different effects across groups. The literature suggests that stress and sleep deprivation may mediate the mental health impact after childbirth. Employed women who juggle both work and household responsibilities may experience a greater mental health penalty due to a higher overall workload. Similarly, women who undergo cesarean sections face longer recovery times, potentially leading to more stress. Lower-income women may also face greater financial stress, exacerbating the mental health burden. The heterogeneity analysis focuses on antidepressant prescriptions as the primary outcome. The heterogeneities for the number of visits to the psychiatrist and the GP or HMO and additional heterogeneities by language, nationality, age at first birth or number of children are shown in Appendix \ref{appendix: heterogeneity_analysis}.

\subsubsection{Employment} \label{staggered_diff_diff_accident}

As explained in Section \ref{section: covariates}, not having accident insurance is a good proxy for being employed. As noted in Section \ref{family_policies}, the percentage of women working decreases following childbirth. Despite equality measures, 10\% of the women employed at childbirth leave the labour force in the first two years after childbirth, and many women transition to part-time work and reduce their work hours, whereas few fathers do the same. This disparity may be due to womens' desire to spend more time with their children or the high costs of childcare in combination with a gender pay gap. However, even when women outearn their husbands, they spend more time doing household chores \citep*{Bertrand:2015}. Therefore, mothers who continue to be employed after childbirth face a higher workload as they manage both home and work responsibilities, which could negatively impact their mental health. Hence, if part of the mental health penalty arises from a change in circumstances, such as increased stress and sleep deprivation, the mental health penalty should be higher for women who were employed before having a child and, hence, are more likely to be employed after childbirth.

\begin{figure}[h!]
    \centering
    \begin{minipage}{15cm}
    \caption{Heterogeneity by employment status: Antidepressants prescriptions}
    \label{fig:antidepressants_stagg_diff_diff_accident}
    \includegraphics[width=15cm]{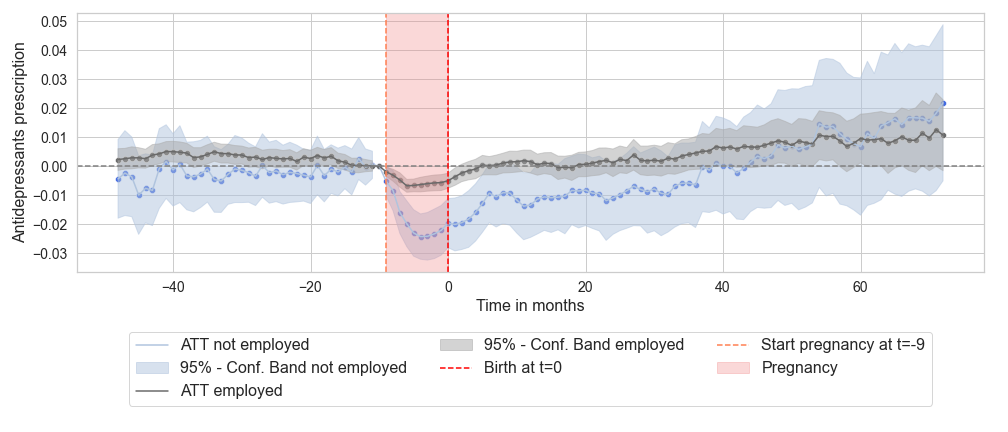}
    \footnotesize \textit{Note:} The figure shows the percentage point change of women taking antidepressants in comparison to the women who do not have a child yet. On average, 1.4\% of women that are employed and 4\% of women that are not employed have an antidepressant prescription before childbirth. The red-shaded area illustrates the pregnancy, and the grey-shaded area represents the 95\% confidence band. The slow decrease in antidepressants at the start of the pregnancy and also the slow increase after giving birth are partly coming from the smoothing of the antidepressant prescriptions. (Not employed: $N=4,927$, employed: $N=20,193$)
    \end{minipage}
\end{figure}

Figure \ref{fig:antidepressants_stagg_diff_diff_accident} shows the change in antidepressant prescriptions for the sample of women employed and not employed at the time of childbirth.

There is a statistically significant increase of 1 p.p. after four years for women employed, which is a 70\% increase compared to before childbirth.\footnote{Only 1.4\% of employed women had an antidepressant prescription ten months before childbirth.} This further increases to a 1.2 p.p. increase after six years. Conversely, there is no increase for women not employed before childbirth until the fourth year after childbirth. Afterwards, they catch up, and six years after childbirth, they encounter an increase of 2 p.p, which is a 50\% increase compared to before childbirth.\footnote{4\% of women not employed have an antidepressant description ten months before childbirth. Hence, some of the women who do not work before childbirth do this due to mental health problems.} Women who are not employed are not more likely to have a second pregnancy. Hence, the lower mental health penalty for not employed women cannot solely be explained by differences in antidepressant use due to a subsequent pregnancy. In line with the expectations, there is an indication that the mental health penalty is higher for employed women (measured in percent) than for women who are not employed.

\subsubsection{Modes of Child Delivery}

The insurance data available shows if a woman has a cesarean section or a vaginal delivery. A cesarean section is a major surgery that leads to a longer recovery time. There has already been research on this topic in the medical and epidemiological literature looking at associations between mode of delivery and postpartum depression. Different meta-analyses conclude that emergency and planned cesarean sections increase the risk for postpartum depression. However, emergency cesarean sections increase it more \citep*{Xu:2017, Moameri:2019}. We are not able to differentiate between emergency and planned cesarean sections, but we can compare cesarean sections to vaginal deliveries.\footnote{Cesarean sections increased over time in Switzerland due to the increased age of women giving birth \citep{SRF:2020a}. This has also been seen in other countries. However, other factors, such as employment, also increase the demand for cesarean sections \citep{Grant:2022}. Unfortunately, we are not able to further explore the planned cesarean section, as we cannot identify them.}

\begin{figure}[h!]
    \centering
    \begin{minipage}{15cm}
    \caption{Heterogeneity by mode of delivery: Antidepressants prescriptions}
    \label{fig:antidepressants_stagg_diff_diff_Kaiserschnitt}
    \includegraphics[width=15cm]{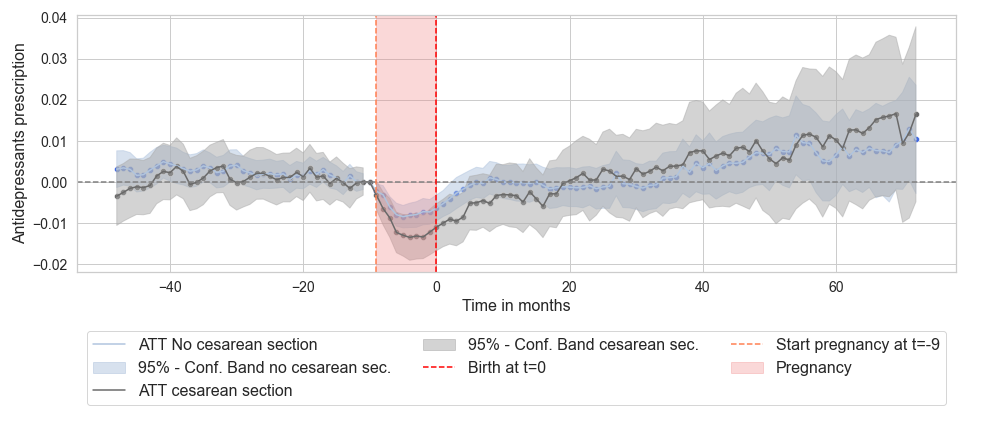}
    \footnotesize \textit{Note:} The figure shows the percentage point change of women taking antidepressants in comparison to the women who do not have a child yet. On average, 2.4\% of women having a cesarean section and 1.7\% having a vaginal delivery have an antidepressant prescription before childbirth. The red-shaded area illustrates the pregnancy, and the grey-shaded area represents the 95\% confidence band. The slow decrease in antidepressants at the start of the pregnancy and also the slow increase after giving birth are partly coming from the smoothing of the antidepressant prescriptions. (Cesarean section: $N=7,396$, no cesarean section: $N=17,724$)
    \end{minipage}
\end{figure}

Figure \ref{fig:antidepressants_stagg_diff_diff_Kaiserschnitt} shows only a slight difference in the increase in antidepressant prescriptions by the mode of delivery. 

The increase in antidepressant prescriptions lies at around 1.5 p.p. six years post-childbirth for women having a cesarean section ($\sim$ 60\% increase) and at one p.p. for women having a vaginal delivery ($\sim$ 60\% increase). Women having a cesarean section have a higher prevalence of taking antidepressants before childbirth. Therefore, women who have more fears before childbirth are probably more likely to select a cesarean section, which could partly explain the association between cesarean sections and postpartum depression found in other studies, as most studies only control for age and complications during pregnancy but not for pre-pregnancy depression.

\subsubsection{Getting Subsidies for Health Insurance and Income index}

As noted in Section \ref{1_health_insurance}, the health insurance costs do not depend on a person's income. Therefore, some people with low incomes get subsidies for their health insurance. Whether someone gets a subsidy or not can be used as a proxy for a family's income. Additionally, we have an income index available for the individuals in the CSS data. Hence, we can split the sample into individuals with an income below and above the median. Because there is some evidence that women with low socioeconomic status are more likely to suffer from mental health problems after childbirth \citep[e.g.][]{Goyal:2010, Rich:2006}, we would expect a larger mental health penalty for individuals receiving a subsidy and having an income below the median. We look at the year before childbirth to assess whether someone receives a subsidy and if someone's income is above or below the median.

\begin{figure}[h!]
    \centering
    \begin{minipage}{14cm}
    \caption{Heterogeneity by health subsidies: Antidepressants prescriptions}
    \label{fig:antidepressants_stagg_diff_diff_subsidy}
    \includegraphics[width=14cm]{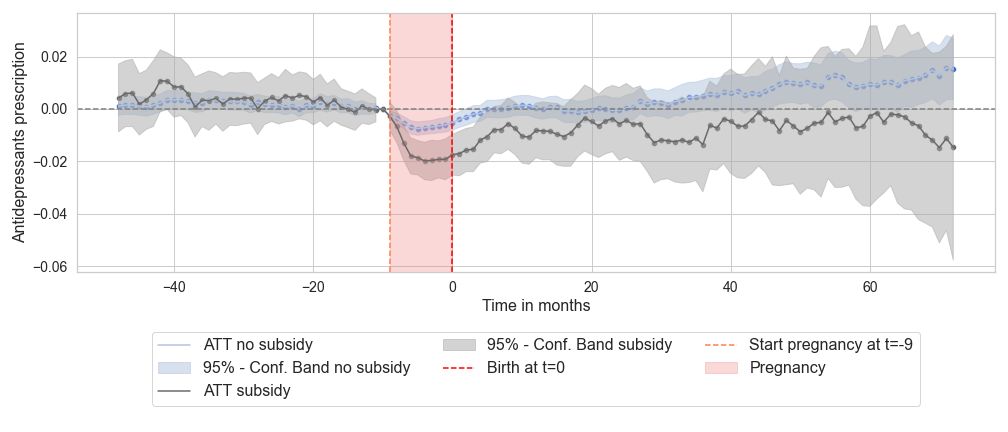}
    \footnotesize \textit{Note:} The figure shows the percentage point change of women taking antidepressants in comparison to the women who do not have a child yet. On average, 3.3\% of women receiving health subsidies and 1.5\% not receiving health subsidies have an antidepressant prescription before childbirth. The red-shaded area illustrates the pregnancy, and the grey-shaded area represents the 95\% confidence band. The slow decrease in antidepressants at the start of the pregnancy and also the slow increase after giving birth are partly coming from the smoothing of the antidepressant prescriptions. (Subsidy: $N=4,652$, no subsidy: $N=20,468$)
    \end{minipage}
\end{figure}

Figure \ref{fig:antidepressants_stagg_diff_diff_subsidy} shows the change in antidepressant prescriptions for the sample of women receiving and not receiving health subsidies at the time of childbirth.

Before childbirth, approximately 3.3\% of women receiving a health subsidy have an antidepressant prescription, compared to only 1.5\% of women not receiving a subsidy. Hence, the mental health of women with a lower socioeconomic status is worse before childbirth. After childbirth, there is a larger mental health penalty for women who do not receive health subsidies, namely an 1.5 p.p. increase in antidepressant prescriptions six years after childbirth, which is a 100\% increase compared to before childbirth.

Looking at the income index, we draw similar conclusions. However, the results are not statistically significantly different from zero due to the smaller sample sizes. Disregarding the non-significance, the mental health penalty is higher for individuals with an income above the median, which is in line with the results using the health subsidy as a proxy. 

\begin{figure}[h!]
    \centering
    \begin{minipage}{14cm}
    \caption{Heterogeneity by income: Antidepressants prescriptions}
    \label{fig:antidepressants_stagg_diff_diff_income_index}
    \includegraphics[width=14cm]{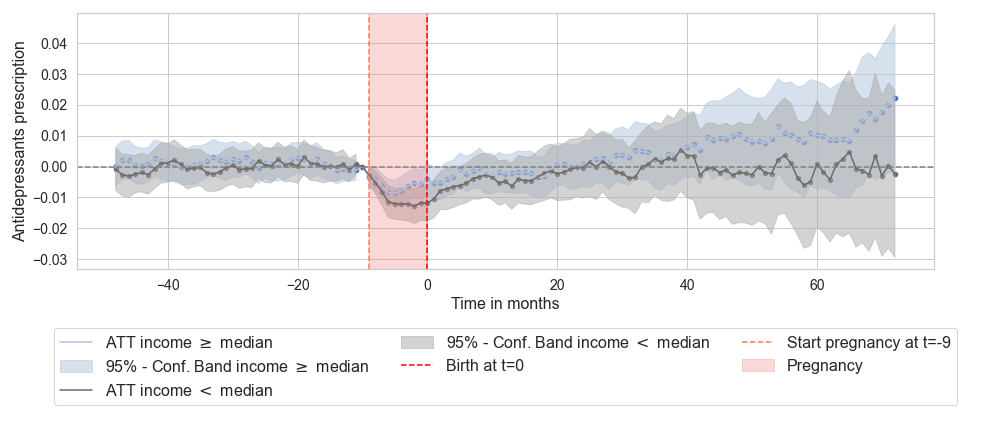}
    \footnotesize \textit{Note:} The figure shows the percentage point change of women taking antidepressants in comparison to the women who do not have a child yet. On average, 1.7\% of women having an income above the median and 1.9\% of women having an income below the median have an antidepressant prescription before childbirth.  The red-shaded area illustrates the pregnancy, and the grey-shaded area represents the 95\% confidence band. The slow decrease in antidepressants at the start of the pregnancy and also the slow increase after giving birth are partly coming from the smoothing of the antidepressant prescriptions. ($\geq$ median: $N = 6, 950$, $<$ median: $N = 6, 943$)
    \end{minipage}
\end{figure}

Interestingly, this result is not in line with the literature, which suggests that individuals with a lower socioeconomic status have a higher mental health penalty after childbirth. As receiving health subsidies is highly correlated with not being employed, an explanation could be that the mental health penalty does not come from income but from employment status. This explanation is supported by Figures \ref{fig:antidepressants_descriptives_health_subsidy} and \ref{fig:antidepressants_descriptives_health_subsidy_employed}  in Appendix \ref{appendix: heterogeneity_analysis}.\footnote{Due to the small sample size, the results are only descriptive.} If we only look at individuals that are employed, there is no difference in the mental health penalty between individuals receiving health subsidies and individuals not receiving health subsidies. Hence, employment status can explain this unexpected result.

\section{Discussion and Potential Mechanisms} \label{discussion}

We conclude that there is an increase in antidepressant prescriptions for women in the years after childbirth. There is an increase of 50\% four years after childbirth. This is a substantial economic effect, as mental illnesses lead to high healthcare costs, and good health is a prerequisite for labour market participation. This effect is similar to the one found by \cite{Ahammer:2023}. They find a two p.p. and one p.p. increase four years after childbirth in Austria and Denmark, which is approximately a 30\% increase.\footnote{In Austria, 6.62\% of women have an antidepressant prescription two years before birth, and in Denmark, 3.23\%.} The mental health penalty further increases over time and reaches a 75\% increase six years post-childbirth. Similarly, a 50\% increase in Austria and a 70\% increase in Denmark were found \citep{Ahammer:2023}.

The results for the number of visits to the psychiatrist are inconclusive. While some specifications suggest a slight positive trend in psychiatric visits, the main specification shows no statistically significant effect. This indicates that women with limited time, particularly those with young children, might opt for antidepressants instead of regularly visiting a psychiatrist. There is also no observed increase in visits to the GP after childbirth, ruling out the possibility that the increased number of GP visits contributes to a higher detection rate of depression. In conclusion, the evidence indicates a substantial mental health penalty for women but not for men after having a child. The next step is to look at the potential mechanisms underlying this mental health penalty.

As noted by \cite{Ahammer:2023}, the decline in womens' mental health after childbirth could be attributed to the biological effects of childbirth or the responsibilities of raising a child. Evidence supporting the latter is found in the pattern of antidepressant prescriptions, which begin to increase two years after childbirth and continue to rise over time rather than immediately postpartum. This suggests that the worsening mental health is a combination of postpartum depression and the accumulating mental and physical burden women face after becoming mothers. Similarly, as shown in Section \ref{section: outcone_variables}, first-time prescriptions stabilise at similar levels in the years following childbirth as they do in the first few months post childbirth. Hence, the increase can not only be attributed to women taking antidepressants directly after childbirth and then developing a drug addiction.

Part of the mental health penalty likely stems from an increased work burden after childbirth, as women still carry the primary responsibilities for childcare and household duties. This higher burden could be a key mechanism linking parenthood to deteriorating mental health. \cite{Ruppanner:2019} suggests that time pressure diminishes mental health by reducing sleep and leisure time. In particular, mothers suffer from poor sleep quality and quantity, often prioritising family needs over self-care \citep*{Craig:2017, Maume:2010}. \cite{Dehos:2024} also support this argument in a complementary analysis by showing that reasons for the increase in antidepressant prescriptions are most likely less sleep, sport and other leisure activities combined with extensive involvement in childcare. The more pronounced mental health effects observed in women who worked before having a child and are also more likely to work after childbirth support this hypothesis. These women face responsibilities both at home and in the workplace, leading to a double burden that exacerbates mental health challenges.

Mental health problems could be tackled by reducing the time pressure women face when they have children. Some possibilities from the government's perspective are introducing cheaper childcare or longer maternal or paternal leave. Which policy would be most effective is country specific and should be evaluated carefully. For example, \cite{Chuard:2023} or \cite{Ahammer:2023} point out that too long maternal leave had adverse effects in Austria when they expanded the maximum duration of maternity leave from 18 to 30 months. Switzerland has only a 14-week maternity leave, one of the shortest in Europe. Hence, an increase in maternity leave in Switzerland would likely not have the same consequences on mental health as the parental leave reform in 2000 in Austria. There is also some evidence that paternity leave leads to an increase in fathers helping with childcare and homework \citep[e.g.][]{Bunning:2015, Gonzalez:2021, Kotsadam:2011}, which could reduce the mental and physical burden for the mother. As explained in Section \ref{family_policies}, a two-week paternal leave was introduced in 2021 in Switzerland. We use this natural experiment to look at the effect of this policy on the mental health of mothers but do not find a statistically significant effect on the 5\% significance level.

This non-significant result can be interpreted in different ways. First, the small number of births per month and the low number of women taking antidepressants around childbirth lead to low power. Hence, the confidence intervals are large, and we cannot detect statistically significant effects. Second, a two-week paternity leave might have no effect as it is relatively short, but we cannot claim to have found a null result due to the lack of power. Moreover, an effect might only be visible in the medium or long-run, as not many women take antidepressants only two years post-childbirth. The extensive analysis can be found in Appendix \ref{appendix: paternal_leave}.
 
\section{Conclusion}  \label{1_conclusion}

This paper analyses the effect of childbirth on the mental health of women in Switzerland. A staggered difference-in-difference approach is used to estimate the effect of childbirth on the probability of having an antidepressant prescription, the number of visits to a psychiatrist, and the number of visits to a GP or HMO. The probability of having an antidepressant prescription increases by one p.p. four years after childbirth, which is a 50\% increase compared to before childbirth, and by 1.7 p.p six years post-childbirth. We do not find a statistically significant effect on the number of visits to a psychiatrist and GP or HMO.

The mental health penalty is higher for women that are employed. This supports the hypothesis that part of the mental health penalty comes from a change of circumstances after childbirth and not only from the biological effect. Moreover, women not receiving health subsidies (or having an income above the median) have a higher mental health penalty than women receiving health subsidies (or having an income below the median). However, this difference can most likely be explained by the strong correlation between receiving mental health subsidies and not being employed. Additionally, there is no difference in the mental health penalty between women having a cesarean section and a vaginal delivery.

Last, we analyse the introduction of paternity leave in Switzerland and do not find any statistically significant effect on mothers' antidepressant prescriptions. Due to relatively larger confidence intervals, it is also impossible to claim that there is no effect.

A limitation of this paper is the length of the data. We cannot examine time horizons longer than six years after childbirth, which makes it impossible to analyse the mental health penalty in the long-term. Similarly, we are not able to consistently identify the first time an individual has an antidepressant prescription. Therefore, future research with data that spans more years should be conducted.  Moreover, diving deeper into the potential mechanisms for the mental health penalty and further disentangling the heterogeneous treatment effects, as shown by \cite{Bearth:2024} in a selection-on-observables setting, would be beneficial in further pining down policy recommendations. Last, it would be interesting to analyse the effect of the new paternal leave policy in Switzerland on the medium to long-term mental health of mothers, as the mental health penalty increases over the years post-childbirth.

\newpage
\bibliographystyle{apacite}
\bibliography{library}
\newpage
\begin{appendices}
\renewcommand{\theequation}{\thesection\arabic{equation}}
\setcounter{equation}{0}
\renewcommand{\thelemma}{\thesection\arabic{lemma}}
\setcounter{lemma}{0}
\renewcommand{\thetable}{\thesection\arabic{table}}
\setcounter{table}{0}
\renewcommand{\thefigure}{\thesection\arabic{figure}}
\setcounter{figure}{0}

\section{Appendix: Child Penalty Staggered Difference-in-Difference} \label{appendix: staggered_diff_diff}

\subsection{Data Descriptives} \label{appendix: data_descriptives}

Table \ref{Table:bund_female_population} indicates that the data represents the Swiss population to some extent. Compared to the insurance data, the average female in Switzerland is slightly less likely to be Swiss, similarly likely to be employed and more likely to be married. This discrepancy in civil status arises because the insurance companies only record the civil status at the initial sign-up, meaning more women are probably married by the time of birth. Notably, the civil status of 20\% of women in the insurance data is unknown. Due to sample restrictions, the average Swiss woman is older at the time of her first birth compared to the average woman in the insurance data. Lastly, the region lemanique is underrepresented in the insurance data, whereas Eastern and Central Switzerland are overrepresented.

\begin{table}[h!]\centering
    \caption{Descriptives: Female population in Switzerland}
    \label{Table:bund_female_population}
    \begin{adjustbox}{max width=0.5\textwidth}
    \begin{threeparttable}
    \begin{tabular}{lrr} \toprule
        & percentage of mothers & percentage of fathers \\\midrule
    \textbf{Nationality Regions}\\
    EU or EFTA & 0.21 & 0.23\\
Switzerland & 0.62 & 0.62\\
other & 0.09 & 0.07\\
other European country & 0.08 & 0.08\\
 \\
    \textbf{Civil status} \\
    divorced & 0.03 & 0.03\\
single & 0.24 & 0.20\\
married & 0.73 & 0.77\\
 \\
    \textbf{Region} \\
    Espace Mittelland & 0.21&\\
Region lemanique & 0.20&\\
Zurich & 0.19&\\
Eastern Switzerland & 0.14&\\
Northwestern Switzerland & 0.13&\\
Central Switzerland & 0.10&\\
Ticino & 0.03&\\
 \\ 
    \textbf{Employed} \\
    employed & 0.80 & 0.92\\
not employed & 0.20 & 0.08\\
 \\ 
    \textbf{Age first birth} & 31.7 & 34.5 \\ \bottomrule
    \end{tabular}
    \begin{tablenotes}
        \small \item \textit{Note:} The table shows some covariates for the Swiss female population and uses administrative data from the Swiss government, namely income compensation data (ZAS - EO), population and household statistics (BFS - STATPOP), vital statistics (BFS - BEVNAT) and social security data (ZAS - IK).
    \end{tablenotes}
    \end{threeparttable}
    \end{adjustbox}
\end{table}

The selected male sample from the insurance is quite different compared to the full population. The number of Swiss individuals is significantly lower in the full population than in the selected sample. Also, the average age at first birth is 5 years higher in the full population than in the selected sample of fathers from the insurance companies.\footnote{The descriptive statistics for the insurance data for men can be found in Table \ref{Table:samples_staggered_diff_diff_men} in Appendix \ref{covariates_men}.} Therefore, the selected sample of fathers is not representative of the Swiss population, and we do not focus on them in this paper.

\clearpage

\subsection{Identification Assumptions} \label{appendix: identification_assumptions}

\begin{figure}[h!]
    \centering
    \begin{minipage}{15cm}
    \caption{Number of visits to psychiatrist over the years by birthyear of mother}
    \label{fig:Psychiater_visits_outcome_years_by_group_mother}
    \includegraphics[width=15cm]{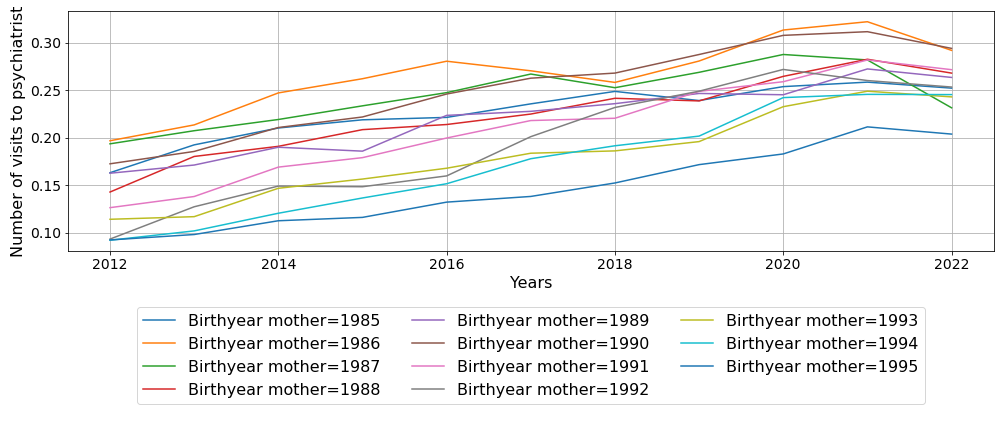}
    \footnotesize \textit{Note:}  This figure shows the number of visits to a psychiatrist for women over the years (2012 to 2022) grouped by the different birthyears of the women. Only women without children have been used. ($N = 31,612$)
    \end{minipage}
\end{figure}

Figure \ref{fig:Psychiater_visits_outcome_years_by_group_mother} shows the parallel trends for women not having a child between 2012 and 2022 across different birthyear cohorts for visits to the psychiatrist.

\begin{figure}[h!]
    \centering
    \begin{minipage}{15cm}
    \caption{Number of visits to GP or HMO over the years by birthyear of mother}
    \label{fig:GP_visits_outcome_years_by_group_mother}
    \includegraphics[width=15cm]{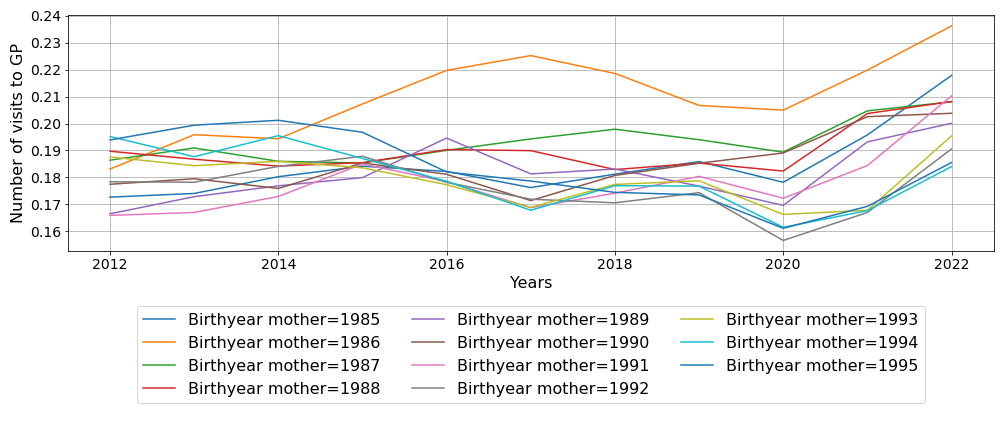}
    \footnotesize \textit{Note:}  This figure shows the number of visits to a GP or HMO for women over the years (2012 to 2022) grouped by the different birthyears of the women. Only women without children have been used. ($N = 31,612$)
    \end{minipage}
\end{figure}

Figure \ref{fig:GP_visits_outcome_years_by_group_mother} shows the parallel trends  for visits to the GP or HMO across different birthyear cohorts of women who do not have a child between 2012 and 2022.

\newpage
\subsection{Robustness Checks}\label{staggered_diff_diff_robustness_checks}

\subsubsection{Unbalanced Sample}

Using an unbalanced sample changes the composition of individuals over time, as some switch to different health insurance providers while others join the health insurance company. This leads to more observations, increasing the analysis's power. However, it also leads to a higher risk of misclassifying some second births as first births because we have not seen the whole history since 2010 or 2012.

\begin{figure}[h!]
    \centering
    \begin{minipage}{14cm}
    \caption{Average effect in unbalanced sample: Antidepressants prescriptions}
    \label{fig:antidepressants_stagg_diff_diff_unbalanced}
    \includegraphics[width=14cm]{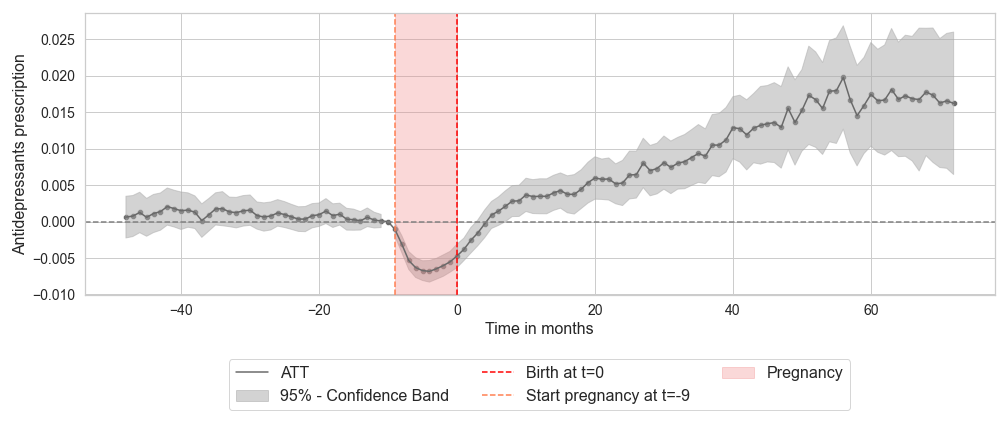}
    \footnotesize \textit{Note:} The figure shows the percentage change of women taking antidepressants compared to women who have not had a child yet. On average, 1.4\% of women in the sample have an antidepressant prescription before childbirth. The red-shaded area illustrates the pregnancy, and the grey-shaded area represents the 95\% confidence band. The slow decrease in antidepressants at the start of the pregnancy and also the slow increase after giving birth are partly coming from the smoothing of the antidepressant prescriptions. ($N= 83,841$)
    \end{minipage}
\end{figure}

Figure \ref{fig:antidepressants_stagg_diff_diff_unbalanced} shows that antidepressant prescriptions increase by 1.5 p.p. after four years. The results are similar to the ones obtained using the balanced sample, but they are statistically significant already one year post-childbirth due to more observations. After six years, there is a 1.7 p.p. increase, which is a 120\% increase compared to before childbirth.

\begin{figure}[h!]
    \centering
    \begin{minipage}{14cm}
    \caption{Average effect in unbalanced sample: Number of visits to psychiatrist}
    \label{fig:Psychiater_stagg_diff_diff_unbalanced}
    \includegraphics[width=14cm]{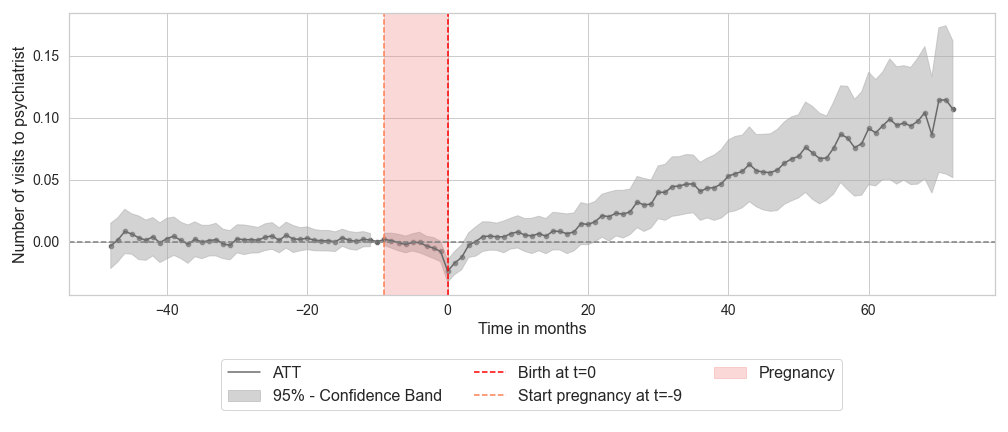}
    \footnotesize \textit{Note:} The figure shows the number of visits to a psychiatrist compared to the women who have not had a child yet. On average, women visit a psychiatrist 0.09 times per month before childbirth. The red-shaded area illustrates the pregnancy, and the grey-shaded area represents the 95\% confidence band. ($N= 83,841$)
    \end{minipage}
\end{figure}

Figure \ref{fig:Psychiater_stagg_diff_diff_unbalanced} depicts a statistically significant increase in the number of visits to the psychiatrist by 0.05 four years after childbirth. Furthermore, after six years, the number of visits to a psychiatrist increases by 0.1, more than a 100\% increase compared to before childbirth. This is a different result than the one obtained with the balanced sample, where we did not find any statistically significant effect on the number of visits to a psychiatrist.

\begin{figure}[h!]
    \centering
    \begin{minipage}{14cm}
    \caption{Average effect in unbalanced sample: Number of visits to GP or HMO}
    \label{fig:Hausarzt_stagg_diff_diff_unbalanced}
    \includegraphics[width=14cm]{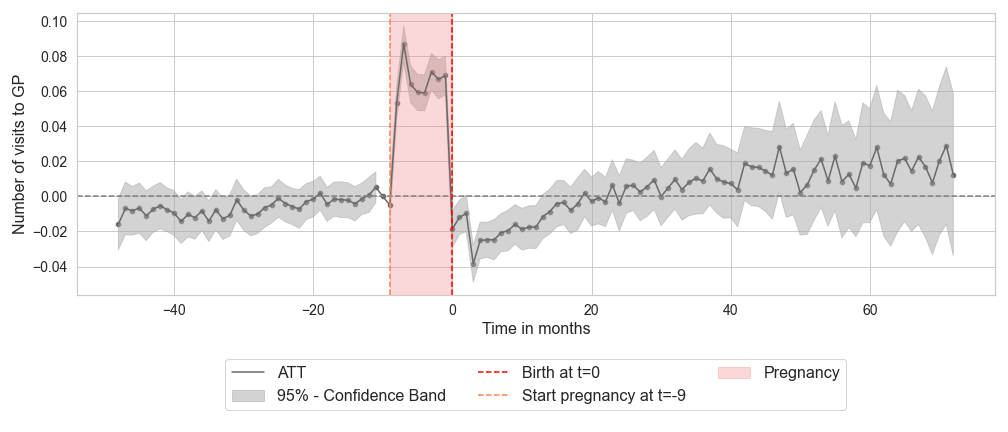}
    \footnotesize \textit{Note:} The figure shows the number of visits to a GP or HMO compared to women who have not had a child yet. On average, women go to a GP or HMO 0.17 times per month before childbirth. The red-shaded area illustrates the pregnancy, and the grey-shaded area represents the 95\% confidence band. ($N= 54,886$)
    \end{minipage}
\end{figure}

Figure \ref{fig:Hausarzt_stagg_diff_diff_unbalanced} shows the effect on the number of visits to a GP or HMO. There is an increase during pregnancy, as some gynaecologists are located at HMOs, which are also included in the GP outcome. We find the same decrease in the first year after childbirth as in the balanced sample. After approximately one year, it reaches the same level as before birth and does not increase further. This result does not differ from the one obtained in the balanced sample.

\subsubsection{Excluding Years in Corona Pandemic} \label{appendix: longer_time_period}

This robustness check shows the results if we disregard the years 2020 and 2021, as the pandemic might have affected the results. Because of the shorter timeframe, we can only look at the first four years after childbirth.

\begin{figure}[h!]
    \centering
    \begin{minipage}{14cm}
    \caption{Average effect without coronavirus period: Antidepressants prescriptions}
    \label{fig:antidepressants_prescription_excluding_corona_period} 
    \includegraphics[width=14cm]{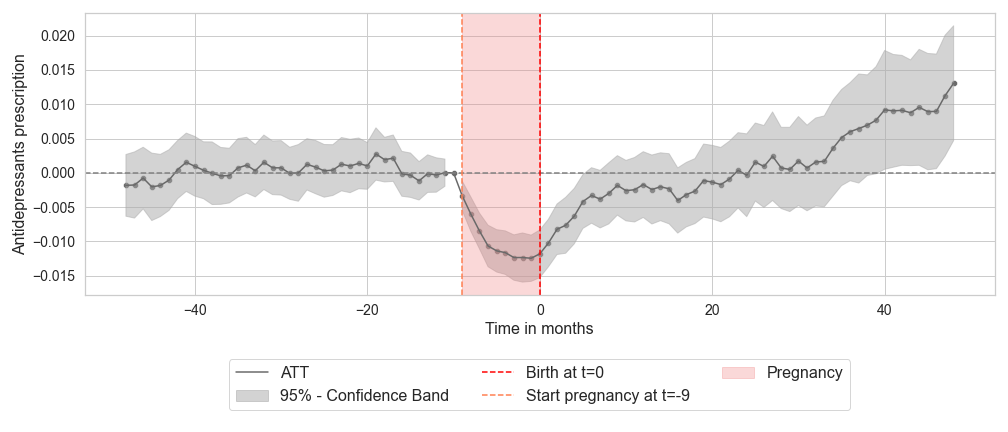}
    \footnotesize \textit{Note:} The figure shows the percentage change of women taking antidepressants compared to women who have not had a child yet. On average, 1.9\% of women in the sample had an antidepressant prescription before childbirth. The red-shaded area illustrates the pregnancy, and the grey-shaded area represents the 95\% confidence band. The slow decrease in antidepressants at the start of the pregnancy and also the slow increase after giving birth are partly coming from the smoothing of the antidepressant prescriptions. ($N= 16,696$)
    \end{minipage}
\end{figure}

Figure \ref{fig:antidepressants_prescription_excluding_corona_period} shows a similar result as including the whole sample. We find a one p.p. increase four years after childbirth for women, which is a 50\% increase compared to before childbirth.

\begin{figure}[h!]
    \centering
    \begin{minipage}{14cm}
    \caption{Average effect without coronavirus period: Number of visits to psychiatrist}
    \label{fig:psychiatrist_excluding_corona_period}
    \includegraphics[width=14cm]{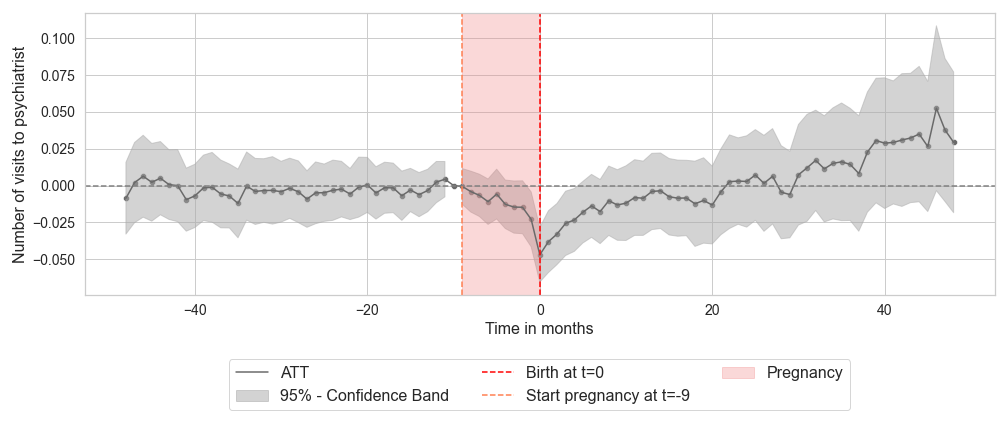}
    \footnotesize \textit{Note:} The figure shows the number of visits to a psychiatrist compared to women and men who have not had a child yet. On average, women visit a psychiatrist 0.1 times per month before childbirth. The red-shaded area illustrates the pregnancy, and the grey-shaded area represents the 95\% confidence band. ($N= 16,696$)
    \end{minipage}
\end{figure}

Figure \ref{fig:psychiatrist_excluding_corona_period} shows the effect on the number of psychiatrist visits. In contrast to the balanced sample, by excluding the coronavirus pandemic years, we see an indication of an effect on the number of visits to a psychiatrist. However, the effect is also not statistically  significantly different from zero.

\begin{figure}[h!]
    \centering
    \begin{minipage}{14cm}
    \caption{Average effect without coronavirus period: Number of visits to GP or HMO}
    \label{fig:Hausarzt_excluding_corona_period}
    \includegraphics[width=14cm]{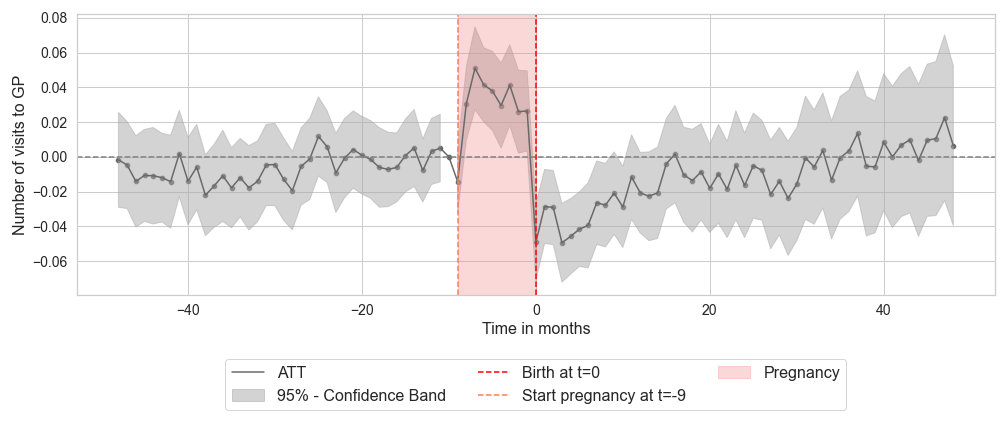}
    \footnotesize \textit{Note:} The figure shows the number of visits to a GP or HMO compared to women who have not had a child yet. On average, women go to a GP or HMO 0.2 times per month before childbirth. The red-shaded area illustrates the pregnancy, and the grey-shaded area represents the 95\% confidence band. ($N= 11,067$)
    \end{minipage}
\end{figure}

Figure \ref{fig:Hausarzt_excluding_corona_period} shows the effect on the number of visits to a GP or HMO. The number of visits to a GP or HMO increases during the pregnancy and decreases in the first month post-childbirth. It reaches the same level as before birth after two years and remains at this level until four years post-childbirth. This result does not differ from the one obtained in the balanced sample.

\subsubsection{Non-smoothed Antidepressant Prescriptions} \label{appendix_non_smoothed}

\begin{figure}[h!]
    \centering
    \begin{minipage}{15cm}
    \caption{Average effect: Antidepressants prescriptions non-smoothed}
    \label{fig:antidepressants_stagg_diff_diff_balanced_non_smoothed}
    \includegraphics[width=15cm]{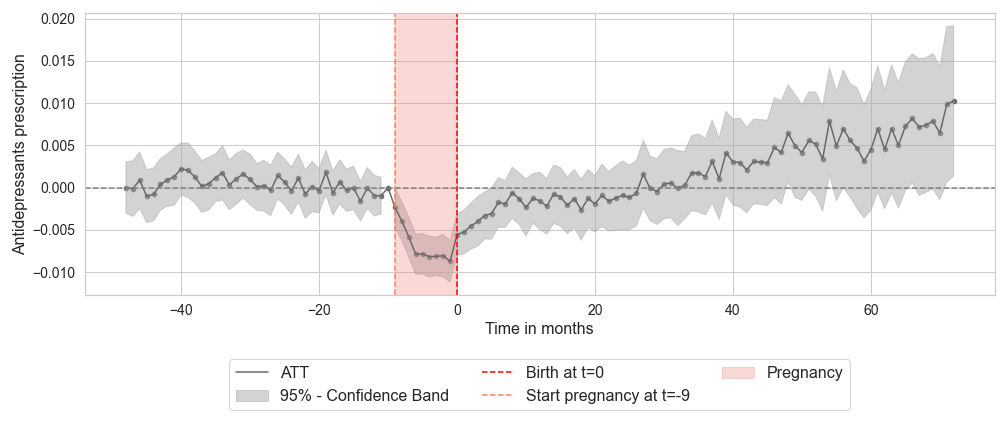}
    \footnotesize \textit{Note:} The figure shows the percentage point increase of women having an antidepressant prescription in comparison to the women who are ``not yet treated''. On average, 1.4\% of women in the sample have an antidepressant prescription before childbirth. The red-shaded area illustrates the pregnancy, and the grey-shaded area represents the 95\% confidence band.  ($N= 25,120$)
    \end{minipage}
\end{figure}

Figure \ref{fig:antidepressants_stagg_diff_diff_balanced_non_smoothed} shows the effect of a first child on the percentage of women having an antidepressant prescription if the outcome variable has not been smoothed. The results are comparable to those using a smoothed outcome variable. However, they are less precise and fluctuate a lot.

\clearpage
\subsection{Heterogeneity Analysis} \label{appendix: heterogeneity_analysis}

\subsubsection{Employment}

\begin{figure}[h!]
    \centering
    \begin{minipage}{14cm}
    \caption{Heterogeneity by emplyment status: Number of visits to psychiatrist}
    \label{fig:Psychiater_stagg_diff_diff_accident}
    \includegraphics[width=14cm]{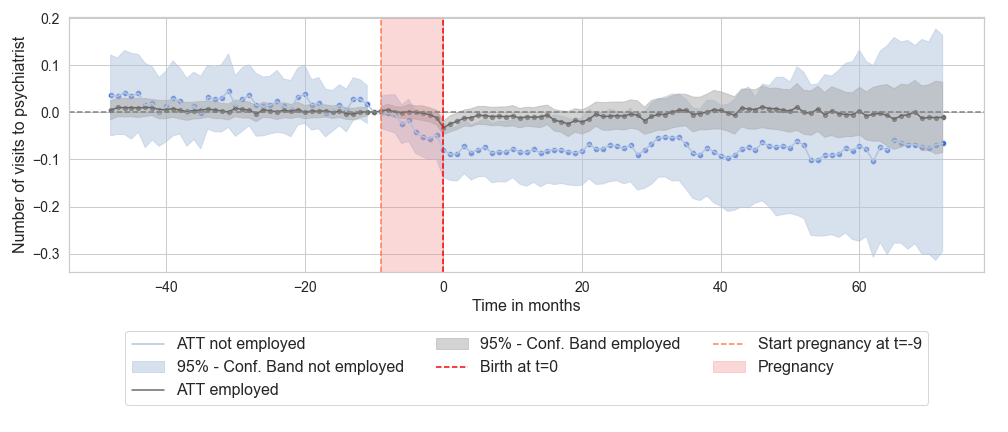}
    \footnotesize \textit{Note:}  The figure shows the number of visits to a psychiatrist compared to the women who do not have a child yet. On average, women that are employed go 0.26 times per month to the psychiatrist before childbirth, whereas women that are not employed only go 0.09 times per month. The red-shaded area illustrates the pregnancy, and the grey-shaded area represents the 95\% confidence band. (Not employed: $N=4,927$, employed: $N=20,193$)
    \end{minipage}
\end{figure}

Figure \ref{fig:Psychiater_stagg_diff_diff_accident} depicts no statistically significant difference in the visits to a psychiatrist for women employed versus women not employed. If any, there is a lower effect for women that are not employed.

\begin{figure}[h!]
    \centering
    \begin{minipage}{14cm}
    \caption{Heterogeneity by employment status: Number of visits to GP or HMO}
    \label{fig:Hausarzt_stagg_diff_diff_accident}
    \includegraphics[width=14cm]{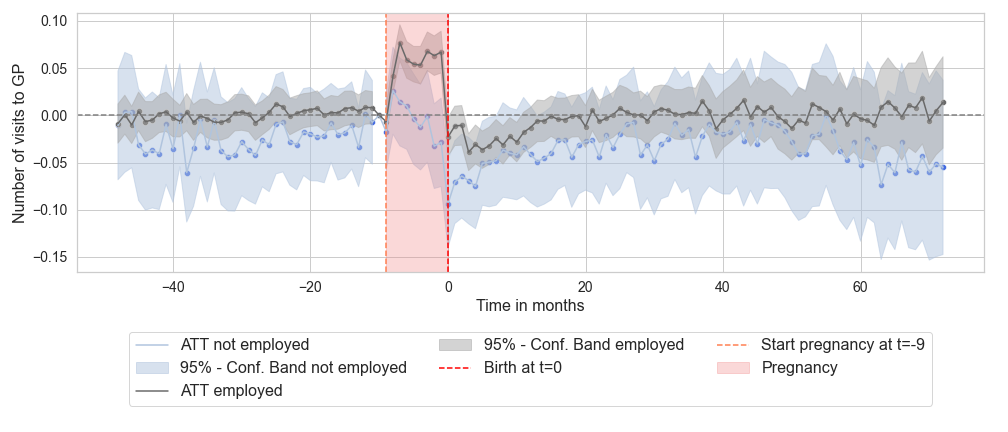}
    \footnotesize \textit{Note:} The figure shows the number of visits to a GP or HMO compared to the women who do not have a child yet. On average, women not employed go 0.25 times per month to the GP or HMO before childbirth, whereas women that are employed only go 0.18 times per month. The red-shaded area illustrates the pregnancy, and the grey-shaded area represents the 95\% confidence band.(Not employed: $N=3,214$, employed: $N=13,154$)
    \end{minipage}
\end{figure}

Figure \ref{fig:Hausarzt_stagg_diff_diff_accident} shows no difference in the effect on the number of visits to a GP or HMO between women employed and not employed.

\subsubsection{Modes of Child Delivery}

\begin{figure}[h!]
    \centering
    \begin{minipage}{14cm}
    \caption{Heterogeneity by mode of delivery: Number of visits to psychiatrist}
    \label{fig:Psychiater_stagg_diff_diff_mode_delivery}
    \includegraphics[width=14cm]{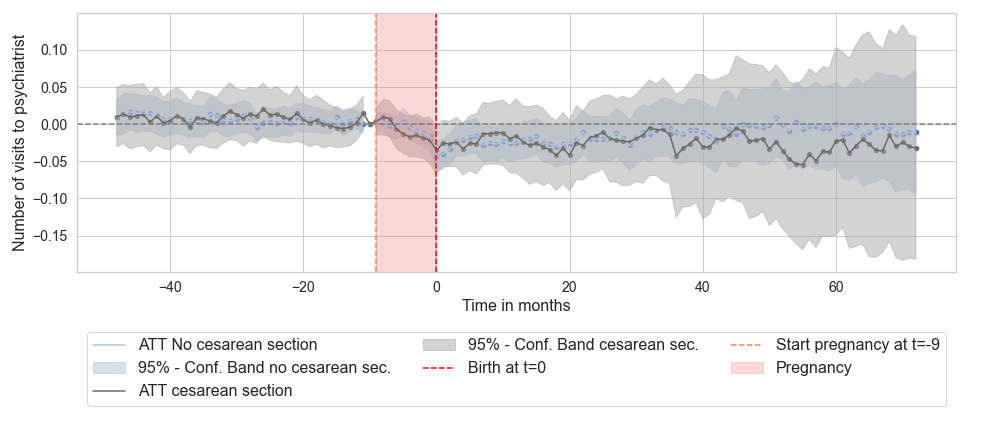}
    \footnotesize \textit{Note:} The figure shows the number of visits to a psychiatrist compared to women who have not had a child yet. On average, women having a cesarean section visit the psychiatrist 0.12 times per month before childbirth, whereas women without a cesarean section only go 0.12 times per month. The red-shaded area illustrates the pregnancy, and the grey-shaded area represents the 95\% confidence band. (Cesarean section: $N=7,396$, no cesarean section: $N=17,724$)
    \end{minipage}
\end{figure}

Figure \ref{fig:Psychiater_stagg_diff_diff_mode_delivery} depicts no difference in the visits to the psychiatrist between women having a cesarean section or not.

\begin{figure}[h!]
    \centering
    \begin{minipage}{14cm}
    \caption{Heterogeneity by mode of delivery: Number of visits to GP or HMO}
    \label{fig:Hausarzt_stagg_diff_diff_mode_delivery}
    \includegraphics[width=14cm]{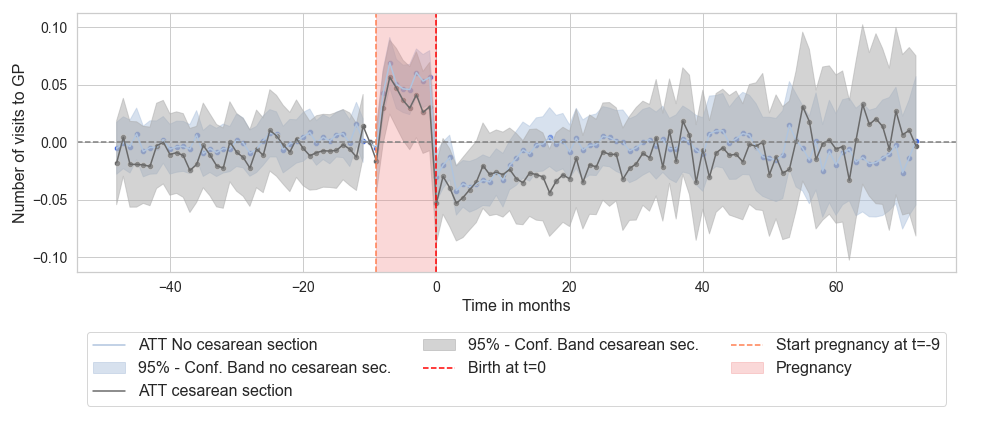}
    \footnotesize \textit{Note:} The figure shows the number of visits to a GP or HMO compared to women who have not had a child yet. On average, women having a cesarean section go to the GP or HMO 0.19 times per month before childbirth, whereas women without a cesarean section only go 0.19 times per month. The red-shaded area illustrates the pregnancy, and the grey-shaded area represents the 95\% confidence band. (Cesarean section: $N=4,813$, no cesarean section: $N=11,555$)
    \end{minipage}
\end{figure}

Figure \ref{fig:Hausarzt_stagg_diff_diff_mode_delivery} shows no difference in the number of visits to a GP or HMO depending on the mode of delivery.

\subsubsection{Getting Subsidies for Health Insurance and Income Index}

There is no difference between women getting and not getting health subsidies for the effect on visits to the psychiatrist, as Figure \ref{fig:Psychiater_stagg_diff_diff_age} depicts.

\begin{figure}[h!]
    \centering
    \begin{minipage}{14cm}
    \caption{Heterogeneity by health subsidies: Number of visits to psychiatrist}
    \label{fig:Psychiater_stagg_diff_diff_subisdy}
    \includegraphics[width=14cm]{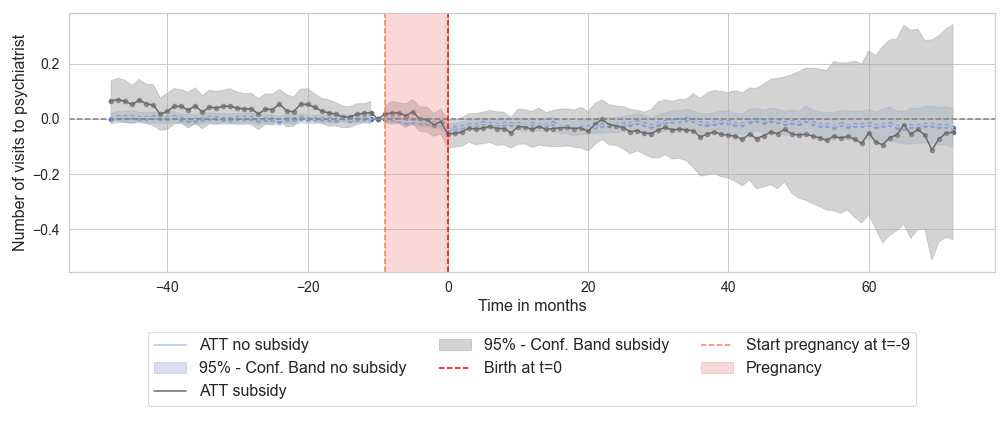}
    \footnotesize \textit{Note:}  The figure shows the number of visits to a psychiatrist compared to women who have not had a child yet. On average, women receiving health subsidies go to the psychiatrist 0.21 times per month before childbirth, whereas women without go only 0.10 times per month. The red-shaded area illustrates the pregnancy, and the grey-shaded area represents the 95\% confidence band. (Subsidy: $N=4,652$, no subsidy: $N=20,468$)
    \end{minipage}
\end{figure}

Figure \ref{fig:Hausarzt_stagg_diff_diff_age} shows the change in the number of visits to a GP or HMO. We do not find any difference between women receiving and not receiving subsidies.

\begin{figure}[h!]
    \centering
    \begin{minipage}{14cm}
    \caption{Heterogeneity by health subsidies: Number of visits to GP or HMO}
    \label{fig:Hausarzt_stagg_diff_diff_subsidy}
    \includegraphics[width=14cm]{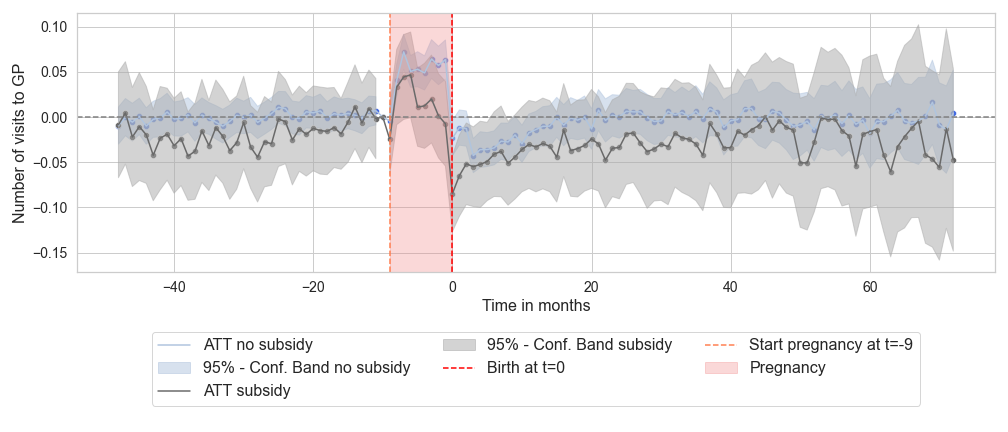}
    \footnotesize \textit{Note:} The figure shows the number of visits to a GP or HMO compared to women who have not had a child yet. On average, women receiving health subsidies go to the GP or HMO 0.26 times per month before childbirth, whereas women without go only 0.17 times per month. The red-shaded area illustrates the pregnancy, and the grey-shaded area represents the 95\% confidence band. (Subsidy: $N= 3,070$, no subsidy: $N=13,298$)
    \end{minipage}
\end{figure}

\begin{figure}[h!]
    \centering
    \begin{minipage}{14cm}
    \caption{Heterogeneity by income: Number of visits to psychiatrist}
    \label{fig:Psychiater_stagg_diff_diff_income_index}
    \includegraphics[width=14cm]{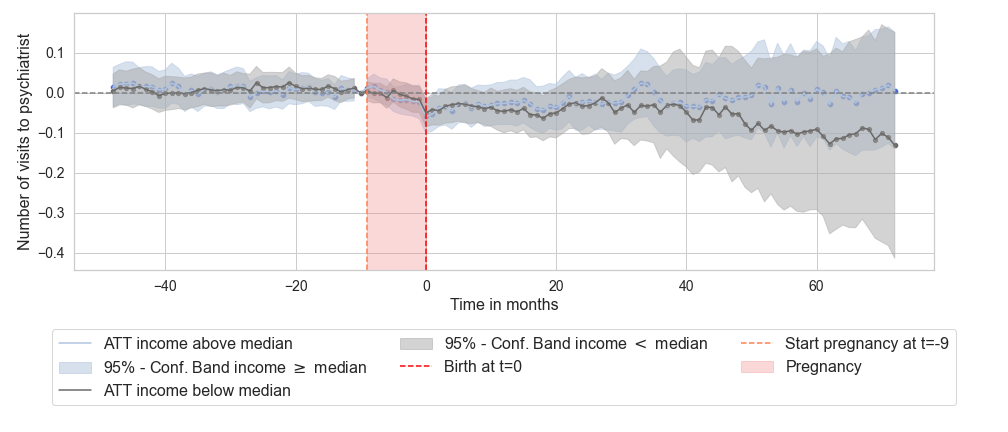}
    \footnotesize \textit{Note:} The figure shows the percentage point change of women taking antidepressants in comparison to the women who do not have a child yet. On average, women having an income above the median go to the psychiatrist 0.14 times per month before childbirth, whereas women below go 0.17 times per month. The red-shaded area illustrates the pregnancy, and the grey-shaded area represents the 95\% confidence band. ($\geq$ median: $N = 6, 950$, $<$ median: $N = 6, 943$)
    \end{minipage}
\end{figure}

There is no statistically significant difference between women with an income above and below the median income for the effect on visits to the psychiatrist, as Figure \ref{fig:Psychiater_stagg_diff_diff_income_index} depicts. If any, there is a lower effect for women having an income below the median.

\begin{figure}[h!]
    \centering
    \begin{minipage}{14cm}
    \caption{Heterogeneity by income: Number of visits to GP or HMO}
    \label{fig:Hausarzt_stagg_diff_diff_income_index}
    \includegraphics[width=14cm]{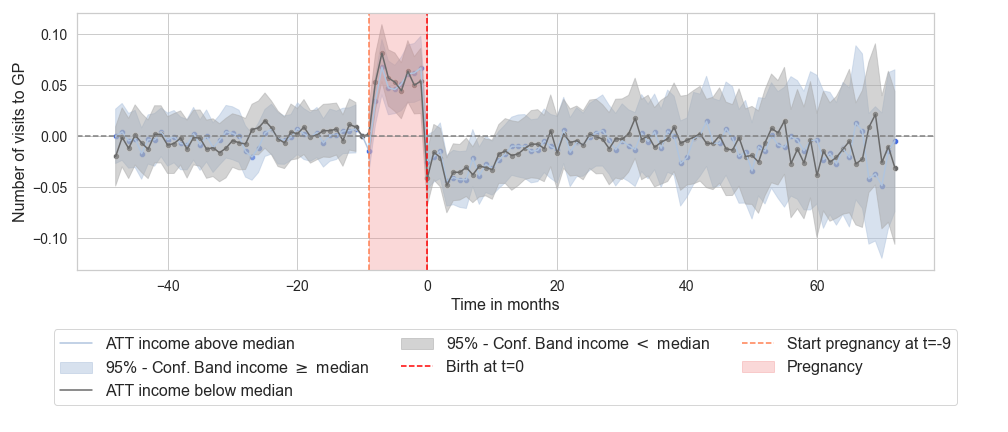}
    \footnotesize \textit{Note:} The figure shows the number of visits to a GP or HMO compared to women who have not had a child yet. On average, women with an income above the median go to the GP or HMO 0.20 times per month before childbirth, whereas women without go 0.17 times per month. The red-shaded area illustrates the pregnancy, and the grey-shaded area represents the 95\% confidence band. ($\geq$ median: $N = 6, 950$, $<$ median: $N = 6, 943$)
    \end{minipage}
\end{figure}

Figure \ref{fig:Hausarzt_stagg_diff_diff_income_index} shows the change in the number of visits to a GP or HMO. We do not find any difference between women having an income above and below the median.

\begin{figure}[h!]
    \centering
    \begin{minipage}{14cm}
    \caption{Descriptives by health subsidy: Antidepressants prescriptions}
    \label{fig:antidepressants_descriptives_health_subsidy}
    \includegraphics[width=14cm]{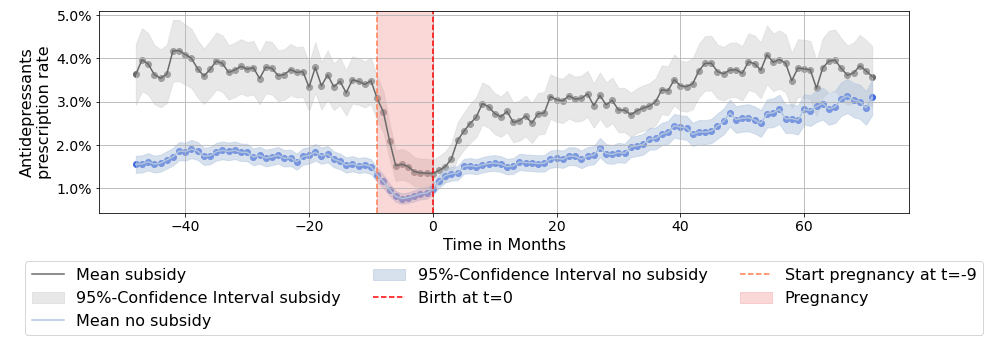}
    \footnotesize \textit{Note:} The figure shows the percentage of women taking antidepressants for women getting health subsidies and women not getting health subsidies.  The red-shaded area illustrates the pregnancy, and the grey- or blue-shaded area represents the 95\% confidence band. The slow decrease in antidepressants at the start of the pregnancy and also the slow increase after giving birth are partly coming from the smoothing of the antidepressant prescriptions. (Subsidy: $N=4,652$, no subsidy: $N=20,468$)
    \end{minipage}
\end{figure}

Figure \ref{fig:antidepressants_descriptives_health_subsidy} shows the percentage of women having an antidepressant prescription, divided into women receiving health subsidies at the time of the pregnancy and women not receiving health subsidies.

\begin{figure}[h!]
    \centering
    \begin{minipage}{14cm}
    \caption{Descriptives employed individuals by health subsidy: Antidepressants prescriptions}
    \label{fig:antidepressants_descriptives_health_subsidy_employed}
    \includegraphics[width=14cm]{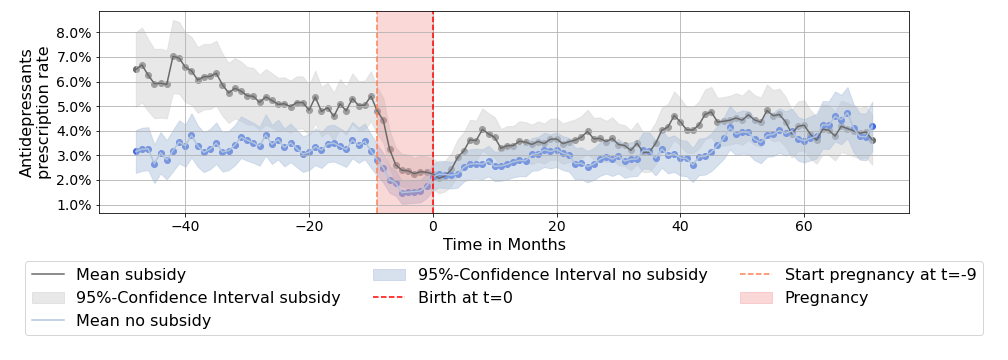}
    \footnotesize \textit{Note:} The figure shows the percentage of women taking antidepressants for women who are employed at the time of childbirth divided by women getting health subsidies and women not getting health subsidies. The red-shaded area illustrates the pregnancy, and the grey- or blue-shaded area represents the 95\% confidence band. The slow decrease in antidepressants at the start of the pregnancy and also the slow increase after giving birth are partly coming from the smoothing of the antidepressant prescriptions. (Subsidy: $N=2,751$, no subsidy: $N=17,442$)
    \end{minipage}
\end{figure}

Figure \ref{fig:antidepressants_descriptives_health_subsidy_employed} shows the same as Figure \ref{fig:antidepressants_descriptives_health_subsidy} but only for women who were employed at the time of childbirth. The pattern of women getting and not getting health subsidies is more similar if we only look at employed women. This supports the hypothesis that the difference in the effect for women getting health subsidies and not getting health subsidies is partly due to the high correlation between being employed and not getting health subsidies.

\newpage
\subsubsection{Language} \label{staggered_diff_diff_language_appendix}

Switzerland has four official languages: German, French, Italian and Romansh. There is one main language border dividing Switzerland into the French and German part.\footnote{This language border does not coincide with the cantonal borders and is often called \textit{Rösti} border.} Several papers have used the language difference as a proxy for cultural difference, as this border divides Switzerland independently from institutional or environmental variations and not only language but also some cultural traits change \citep[e.g.][]{Aepli:2021,Eugster:2017, Faessler:2024}. One example is differences in voting behaviour across the language border \citep{Etter:2014}. On average, the French-speaking part is more left on the political spectrum, while the German-speaking part is more conservative. If more conservative gender norms lead to fathers helping less in the household \citep{Sevilla:2010}, the mental health penalty might be higher in the German-speaking compared to the French-speaking part.

\begin{figure}[h!]
    \centering
    \begin{minipage}{15cm}
    \caption{Heterogeneity by language: Antidepressants prescriptions}
    \label{fig:antidepressants_stagg_diff_diff_language}
    \includegraphics[width=15cm]{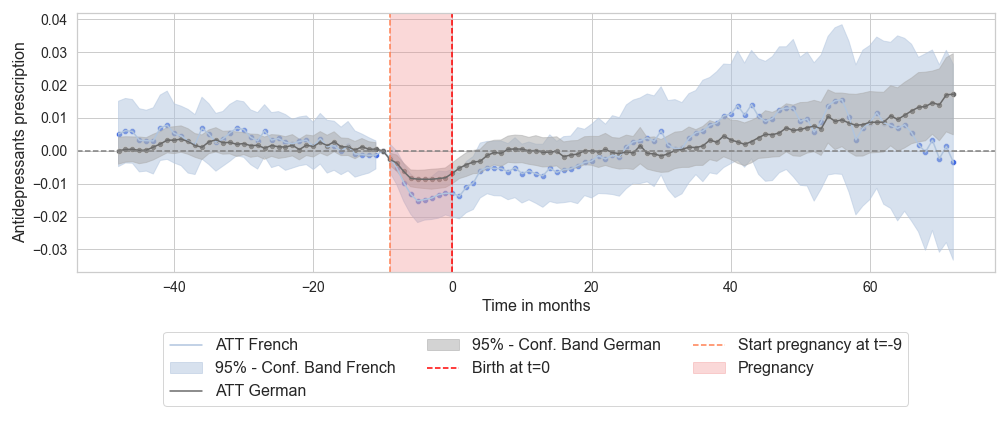}
    \footnotesize \textit{Note:} The figure shows the percentage point change of women taking antidepressants in comparison to the women who do not have a child yet. On average, 1.9\% of German-speaking women and 2.1\% of French-speaking women have an antidepressant prescription before childbirth. The red-shaded area illustrates the pregnancy, and the grey-shaded area represents the 95\% confidence band. The slow decrease in antidepressants at the start of the pregnancy and also the slow increase after giving birth are partly coming from the smoothing of the antidepressant prescriptions. (German: $N=19,552$, French: $N=4,715$)
    \end{minipage}
\end{figure}

According to Figure \ref{fig:antidepressants_stagg_diff_diff_language}, the mental health penalty is similar in the first five years after childbirth for both language groups. Six years after childbirth, the mental health penalty is higher for German-speaking than for French-speaking women, but the confidence intervals are large.

\begin{figure}[h!]
    \centering
    \begin{minipage}{14cm}
    \caption{Heterogeneity by language: Number of visits to psychiatrist}
    \label{fig:Psychiater_stagg_diff_diff_language}
    \includegraphics[width=14cm]{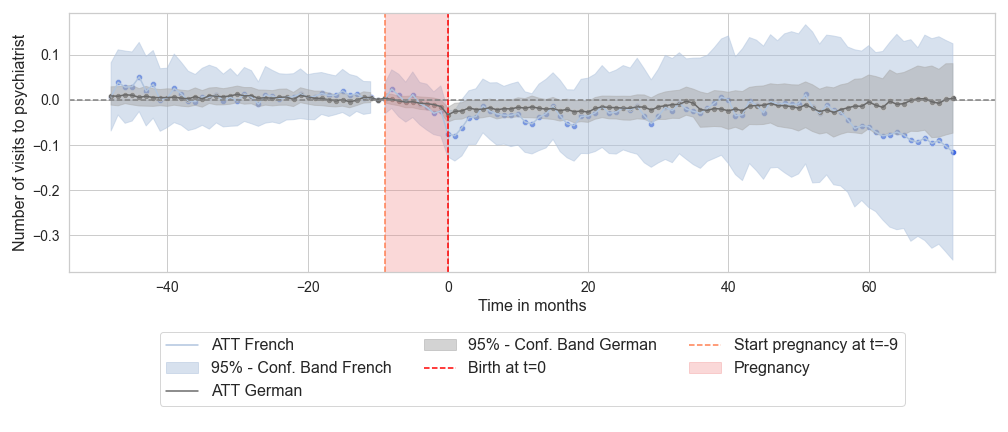}
    \footnotesize \textit{Note:}  The figure shows the number of visits to a psychiatrist compared to the women who do not have a child yet. On average, German-speaking women go 0.10 times per month to the psychiatrist before childbirth, whereas French-speaking women go 0.21 times per month. The red-shaded area illustrates the pregnancy, and the grey-shaded area represents the 95\% confidence band. (German: $N=19,552$, French: $N=4,715$)
    \end{minipage}
\end{figure}

Figure \ref{fig:Psychiater_stagg_diff_diff_language} depicts no statistically significant difference in the number of visits to a psychiatrist for French-speaking compared to German-speaking women. If any, we see the same pattern as for antidepressant prescriptions, with a higher effect for German-speaking women.

\begin{figure}[h!]
    \centering
    \begin{minipage}{14cm}
    \caption{Heterogeneity by language: Number of visits to GP or HMO}
    \label{fig:Hausarzt_stagg_diff_diff_language}
    \includegraphics[width=14cm]{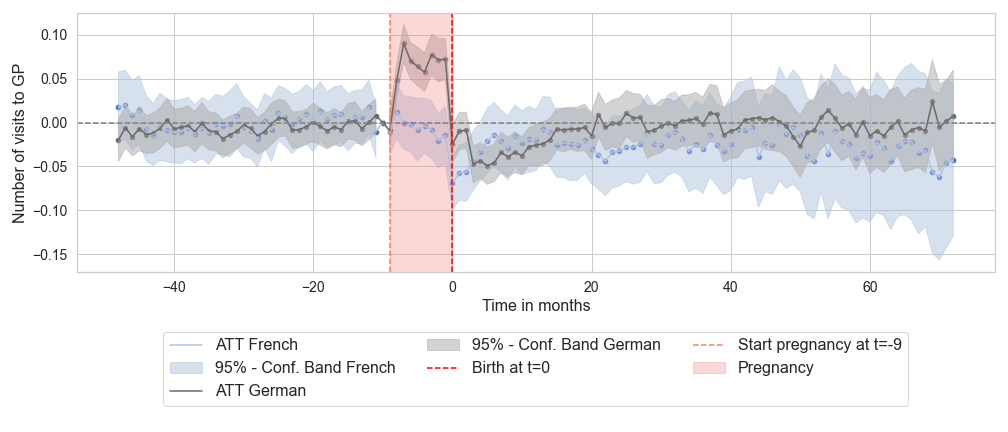}
    \footnotesize \textit{Note:} The figure shows the number of visits to a GP or HMO compared to the women who do not have a child yet. On average, German-speaking women go 0.19 times per month to the GP or HMO before childbirth, whereas French-speaking women go 0.19 times per month. The red-shaded area illustrates the pregnancy, and the grey-shaded area represents the 95\% confidence band.(German: $N=11,471$, French: $N=4,349$)
    \end{minipage}
\end{figure}

Figure \ref{fig:Hausarzt_stagg_diff_diff_language} shows no difference in the effect on the number of visits to a GP or HMO between French- and German-speaking women.

\subsubsection{Nationality} \label{staggered_diff_diff_nationality_appendix}

\begin{figure}[h!]
    \centering
    \begin{minipage}{14cm}
    \caption{Heterogeneity by nationality: Antidepressants prescriptions}
    \label{fig:antidepri_stagg_diff_diff_nationality}
    \includegraphics[width=14cm]{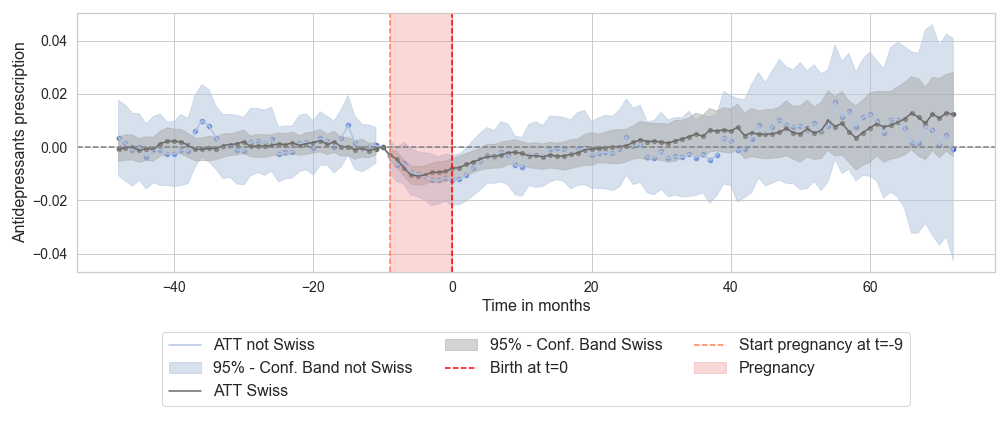}
    \footnotesize \textit{Note:}  The figure shows the percentage point increase of women having an antidepressant prescription in comparison to the women who are ``not yet treated''. On average, 1.8\% of Swiss women in the sample have an antidepressant prescription before childbirth, whereas also 1.6\% of non-Swiss women have one. The red-shaded area illustrates the pregnancy, and the grey-shaded area represents the 95\% confidence band. The slow decrease in antidepressants at the start of the pregnancy and also the slow increase after giving birth are partly coming from the smoothing of the antidepressant prescriptions. (Swiss: $N=13,865$, Non-swiss: $N=2,278$)
    \end{minipage}
\end{figure}

Not only language but also nationality can be used as a proxy for cultural differences that translate into different support of women after childbirth and, hence, a different mental health penalty. A problem with this approach is that we only have nationality for the CSS data, and 80\% of women in the sample are Swiss. First, this leads to results for non-Swiss individuals that are imprecise and second, the non-Swiss individuals are a heterogeneous group coming from Northern countries with more equal gender norms than Switzerland, but also from countries from Africa or Asia with more traditional gender norms.
Figure \ref{fig:antidepri_stagg_diff_diff_nationality} shows no statistically significant difference in antidepressant prescriptions for Swiss compared to non-Swiss individuals. 

\begin{figure}[h!]
    \centering
    \begin{minipage}{14cm}
    \caption{Heterogeneity by nationality: Number of visits to psychiatrist}
    \label{fig:Psychiater_stagg_diff_diff_nationality}
    \includegraphics[width=14cm]{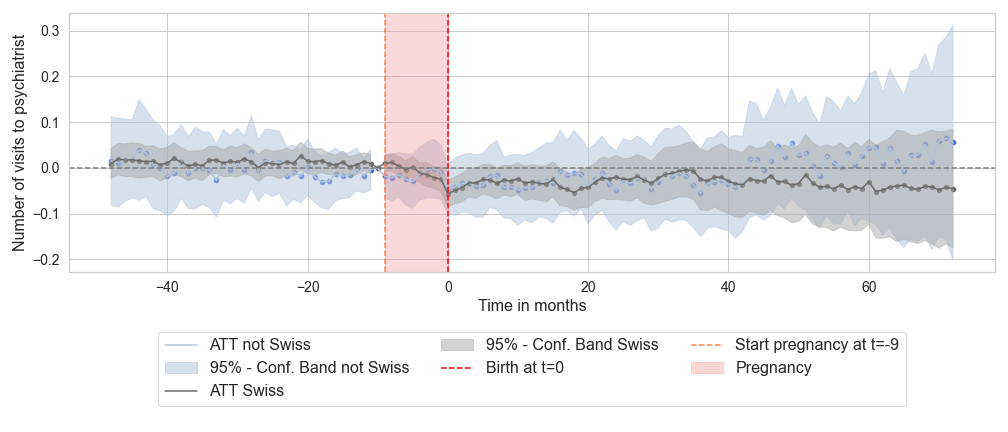}
    \footnotesize \textit{Note:}  The figure shows the number of visits to a psychiatrist compared to the women who do not have a child yet. On average, Swiss women go 0.16 times per month to the psychiatrist before childbirth, whereas non-Swiss women go 0.10 times per month. The red-shaded area illustrates the pregnancy, and the grey-shaded area represents the 95\% confidence band. (Swiss: $N=13,865$, Non-swiss: $N=2,278$)
    \end{minipage}
\end{figure}

Figure \ref{fig:Psychiater_stagg_diff_diff_nationality} depicts no statistically significant difference in the number of visits to a psychiatrist for Swiss compared to non-Swiss women.

\begin{figure}[h!]
    \centering
    \begin{minipage}{14cm}
    \caption{Heterogeneity by nationality: Number of visits to GP or HMO}
    \label{fig:Hausarzt_stagg_diff_diff_nationality}
    \includegraphics[width=14cm]{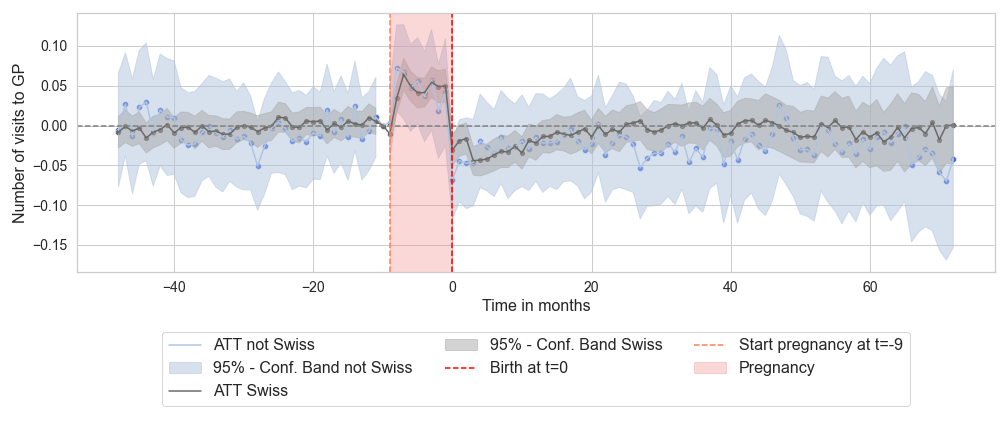}
    \footnotesize \textit{Note:} The figure shows the number of visits to a GP or HMO compared to the women who do not have a child yet. On average, Swiss women go 0.18 times per month to the GP or HMO before childbirth, whereas non-Swiss women go 0.23 times per month. The red-shaded area illustrates the pregnancy, and the grey-shaded area represents the 95\% confidence band. (Swiss: $N=13,865$, Non-swiss: $N=2,278$)
    \end{minipage}
\end{figure}

Figure \ref{fig:Hausarzt_stagg_diff_diff_nationality} shows no difference in the effect on the number of visits to a GP or HMO between Swiss and non-Swiss women.

\subsubsection{Age at First Birth}

Younger parents tend to be healthier and more resilient to stress, but parenthood may be more stressful for younger parents, especially if the pregnancy was not planned. On the other hand, a planned pregnancy does not lead to a lower risk of first-time psychiatric disorders \citep{Munk:2015}. Therefore, the direction is not clear, but most papers find larger mental health penalties for younger parents \citep[e.g.][]{Ahammer:2023, Muller:2018, Aras:2013, Falci:2010}. The short time period of the data introduces certain limitations to this analysis. Looking at longer periods for women below 28 years (young) is impossible because we can only contemplate women with birthdates from 1985 to 1989 for longer than four years.\footnote{As a reminder, we have data from 2012 to 2022. Hence, women born in 1985 are already at least 27 years old - so only a few are in the young women sample. On the other hand, a woman born in 1990 is 32 years old at most.} Therefore, the sample size would need to be bigger to draw any conclusions later than four years after childbirth.

\begin{figure}[h!]
    \centering
    \begin{minipage}{14cm}
    \caption{Heterogeneity by age: Antidepressants prescriptions}
    \label{fig:antidepressants_stagg_diff_diff_age}
    \includegraphics[width=14cm]{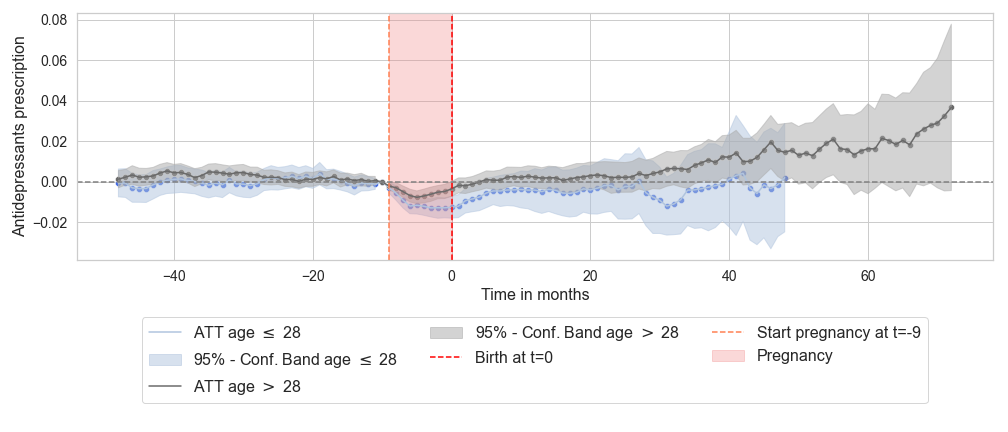}
    \footnotesize \textit{Note:} The figure shows the percentage point change of women taking antidepressants compared to women who have not had a child yet. On average, 1.9\% of women above 28 years and 1.9\% below or equal to 28 years old have an antidepressant prescription before childbirth. The red-shaded area illustrates the pregnancy, and the grey-shaded area represents the 95\% confidence band. The slow decrease in antidepressants at the start of the pregnancy and also the slow increase after giving birth are partly coming from the smoothing of the antidepressant prescriptions. ($> 28$ years old: $N=14,363$, $\leq 28$ years old: $N=10,757$)
    \end{minipage}
\end{figure}

Figure \ref{fig:antidepressants_stagg_diff_diff_age} shows that we cannot confirm the results from the literature. If any difference exists, there is a larger mental health penalty for older than for younger women.

\begin{figure}[h!]
    \centering
    \begin{minipage}{14cm}
    \caption{Heterogeneity by age: Number of visits to psychiatrist}
    \label{fig:Psychiater_stagg_diff_diff_age}
    \includegraphics[width=14cm]{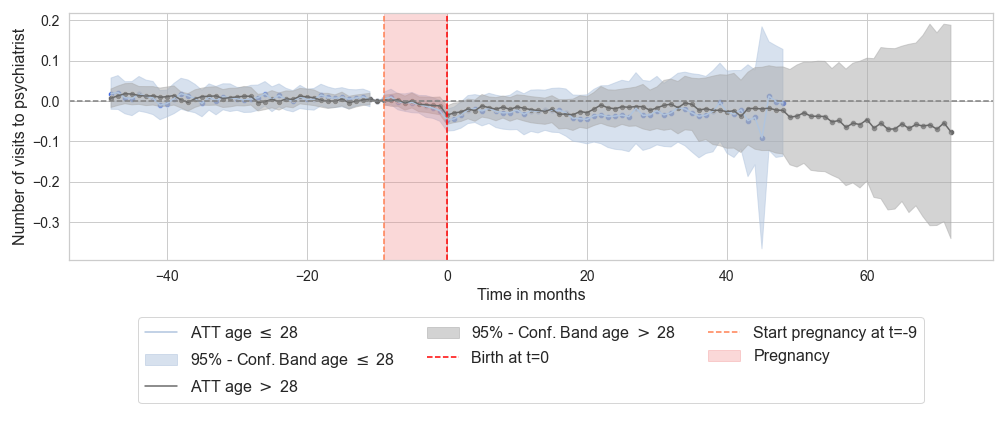}
    \footnotesize \textit{Note:} The figure shows the number of visits to a psychiatrist compared to women who have not had a child yet. On average, women older than 28 years at the time of childbirth go to the psychiatrist 0.12 times per month before childbirth, whereas women younger go only 0.12 times per month. The red-shaded area illustrates the pregnancy, and the grey-shaded area represents the 95\% confidence band. ($> 28$ years old: $N=14,363$, $\leq 28$ years old: $N=10,757$)
    \end{minipage}
\end{figure}

As depicted in Figure \ref{fig:Psychiater_stagg_diff_diff_age}, there is no statistically significant difference in the number of visits to the psychiatrist between the two groups of women with different ages at the birth of their first child.

\begin{figure}[h!]
    \centering
    \begin{minipage}{14cm}
    \caption{Heterogeneity by age: Number of visits to GP or HMO}
    \label{fig:Hausarzt_stagg_diff_diff_age}
    \includegraphics[width=14cm]{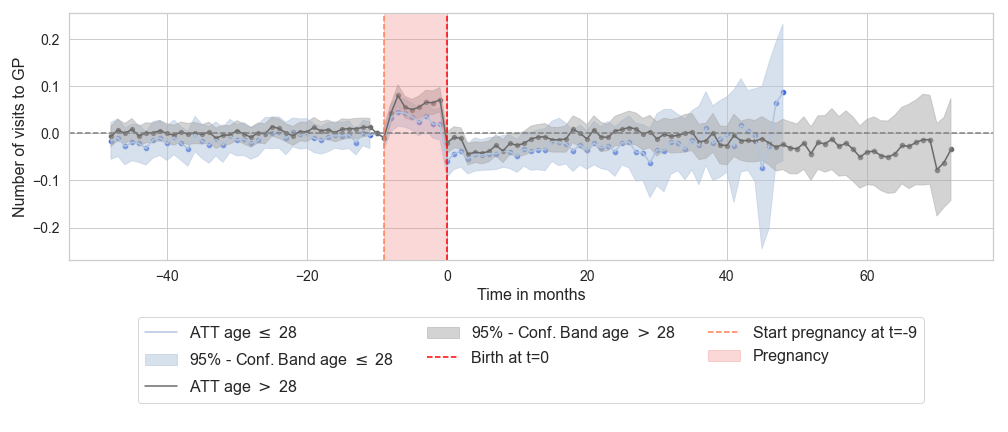}
    \footnotesize \textit{Note:} The figure shows the number of visits to a GP or HMO compared to women who have not had a child yet. On average, women older than 28 years at the time of childbirth go to the GP or HMO 0.17 times per month before childbirth, whereas women younger go only 0.22 times per month. The red-shaded area illustrates the pregnancy, and the grey-shaded area represents the 95\% confidence band.  ($> 28$ years old: $N=8,849$, $\leq 28$ years old: $N=7,519$)
    \end{minipage}
\end{figure}

Similarly, Figure \ref{fig:Hausarzt_stagg_diff_diff_age} shows no difference between the two groups in the number of visits to a GP or HMO after childbirth.

\newpage
\subsubsection{First versus Second Childbirth and Number of Children}

The number of children a woman has is a decision made by the couple. Hence, women with worse mental health might not have a second child, or women who are not able to conceive a second child might be mentally worse off. Therefore, this analysis should be interpreted cautiously, as subsequent fertility is most likely endogeneous.

First, we look at the mental health penalty for the first child versus the second child. Having a second child might influence the mental health of mothers even more, as having two children is probably more stressful than having only one child. On average, women have 2.5 years after the first a second child (if they have a second one). By looking at the first childbirth, we look at the accumulation of the effect of the first and second child some years post-childbirth (for women with a second child).

\begin{figure}[h!]
    \centering
    \begin{minipage}{14cm}
    \caption{Heterogeneity by number of birth: Antidepressants prescriptions}
    \label{fig:antidepressants_stagg_diff_diff_second_birth}
    \includegraphics[width=14cm]{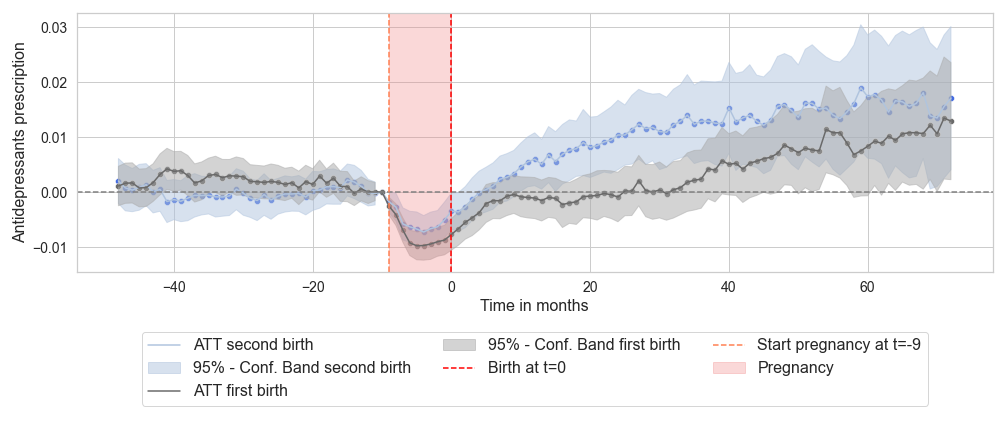}
    \footnotesize \textit{Note:} The figure shows the percentage change of women taking antidepressants in comparison to the women who do not have a child yet. On average, 1.9\% of women have an antidepressant prescription before the first childbirth. Before the second childbirth, only 1.5\% have an antidepressant prescription. The red-shaded area illustrates the pregnancy, and the grey-shaded area represents the 95\% confidence band. The slow decrease in antidepressants at the start of the pregnancy and also the slow increase after giving birth are partly coming from the smoothing of the antidepressant prescriptions. (First childbirth: $N=25,120$, second childbirth: $N=13,169$)
    \end{minipage}
\end{figure}

Figure \ref{fig:antidepressants_stagg_diff_diff_second_birth} shows antidepressant prescriptions increase by one p.p. two years post-childbirth and increase by 1.5 p.p. after four years for the second childbirth. There is no increase in antidepressant prescriptions in the first two years after childbirth for the first child. This can be explained by the fact that antidepressants should not be taken during pregnancy and breastfeeding. For the second child, the increase happens directly after childbirth, as most women do not have more than two children.

\begin{figure}[h!]
    \centering
    \begin{minipage}{14cm}
    \caption{Heterogeneity by number of birth: Number of visits to psychiatrist}
    \label{fig:Psychiater_stagg_diff_diff_first_vs_second}
    \includegraphics[width=14cm]{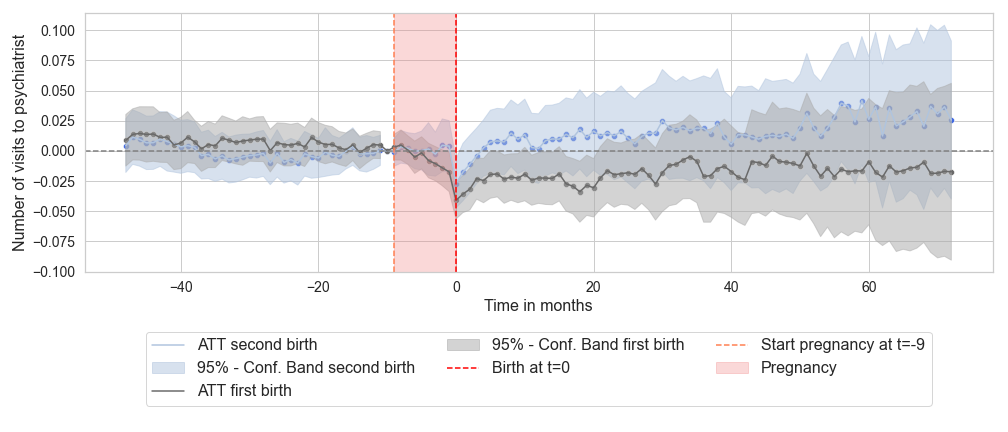}
    \footnotesize \textit{Note:} The figure shows the number of visits to a psychiatrist compared to women who have not had a child yet. On average, women before having the first child go to the psychiatrist 0.12 times per month, whereas women before having the second child only go 0.09 times per month. The red-shaded area illustrates the pregnancy, and the grey-shaded area represents the 95\% confidence band.   (First childbirth: $N=25,120$, second childbirth: $N=13,169$)
    \end{minipage}
\end{figure}

Figure \ref{fig:Psychiater_stagg_diff_diff_first_vs_second} depicts a similar difference for the effect on the number of visits to the psychiatrist. The effect seems bigger for women having a second child, but it is still not statistically significantly different from zero.

\begin{figure}[h!]
    \centering
    \begin{minipage}{14cm}
    \caption{Heterogeneity by number of birth: Number of visits to GP or HMO}
    \label{fig:Hausarzt_stagg_diff_diff_first_vs_second}
    \includegraphics[width=14cm]{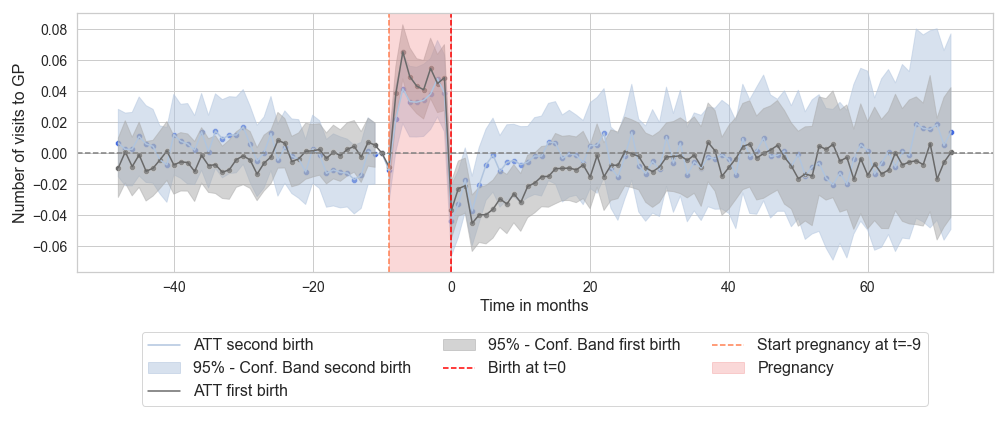}
    \footnotesize \textit{Note:} The figure shows the number of visits to a GP or HMO compared to women who have not had a child yet. On average, women before having the first child go to the GP or HMO 0.19 times per month, whereas women before having the second child also go 0.19 times per month. The red-shaded area illustrates the pregnancy, and the grey-shaded area represents the 95\%  confidence band. (First childbirth: $N=16,368$, second childbirth: $N=8,695$)
    \end{minipage}
\end{figure}

Figure \ref{fig:Hausarzt_stagg_diff_diff_first_vs_second} shows no statistically significant difference in the number of visits to a GP or HMO depending on the number of childbirths.

\begin{figure}[h!]
    \centering
    \begin{minipage}{14cm}
    \caption{Heterogeneity by number of children: Antidepressants prescriptions}
    \label{fig:antidepressants_stagg_diff_diff_one_child}
    \includegraphics[width=14cm]{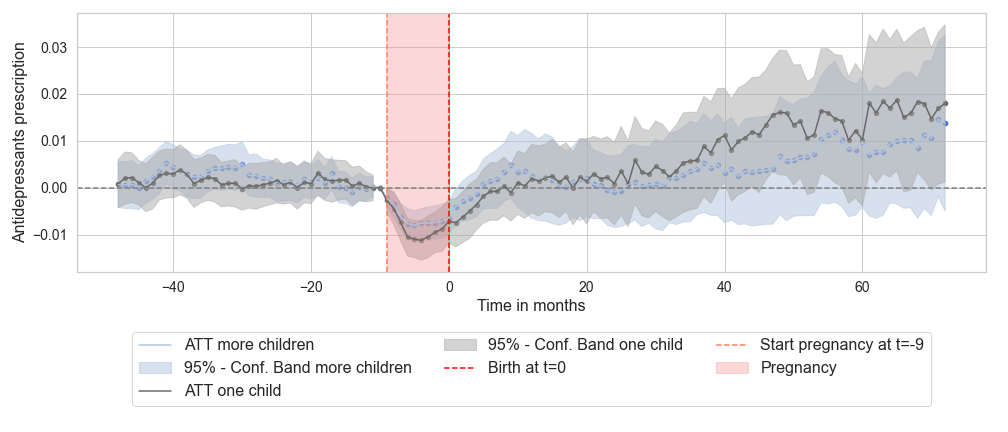}
    \footnotesize \textit{Note:} The figure shows the percentage change of women taking antidepressants in comparison to the women who do not have a child yet.  On average, 1.5\% of women with more than one child and 2.3\% with one child (at the end of the period available) have an antidepressant prescription before childbirth. The red-shaded area illustrates the pregnancy, and the grey-shaded area represents the 95\% confidence band. The slow decrease in antidepressants at the start of the pregnancy and also the slow increase after giving birth are partly coming from the smoothing of the antidepressant prescriptions. (One child: $N=11,951$, several children: $N=13,169$)
    \end{minipage}
\end{figure}

Second, we can separate the sample by the number of children. \cite{Ahammer:2023} find a larger mental health penalty for women with two or three children than for women with only one child. This is in contrast to a study by \cite{Kravdal:2017} who find that individuals with only one child face a higher probability of taking antidepressants than individuals with more than one child also in the long-run. 
Figure \ref{fig:antidepressants_stagg_diff_diff_one_child} shows that there is no difference between women having one or more children. If any, we find a higher percentage point increase for women only having one child. 

As women with only one child are already pre-childbirth more likely to consume antidepressants, the penalty in percentage terms is similar for both groups. This is different from the results of \cite{Ahammer:2023}, who find a larger mental health penalty for women with two or three children.

\begin{figure}[h!]
    \centering
    \begin{minipage}{14cm}
    \caption{Heterogeneity by number of children: Number of visits to psychiatrist}
    \label{fig:Psychiater_stagg_diff_diff_one_child}
    \includegraphics[width=14cm]{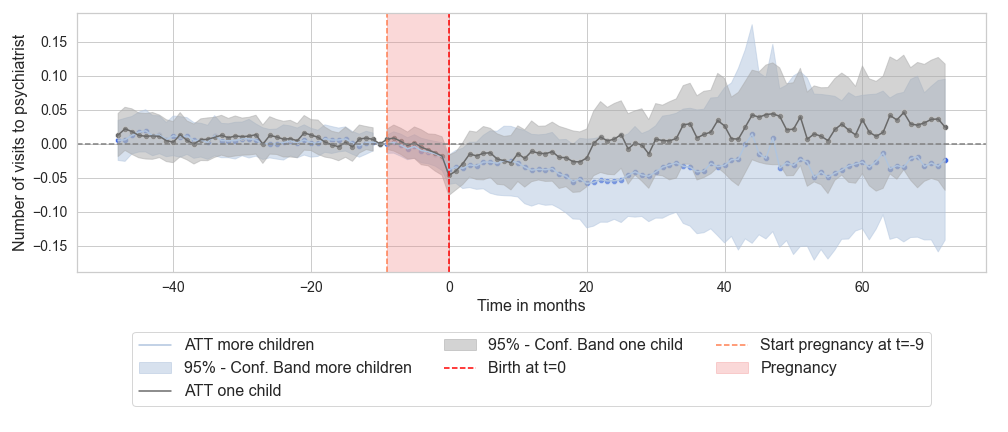}
    \footnotesize \textit{Note:} The figure shows the number of visits to a psychiatrist compared to women who have not had a child yet. On average, women with one child go to the psychiatrist 0.14 times per month before childbirth, whereas women with more children only go 0.09 times per month. The red-shaded area illustrates the pregnancy, and the grey-shaded area represents the 95\% confidence band.  (One child: $N=11,951$, several children: $N=13,169$)
    \end{minipage}
\end{figure}

Figure \ref{fig:Psychiater_stagg_diff_diff_one_child} depicts that there is an indication of an effect in the number of visits to the psychiatrist for women that have only one child but again no effect for women having more children (but no statistically significant effect).

\begin{figure}[h!]
    \centering
    \begin{minipage}{14cm}
    \caption{Heterogeneity by number of children: Number of visits to GP or HMO}
    \label{fig:Hausarzt_stagg_diff_diff_one_child}
    \includegraphics[width=14cm]{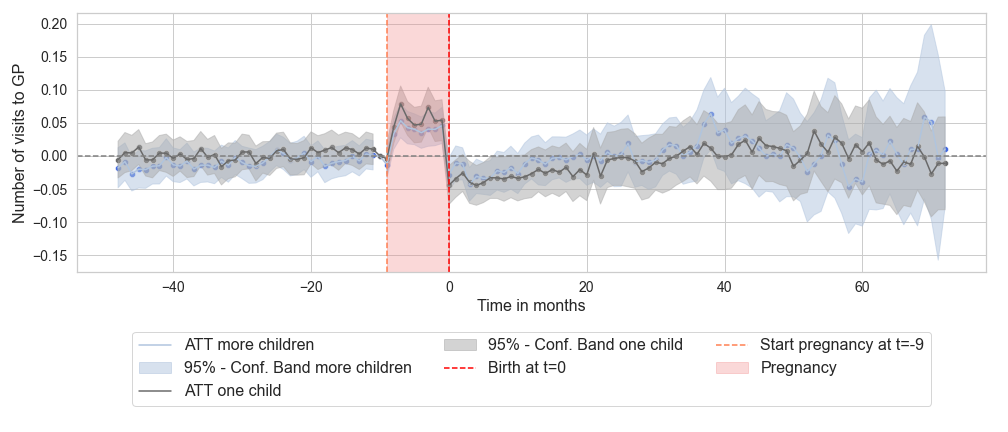}
    \footnotesize \textit{Note:} The figure shows the number of visits to a GP or HMO compared to the women who do not have a child yet. On average, women with one child go 0.19 times per month to the GP or HMO before childbirth, whereas women with more children only go 0.19 times per month. The red-shaded area illustrates the pregnancy, and the grey-shaded area represents the 95\% confidence band. (One child: $N=7,673$, several children: $N=8,695$)
    \end{minipage}
\end{figure}

Figure \ref{fig:Hausarzt_stagg_diff_diff_one_child} shows no difference in the number of visits to a GP or HMO depending on how many children a woman has.

\newpage

\clearpage
\section{Appendix: Effect of Having a Child on Mental Health of Men} \label{appendix: effect_on_men}

\subsection{Sample Selection}

We can only include men insured at the same insurance as the mother, as this is necessary to establish the link between them, given that men do not experience childbirth. We restrict the sample of men in the same way as the sample of women; namely, we choose men born between 1985 and 1995 and having a child between 20 and 40 years of age. This is done to ensure that the samples of men and women are comparable.

\subsection{Data Descriptives}
\subsubsection{Outcome Variables}
\begin{figure}[h!]
    \centering
    \begin{minipage}{15cm}
    \caption{Descriptives men: Antidepressants prescriptions}
    \label{fig:antidepressants_prescription_descriptives_men}
    \includegraphics[width=15cm]{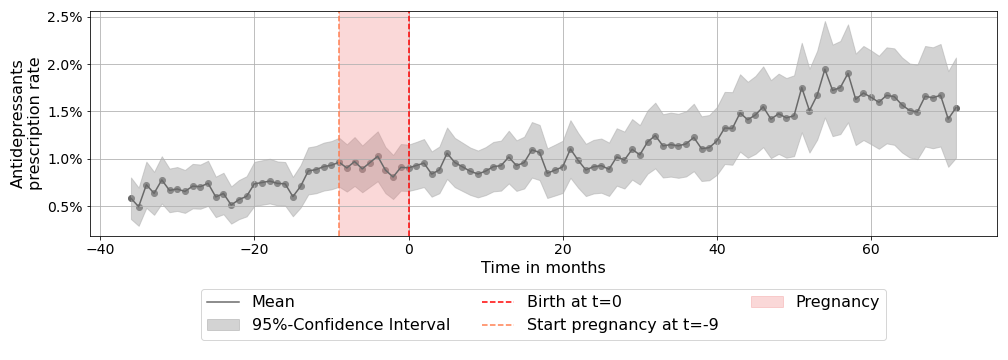}
    \footnotesize \textit{Note:} The figure shows the percentage of men taking antidepressants around the time of birth ($t = 0$: time of first birth). The red-shaded area illustrates the ``pregnancy'', and the grey-shaded area represents the 95\% confidence interval. ($N = 5,749$) 
    \end{minipage}
\end{figure}

Figure  \ref{fig:antidepressants_prescription_descriptives_men} shows that the percentage of men taking antidepressants is lower, namely around 0.5\% to 1\% in a given month before having a child.\footnote{If we look at yearly prescriptions, approximately 2.5\% of men have an antidepressant prescription before childbirth. This number is slightly lower than the average in the Swiss population as antidepressant consumption increases with age. However, also in the Swiss population, women consume more antidepressants than men \citep{Obsan:2022}.} This percentage increases over time to around 1.5\% six years after the birth of the first child. However, for men, it seems to be more of a constant increase over time instead of a steeper increase after childbirth.

\begin{figure}[h!]
    \centering
    \begin{minipage}{15cm}
    \caption{Descriptives men: Visits to psychiatrist}
    \label{fig:psychiatrist_descriptives_men} 
    \includegraphics[width=15cm]{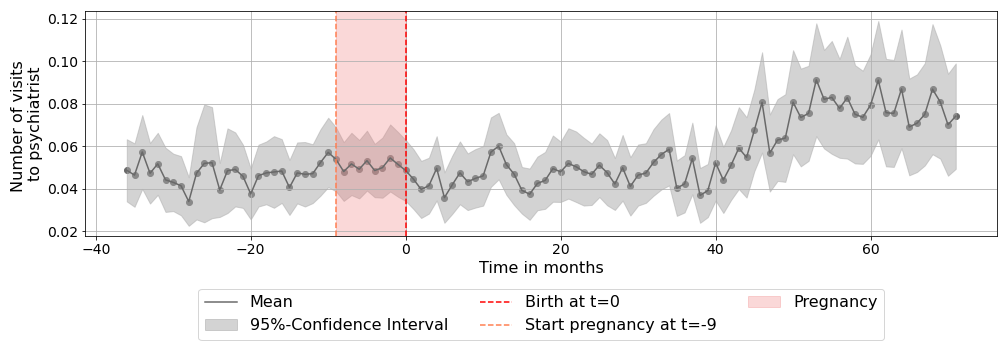}
    \footnotesize \textit{Note:} The figure shows the number of visits to a psychiatrist around the time of birth ($t = 0$: time of first birth). The red-shaded area illustrates the ``pregnancy'', and the grey-shaded area represents the 95\% confidence interval.  ($N = 5,749$)
    \end{minipage}
\end{figure}

Figure \ref{fig:psychiatrist_descriptives_men} show that men are also less often visiting a psychiatrist, with around 0.05 monthly visits. The number of visits increases five years after childbirth to around 0.08 monthly visits. Interestingly, the increase mostly happens around five years after childbirth. Hence, this increase might not be due to the birth of the child but due to other factors.

Similarly, the number of visits to a GP or HMO steadily increases over time, starting at 0.1 visits per month before childbirth and reaching 0.13 visits per month six years after. Since men do not experience pregnancy, there is no noticeable increase in the months before childbirth.

\begin{figure}[h!]
    \centering
    \begin{minipage}{15cm}
    \caption{Descriptives men: Visits to GP or HMO}
    \label{fig:GP_descriptives_men} 
    \includegraphics[width=15cm]{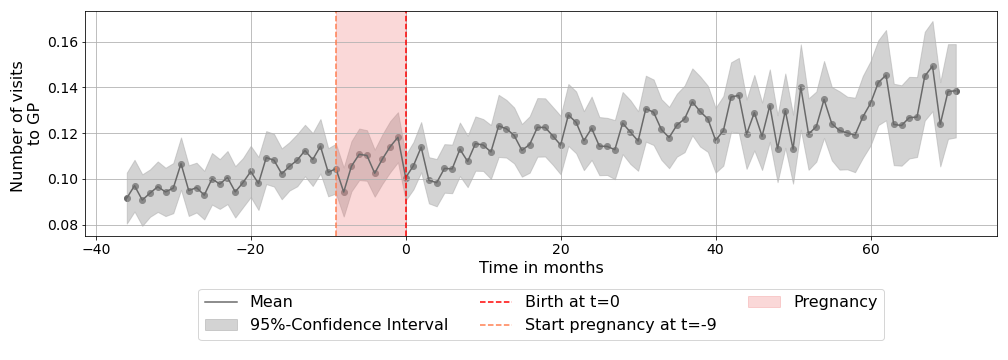}
    \footnotesize \textit{Note:} The figure shows the number of visits to a GP or HMO around the time of birth ($t = 0$: time of first birth). The red-shaded area illustrates the ``pregnancy'', and the grey-shaded area represents the 95\% confidence interval. ($N = 5,749$) 
    \end{minipage}
\end{figure}

\subsubsection{Covariates} \label{covariates_men}

Figure \ref{fig:age_descriptives} shows that most men are also having their first child around the age of 30. They are a bit older on average than the women.

\begin{figure}[h!]
    \centering
    \begin{minipage}{13cm}
    \caption{Descriptives men: Age at first birth}
    \label{fig:age_descriptives_men}
    \includegraphics[width=13cm]{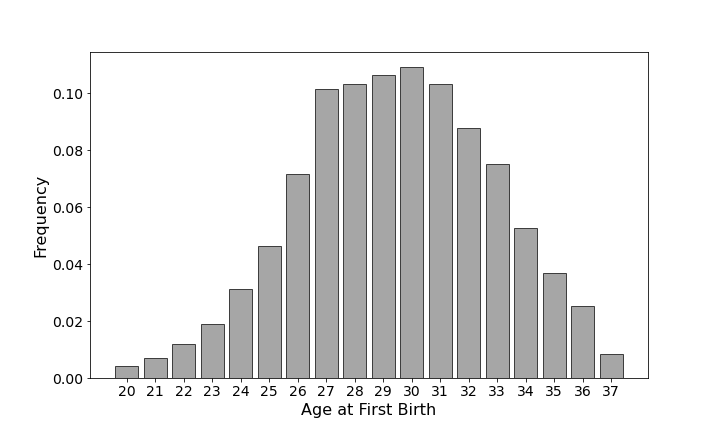}
    \footnotesize \textit{Note:} The figure shows the age distribution of men having a child between 2012 and 2022. ($N = 5,749$) 
    \end{minipage}
\end{figure}

The left part of Table \ref{Table:samples_staggered_diff_diff_men} compares the different samples, whereas the right part compares men who have an antidepressant prescription at least once between 2012 and 2022 with those who do not.

Table \ref{Table:samples_staggered_diff_diff_men} shows that men in the balanced sample are more likely to be Swiss and are less likely to have English as a contact language. 90\% of men have more than one child. This pattern happens because we can only identify fathers insured with the same insurance as the mothers. Hence, families are probably more likely insured at one company when they have multiple children and not directly after the first child. Similarly, men in the balanced sample seem more likely to be single. This is probably the case because the civil status is not updated and is only observed the first time an individual signs up for the insurance. Hence, if men go in and out of the insurance, the civil status is updated, whereas for individuals staying at the same insurance over the years, the civil status is not updated. 

The right part of Table \ref{Table:samples_staggered_diff_diff_men} shows that similar to women, men having an antidepressant prescription are more likely to receive health subsidies and to have accident coverage, namely not to be employed. Men having an antidepressant prescription are also less likely to be Swiss than men not having an antidepressant prescription.

\begin{table}[h!]\centering
    \caption{Descriptives: Comparison of different samples for staggered difference-in-difference analysis}
    \label{Table:samples_staggered_diff_diff_men}
    \begin{adjustbox}{max width=\textwidth}
    \begin{threeparttable}
    \begin{tabular}{l|rrrrr|rrr} \toprule
    Sample & \multicolumn{1}{c}{Balanced} & \multicolumn{2}{c}{Unbalanced} & \multicolumn{2}{c}{w/o Corona years} & \multicolumn{1}{c}{Antidepressants} & \multicolumn{2}{c}{No antidepressants} \vspace{0.1cm}\\ 
        &\textit{Mean} & \textit{Mean} & \textit{Std. diff.} & \textit{Mean}  & \textit{Std. diff.}& \textit{Mean} & \textit{Mean} & \textit{Std. diff.} \\\midrule
       & \multicolumn{8}{c}{\textbf{Discrete variables}} \\ \midrule
    \textbf{Nationality Regions}\\
    East Europe & 0.13 & 0.19 & 17.37 & 0.17 & 11.27 & 0.19 & 0.12 & 18.56\\
South Europe & 0.04 & 0.08 & 20.56 & 0.04 & 4.68 & 0.03 & 0.04 & 2.39\\
Switzerland & 0.80 & 0.61 & 42.59 & 0.74 & 14.31 & 0.72 & 0.81 & 20.10\\
West Europe & 0.01 & 0.05 & 22.17 & 0.02 & 3.89 & 0.02 & 0.01 & 5.11\\
not available & 0.01 & 0.02 & 8.08 & 0.01 & 2.38 & 0.01 & 0.01 & 2.47\\
other & 0.02 & 0.04 & 15.70 & 0.02 & 3.08 & 0.03 & 0.01 & 10.08\\
 \\
    \textbf{Civil status} \\
    
single & 0.42 & 0.27 & 31.65 & 0.37 & 8.91 & 0.40 & 0.42 & 4.70\\
not available & 0.23 & 0.32 & 19.99 & 0.23 & 1.18 & 0.27 & 0.22 & 10.57\\
married & 0.35 & 0.41 & 12.35 & 0.39 & 7.68 & 0.33 & 0.35 & 4.64\\
 \\
    \textbf{Contact language} \\
    English & 0.00 & 0.03 & 20.53 & 0.00 & 0.80 & nan & 0.00 & nan\\
French & 0.23 & 0.19 & 9.87 & 0.24 & 2.75 & 0.24 & 0.22 & 3.80\\
German & 0.74 & 0.73 & 4.04 & 0.73 & 2.67 & 0.73 & 0.74 & 2.20\\
Italian & 0.03 & 0.06 & 15.91 & 0.03 & 0.33 & 0.03 & 0.03 & 2.13\\
 \\
    \textbf{Region} \\
    Central Switzerland & 0.26 & 0.20 & 14.69 & 0.24 & 6.10 & 0.21 & 0.27 & 14.15\\
Eastern Switzerland & 0.14 & 0.16 & 6.43 & 0.15 & 2.16 & 0.15 & 0.14 & 4.24\\
Espace Mittelland & 0.18 & 0.16 & 6.27 & 0.18 & 0.89 & 0.18 & 0.18 & 0.67\\
Northwestern Switzerland & 0.15 & 0.16 & 1.34 & 0.16 & 1.53 & 0.16 & 0.15 & 1.72\\
Region lemanique & 0.13 & 0.11 & 3.62 & 0.14 & 3.86 & 0.14 & 0.12 & 6.18\\
Ticino & 0.03 & 0.04 & 9.56 & 0.03 & 0.13 & 0.02 & 0.03 & 1.61\\
Zurich & 0.11 & 0.16 & 15.19 & 0.11 & 1.21 & 0.13 & 0.11 & 5.60\\
  \midrule
    & \multicolumn{8}{c}{\textbf{Binary variables}} \\ \midrule
    
Employed & 0.88 & 0.91 & 8.59 & 0.87 & 3.27 & 0.80 & 0.89 & 24.91\\

Subsidy & 0.20 & 0.24 & 10.32 & 0.24 & 10.34 & 0.33 & 0.18 & 34.35\\
 
    One child & 0.10 & 0.05 & 19.87 & 0.10 & 1.44 & 0.12 & 0.10 & 6.78\\
\midrule
    & \multicolumn{8}{c}{\textbf{Continuous variables}}  \\ \midrule
    income index & -0.56 & -0.57 & 3.34 & -0.58 & 7.82 & -0.56 & -0.56 & 0.05\\
age at first birth & 29.46 & 29.98 & 15.63 & 28.16 & 40.90 & 29.46 & 29.53 & 2.24\\
  \midrule
    N &   5,749      &  20,642 &   & 4,057 & &  629      &  629 &   \\ \midrule
    \end{tabular}
    \begin{tablenotes}
        \small \item \textit{Note:} The table shows the covariates of men in the month they have a child. The number of observations in the unbalanced sample is lower than that in the unbalanced sample because we cannot observe all men in the month they become fathers. Column (2) shows the mean for the balanced sample. Column (3) shows it for the unbalanced sample, and column (4) shows the standardised difference comparing the unbalanced with the balanced sample. Columns (5) and (6) do the same for a balanced sample, only containing the years until 2019. Column (7) shows descriptives for men taking antidepressants, and column (8) for men not taking antidepressants. The income index is between -1 and 1 and is estimated by the insurance company and only available for the data from the CSS insurance.
    \end{tablenotes}
    \end{threeparttable}
    \end{adjustbox}
\end{table}
\clearpage

\subsubsection{Average Effects of Childbirth on Mental Health}

\begin{figure}[h!]
    \centering
    \begin{minipage}{15cm}
    \caption{Average effect men: Antidepressants prescriptions}
    \label{fig:antidepressants_stagg_diff_diff_balanced_men}
    \includegraphics[width=15cm]{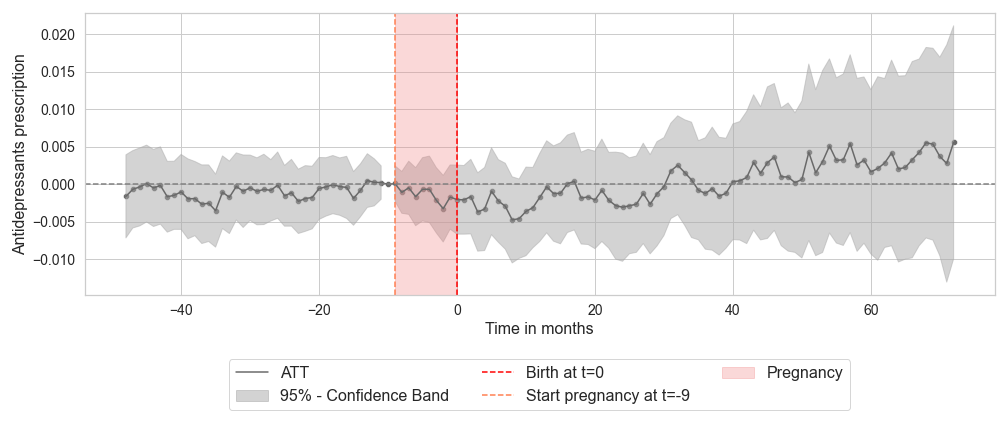}
    \footnotesize \textit{Note:} The figure shows the percentage point increase for men having an antidepressant prescription in comparison to the men who are ``not yet treated''. On average, 1\% of men in the sample have an antidepressant prescription before childbirth. The red-shaded area illustrates the pregnancy, and the grey-shaded area represents the 95\% confidence band. ($N = 5,749$)
    \end{minipage}
\end{figure}

We do not find an increase in antidepressant prescriptions after childbirth for men. The results using a balanced sample are shown in Figure \ref{fig:antidepressants_stagg_diff_diff_balanced_men}.

\begin{figure}[h!]
    \centering
    \begin{minipage}{15cm}
    \caption{Average effect men: Number of visits to psychiatrist}
    \label{fig:Psychiater_stagg_diff_diff_balanced_men}
    \includegraphics[width=15cm]{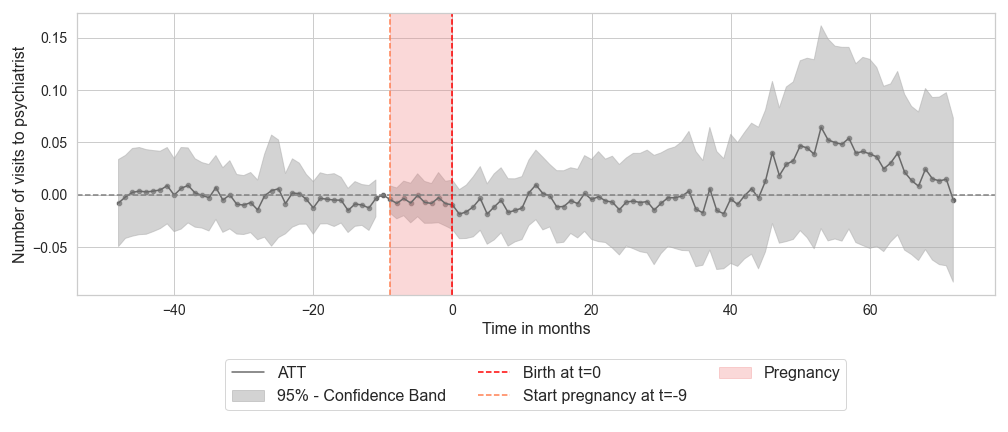}
    \footnotesize \textit{Note:} The figure shows the increase in the number of visits to the psychiatrist compared to the men who are ``not yet treated''. On average, men go to a psychiatrist 0.05 times per month before childbirth. The red-shaded area illustrates the pregnancy, and the grey-shaded area represents the 95\% confidence band.  ($N = 5,749$)
    \end{minipage}
\end{figure}

Similarly, as shown in Figure \ref{fig:Psychiater_stagg_diff_diff_balanced_men}, we also do not find a statistically significant increase in visits to the psychiatrist after childbirth for men.

\begin{figure}[h!]
    \centering
    \begin{minipage}{15cm}
    \caption{Average effect men: Number of visits to GP or HMO}
    \label{fig:Hausarzt_stagg_diff_diff_balanced_men}
    \includegraphics[width=15cm]{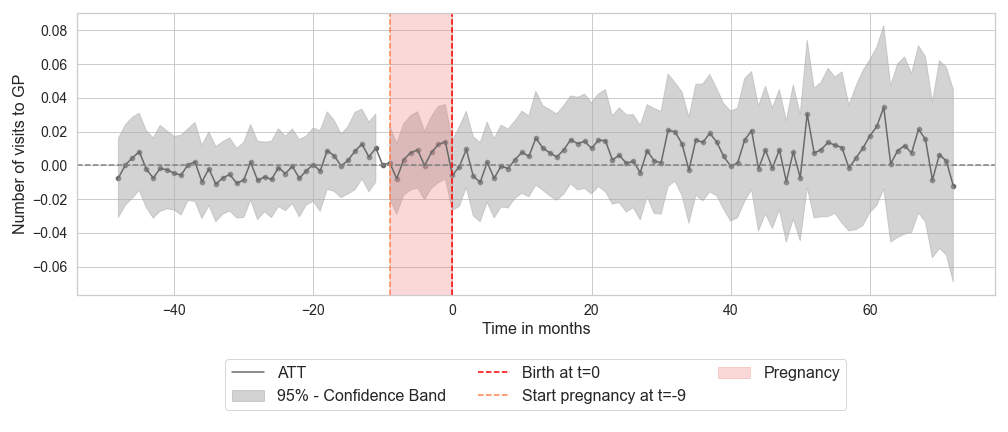}
    \footnotesize \textit{Note:} The figure shows the increase in the number of visits to the GP or HMO in comparison to the men who are ``not yet treated''. On average, men go to a GP or HMO approximately 0.11 times per month before childbirth. The red-shaded area illustrates the pregnancy, and the grey-shaded area represents the 95\% confidence band. ($N = 5,749$)
    \end{minipage}
\end{figure}

Last, there is also no statistically significant increase in visits to the GP or HMO after childbirth for men, as shown in Figure \ref{fig:Hausarzt_stagg_diff_diff_balanced_men}. 

Concluding, it seems that there is no mental health penalty for men after the birth of their first child. However, as the sample size is relatively small and a particular sample, the results should be interpreted cautiously.

\clearpage
\section{Effect of Paternity Leave} \label{appendix: paternal_leave}

\subsection{Insititutional Setting} \label{1_institutional_set_up_paternity}
\subsubsection{Paternity leave} \label{1_institutional_set_up_paternity_leave}
On September 27th 2020, the Swiss population approved the introduction of paternity leave. Although most European countries already have paternity leave, it was not apparent in advance that the Swiss population would agree to introduce a two-week paternity leave. The public debate on paternity leave was triggered by an initiative submitted in 2017.\footnote{The popular initiative is a political right in Switzerland that voters can take at federal, cantonal and municipal levels. Switzerland's direct democracy recognises two fundamentally different instruments of direct political influence. With the instrument of the popular initiative, the sovereign decides to include a new provision in the constitution. With the instrument of the referendum, the sovereign seizes the final decision-making authority over new laws.} It demanded that all working fathers be allowed to take at least four weeks of paternity leave and receive compensation during their absence. The Swiss parliament found this too far-reaching and proposed a counter-proposal of a two-week paternity leave. The Federal Council rejected the petition and the counter-proposal \citep{SRF:2020}. Generally, the French-speaking part of Switzerland more often agrees on introducing family policies (e.g. childcare subsidies) than the German-speaking part of Switzerland. Similarly, the cities are more keen to introduce family policies than the countryside \citep{BFS:2021a}.

The two weeks of paternity leave were introduced on January 1st, 2021. Before 2021, a Swiss father was entitled to have one day of paternity leave. Hence, if the birth endured more than one day, he had to take a day of vacation. \footnote{Some big (international) firms (e.g. Google, Ikea, Migros) and some public administrations of individual cantons have already had paternity leave. For example, civil servants of the canton of Zurich already had five days of paternity leave before 2021. However, most of the fathers in Switzerland did not because two-thirds of the individuals work in companies that are KMUs\footnote{KMU is the abbreviation for ``Kleine und mittlere Unternehmen," which means small and middle-sized firms in English.} \citep{SECO:2022d} which mostly had no paternity leave \citep{Tagblatt:2019}. They may have had the possibility to take vacation around the birth of a child, but this reduced the number of ``regular" vacation weeks in that year.} From January 1st, 2021, employed or self-employed fathers can take two weeks (10 days) of paternity leave. If they take it, they receive 80\% of their salary.\footnote{This also applies to self-employed.} They must take these ten days in the first six months after the birth of their child. The government pays 80\% of their regular salary, and the remaining 20\% is often paid by the employer \citep{BfS:2024b}. It is unclear if firms that already had paternity leave should increase the leave by two weeks. Important is that the employers can not be worse off than before \citep{Geiser:2021}. In conclusion, most of the fathers in Switzerland were eligible for paternity leave if they became fathers on or after January 1st, 2021. No other important and related reforms were introduced in Switzerland on January 1st 2021 \citep*{Eigensatz:2021}.

In 2021, 73\% of the fathers living and working in Switzerland took paternity leave.\footnote{Fathers working in Switzerland and living in another country are also entitled to take paternity leave. However, they are excluded from this analysis.} In 2022, the take-up rate increased to 76\%. There are differences in take-up rates across different sociodemographic characteristics. Swiss fathers are seven percentage points more likely to use the two weeks of paternity leave than fathers of European nationality. Fathers with another nationality are even less likely to take paternity leave. The civil status does not matter too much, but it seems that divorced fathers are less likely to take paternity leave than single and married fathers. Interestingly, the take-up rates across age groups are different. Individuals below 26 have a take-up rate of 66\% in 2021, whereas individuals between 31 and 35 years have a take-up rate of 76\%. The take-up rate decreases by age to 56\% for fathers above 50 years. Last, the father's yearly income also influences the take-up rates of paternity leave. For individuals that earn less than CHF 50,000 per year, the take-up rate lies around 50\%. This might be explained by the fact that for those families, the 80\% of the income fathers receive when they take paternity leave is not enough. Hence, they keep working. For individuals earning between 50,000 and 250,000 CHF per year, the take-up rate is around 71 to 78\% in 2021. For individuals that earn more, the take-up rate again drops to 66\%. Those individuals might work in jobs that other individuals can not do. Hence, it is difficult for them to miss two weeks of work.\footnote{This paragraph uses information from administrative data from the Swiss government, namely income compensation data (ZAS - EO), population and household statistics (BFS - STATPOP), vital statistics (BFS - BEVNAT) and social security data (ZAS - IK).}

\subsubsection{The Coronavirus Pandemic in Switzerland} \label{1_corona_virus_pandemic}

Paternity leave was introduced during the coronavirus pandemic. The first lockdown was implemented on March 16th 2020. For example, restaurants, schools and sports facilities were closed. From then on, it was strongly recommended to stay home and work from home, if possible \citep{Bundesrat:2020}. To prevent a recession, the federal government set up a program for companies in need. They could apply for short-time work compensation \citep{SECO:2022e}. This has led to only a slight increase in the unemployment rate from 2.3\% in 2019 to 3.1\% and 3\% in 2020 and 2021, respectively. In 2017, the unemployment rate was already at 3.1\% \citep{SECO:2022a}. Hence, the pandemic did not lead to an exceptionally high unemployment rate. 

In autumn 2020, the second wave of the Coronavirus hit Switzerland. From October 19th, 2020, meeting more than 15 people at the time was forbidden, and it was again recommended to work from home, which led to 25\% of the individuals primarily working in the home office \citep{SocialMonitor:2022}. From December 22nd, restaurants were closed again, and from January 18th, 2021, the government implemented a mandatory home office rule and closed stores for goods of non-daily use and sports facilities. This led to an increase of around five percentage points (from 25\% to 30\%) of individuals primarily working from home \citep{SocialMonitor:2022}. The schools and childcare facilities were open during the second lockdown. Individuals receiving short-time work compensations increased from around 6\% in December 2020 to around 8\% in January 2021 \citep{SECO:2022e}. From March 1st, 2021, the government started to reopen the economy again. 

The coronavirus pandemic could lead to two potential issues in this paper. First, it could lead to threats to the identification strategy. To discuss these potential problems, we refer to Section \ref{1_identification}. A second concern could be that the effect of the paternity leave introduction on the mental health of mothers cannot be translated into non-pandemic times. During the pandemic, around one-third of the individuals worked partly from home \citep{BFS:2022c}. The fathers working in the home office and having a child could potentially support their wives better without paternity leave. However, according to the \cite{SocialMonitor:2022}, individuals claimed they only lost 5\% of working time due to the pandemic. This is anecdotal evidence that individuals in the home office work and do not spend the whole day with their children.
Furthermore, some papers have examined the Coronavirus pandemic's effect on gender equality in other countries. Evidence from the UK suggests that fathers help slightly more, but the main burden is still carried by the mothers \citep{Chung:2021}. \cite{Hank:2021} did not change the couple's division of labour during the pandemic in Germany on the aggregate level, but they found that the proportion of couples where one person does all the housework increased. However, the proportion of men taking care of all house duties is still low (an increase from 0.8\% to 2\%). The problem with these studies is that the schools and childcare facilities were closed in the period they are looking at. Hence, the burden for both parents is higher when the children are also at home. Therefore, it is not easy to translate those findings to a period where the schools and childcare facilities are open. In conclusion, the home office should have a different effect than having two weeks off to support the wife or life partner. However, we estimate a lower bound, and the effect would be larger in non-pandemic times. If we assume that the pandemic also leads to more home office in the future, the effect we might find can directly be translated into the future. Individuals staying at home due to short-term work even had more time to support their wives or life partners without having paternity leave. Also, in this scenario, the effect we estimate is a lower bound.

\subsection{Empirical Strategy} \label{1_empirical_strategy}

\subsubsection{Identification} \label{1_identification}

The identification strategy relies on introducing paternity leave in Switzerland as explained in Section \ref{1_institutional_set_up_paternity_leave}. We use this quasi-experimental setting to identify a causal effect, as this reform can be seen as an exogenous variation. We are aware of self-selection in the decision of whether to take up paternity leave or not. It could happen that fathers will more often take paternity leave if they suffer from mental health issues or if the mother suffers from mental health issues. Furthermore, take-up rates probably differ by other covariates, such as household income. Therefore, the treatment variable for the analysis is the eligibility for paternity leave rather than whether it was taken. This resembles an "Intention-To-Treat"-design. We define an indicator variable $D_i \in \{0,1\}$ with $D_i = 1$ if a mother gave birth between the 1st of January 2021 and the 31st of March 2021 and $D_i = 0$ if a mother gave birth between the 1st of October 2020 and the 31st of December 2020.\footnote{Number of births in Switzerland between the 1st of January 2021 and the 31st of March 2021: 20,077 (2,500 in this sample) and Number of births between the 1st of October 2020 and the 31st of December 2020: 20,619 (2,233 in this sample)} By only comparing these two groups with each other, the problem that we do not account for seasonalities could arise.\footnote{Some research shows that the characteristics of mothers with children born in different months are different \citep*{Buckles:2013}.} Therefore, we will compare this difference with the difference around the same cut-off one year before the reform. Hence, we have to define a second indicator variable $T_i \in \{0,1\}$ with $T_i = 1$ if a mother gave birth between the 1st of October 2020 and the 31st of March 2021 and $T_i = 0$ if a mother gave birth between the 1st of October 2019 and the 31st of March 2020.\footnote{Number of births between the 1st of October 2019 and the 31st of December 2019: 20,797 (2,173 in this sample) and Number of births between the 1st of January 2020 and the 31st of March 2020: 20,917 (2,086 in this sample)} This difference-in-difference identification strategy leads to an average treatment effect of the treated (ATET) in $T_i = 1$ if the necessary assumptions are fulfilled.

\begin{table}[H]
    \centering
    \begin{tabular}{lll} \toprule
    Cut-off 01/01/2021 & Control group & Treatment group \\ \midrule
     $T_i = 0$ & 01/10/2019 - 31/12/2019 & 01/01/2020 - 31/03/2020\\
     $T_i = 1$ & 01/10/2020 - 31/12/2020 & 01/01/2021 - 31/03/2021\\ \bottomrule
    \end{tabular}
    \caption{Overview treatment definitions}
    \label{tab:overview groups}
\end{table}

The analysis is based on the potential outcome framework by \cite{Rubin:1974}. The well-known problem is that we only observe one realised outcome $D_i = d$ for each observation, hence, one potential outcome. The potential outcomes are defined for two time periods, namely $Y_{i,0}^d$ and $Y_{i,1}^d$. Additionally, we can observe several covariates $X$. Following \cite*{Heckman:1997} we can estimate the $ATET = \E \left[Y_{i,1}^1 - Y_{i,1}^0 | D_i = 1 \right]$ if the following assumptions hold.

\textbf{Assumption 1: Common trends} 

\begin{equation}
    \E\left[Y_{i,1}^0 - Y_{i,0}^0 | D_i = 0\right] = \E\left[Y_{i,1}^0 - Y_{i,0}^0 | D_i = 1 \right]
\end{equation}

The average outcomes of the treatment and control group have followed parallel trends without treatment. In this setting, if there had not been the paternity leave reform starting in 2021, there would be no difference in the outcomes of the mothers giving birth before the 1st of January 2021 and afterwards. This assumption is not testable, but following the standard literature, some tests can be done to assess the credibility of the assumption. A placebo reform three and six months before the actual reform is done in Section \ref{robustness_checks_paternity_leave}. We do not find any statistically significant effect.

A potential threat to the common trends assumption could be the coronavirus pandemic. As explained in Subsection \ref{1_corona_virus_pandemic}, the second lockdown happened during the introduction of paternity leave. However, we show in Section \ref{1_data_paternity_leave} that we do not see any effect of the pandemic on the number of antidepressant prescriptions. A further issue could be an increase of 5 percentage points in individuals doing home office between December 2020 and January 2021 and an increase of 2 percentage points in individuals being on short-term work. Fathers working from home have potentially more time to help their wives or life partners. Since both groups still (partly) work, the slight increase should not threaten the identification.

\textbf{Assumption 2: No anticipation}

Individuals are not allowed to anticipate the reform such that the reform already has an effect before the implementation. Formally, this means 
\begin{equation}
    \E\left[Y_{i,0}^1 - Y_{i,0}^0 | D_i = 1\right] = 0
\end{equation}
This assumption could be violated if mothers planned to conceive a child later to benefit from the reform. The Swiss population decided to introduce the reform on September 27th, 2020. If (to-be) parents had waited to conceive a child until the vote, they would have had their child approximately in June 2021, which is not in the sample anymore. Hence, this assumption should hold.

\textbf{Assumption 3: Stable unit treatment value assumption (SUTVA)}

The SUTVA requires that each individual is allocated to the treatment or control group and that if one mother is treated ($D_i = 1$), this does not affect the outcome of another mother. This would, for example, be the case if one mother who gave birth before the reform gets mental health problems when seeing that another mother who gave birth after the reform has a husband who can take paternity leave. We can not completely rule out this peer effect, but it is unlikely.

\textbf{Assumption 4: Common support}
    
The common support assumption ensures that the support of the propensity score of the treatment group is a subset of the support for the control group. Formally, this assumption looks as follows:
\begin{equation}
         P(D_i = 1 | X_i) < 1
\end{equation} 
To ensure this assumption, all observations with no overlap are dropped.

\subsubsection{Estimation} \label{1_estimation}
    
Several parametric and non-parametric estimation strategies could be used. One drawback of using a simple two-way fixed effects regression is that it leads to biased results if the treatment is heterogeneous across $X_i$'s \citep*{Roth:2022}. Therefore, a possible estimation strategy would be to rely on a semi-parametric difference-in-difference approach \citep{Heckman:1997, Abadie:2005}. A double robust approach additionally has the advantage of being still consistent when either the propensity score or the outcome model is misspecified \citep{SantAnna:2020, Zimmert:2018, Chang:2020}. \cite{Abadie:2005} identifies the ATET as: 
\begin{align*}
        \theta_0 = \E\left[\frac{T_i - P(T_i= 1)}{P(T_i= 1)(1-P(T_i= 1))}\frac{Y_i}{P(D_i = 1)}\frac{D_i - P(D_i = 1|X_i)}{1-P(D_i = 1|X_i)}\right]
\end{align*}

\cite*{Chernozhukov:2018} developed a double/debiased machine learning approach that allows using machine learning methods and still getting a $\sqrt{N}$-consistent estimator and, therefore, valid inference. They managed to achieve this by using cross-fitting and Neyman-orthogonal scores. Based on this approach by \cite{Chernozhukov:2018} and the estimator by \cite{Abadie:2005}, \cite{Chang:2020} shows how to build a score function that is Neyman-orthogonal and identifies the ATET in a difference-in-difference setting. The score function looks the following: 
\begin{align*}
        \phi(y,d,x,t, \theta_0, p_0, \lambda_0, \eta_{00})  &= \frac{t - \lambda_0}{\lambda_0(1-\lambda_0)}\frac{y}{p_0}\frac{d-P(D_i=1|X_i=x)}{1-P(D_i=1|X_i=x)} \\&- \theta_0 -\frac{d - P(D_i=1|X_i=x)}{\lambda_0(1-\lambda_0)p_0(1-P(D_i = 1|X_i=x))}\E[(T_i-\lambda_0)Y_i|X_i=x, D_i = 0]
\end{align*}
The nuisance parameters that have to be estimated are $p_0 = P(D_i = 1)$, $\lambda_0 = P(T_i= 1)$ and $\eta_{00} = [P(D_i = 1|X_i=x), \E[(T_i-\lambda_0)Y_i|X_i=x, D_i = 0]] = (g_0,l_0)$. Machine learning methods can be used to estimate the nuisance parameters as long as the propensity score and the outcome model are consistent and the product of the errors converges with a convergence rate of $\sqrt{N}$. For example, Lasso or Random forests satisfy this condition under sparsity forms \citep{Chernozhukov:2018}. We use a Random Forest with 1000 trees. Furthermore, the maximum depth and the minimum leaf size were tuned using a grid search. The package DoubleML has been used for the analysis.\footnote{https://docs.doubleml.org/stable/intro/install.html}

\subsection{Data} \label{1_data_paternity_leave}
\subsubsection{Outcome variable}
In this part of the paper, we look at yearly estimates. The outcome variables are measured in relation to the time of birth $t$, hence, $t+1, t+2, \dots$ with the time unit being years. Event-time $t = 0$ means 11 months before birth until the month of birth, and event-time $t = 1$ means one month after birth until 12 months after birth. The antidepressant prescription is a binary variable that takes the value one when women had an antidepressant prescription in a year and zero otherwise. Furthermore, we only look at antidepressant prescriptions.

\begin{figure}[h!]
    \centering
    \begin{minipage}{14cm}
    \caption{Descriptives: Antidepressant prescriptions over the years}
    \label{fig:descriptives_antidepri_over_years}
    \includegraphics[width=14cm]{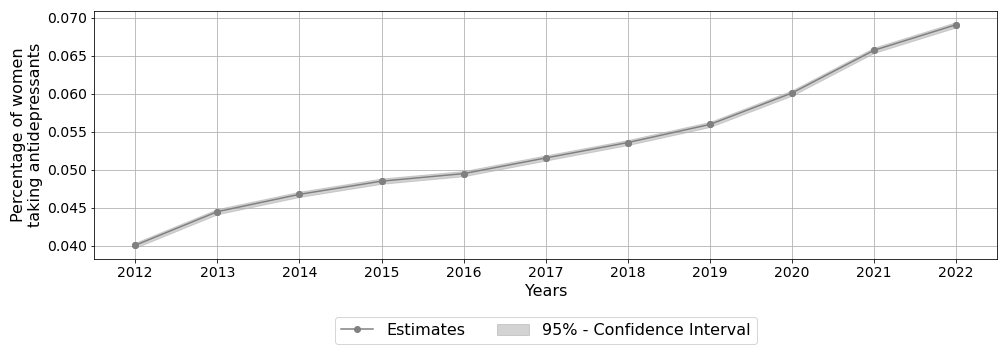}
    \footnotesize \textit{Note:} The figure shows the percentage of women born between 1985 and 1995 having an antidepressant prescription over the years. ($N = 308,987$)
        \end{minipage}
\end{figure}

The reason is that we had a pandemic. The pandemic might have influenced the monthly visits to a GP and HMO or a psychiatrist, and hence, also the number of months women had an antidepressant prescription. However, we assume that if we use a yearly variable that takes the value of one if a woman had at least one month of an antidepressant prescription in one year and zero otherwise, the effect of the pandemic is negligible for antidepressant prescriptions. As Figure \ref{fig:descriptives_antidepri_over_years} points out, the number of antidepressant prescriptions has been increasing over the years, and we do not see any effect of the pandemic on the number of antidepressant prescriptions. Therefore, we also only focus on this outcome variable.

\subsubsection{Control Variables}

Table \ref{Table:control_variables_diff_diff} shows the different characteristics of women in the treatment group versus women in the control group. We include covariates such the region in Switzerland, civil status, contact language, employment, if someone receives subsidies from the government, and the age at first birth. Most covariates do not change over time and are measured the first time someone signs up at the insurance. However, the variables that change over time are measured in the last month before childbirth. The covariates are not needed for the parallel trend assumption to hold but for potential more precision in the estimates. It seems that the treatment and control groups are not statistically significantly different. All values, despite age at first birth, are pretty small. Hence, the sample is well-balanced. Age at first birth is by construction different between the two groups because we only have the mother's birth year. Hence, the age at first birth should be around one year higher for the treatment group.

\begin{table}[h!]\centering
        \caption{Descriptives paternal leave: Comparison of controls variables for treatment and control group}
        \label{Table:control_variables_diff_diff}
        \begin{adjustbox}{max width=0.55\textwidth}
        \begin{threeparttable}
        \begin{tabular}{lrrr} \toprule
         & \multicolumn{1}{c}{Treated} & \multicolumn{1}{c}{Control} & \\ 
         & Mean  & Mean  & Std. diff. \\\midrule
        \textbf{Discrete variables}  &  & & \\ \midrule
        \textbf{Nationality Regions}\\
        Switzerland & 0.04 & 0.05 & 1.93\\
East Europe & 0.08 & 0.09 & 3.36\\
South Europe & 0.00 & 0.00 & 1.89\\
West Europe & 0.08 & 0.07 & 1.41\\
not available & 0.70 & 0.69 & 1.11\\
other & 0.07 & 0.06 & 4.25\\
North Europe & 0.02 & 0.03 & 2.99\\
 \\
        \textbf{Civil status} \\
        married & 0.01 & 0.01 & 0.32\\
not available & 0.32 & 0.36 & 9.14\\
single & 0.22 & 0.20 & 4.84\\
divorced & 0.46 & 0.43 & 4.73\\
widowed & 0.00 & 0.00 & 0.17\\
 \\
        \textbf{Contact language} \\
        German & 0.05 & 0.05 & 1.81\\
French & 0.17 & 0.18 & 3.37\\
Italian & 0.72 & 0.71 & 2.85\\
English & 0.06 & 0.06 & 1.64\\
 \\
        \textbf{Region} \\
        Ticino & 0.14 & 0.14 & 0.16\\
Zurich & 0.18 & 0.18 & 0.72\\
Region lemanique & 0.16 & 0.17 & 3.99\\
Espace Mittelland & 0.12 & 0.12 & 2.12\\
Northwestern Switzerland & 0.13 & 0.12 & 2.00\\
Eastern Switzerland & 0.05 & 0.05 & 2.07\\
Central Switzerland & 0.23 & 0.21 & 4.17\\
not available & 0.00 & 0.00 & 0.00\\
 \\
        \textbf{Employment} \\
        no & 0.15 & 0.18 & 7.46\\
yes & 0.85 & 0.82 & 7.46\\
 \\
        \textbf{Subsidy} \\
        no & 0.85 & 0.85 & 0.08\\
yes & 0.15 & 0.15 & 0.08\\
 \midrule
        \textbf{Continuous variables} & & &   \\ \midrule
        income index & -0.50 & -0.52 & 8.03\\
age at first birth & 31.23 & 30.38 & 30.51\\
 \midrule
        N & 4,673 & 4,319 & \\ \midrule
        \end{tabular}
        \begin{tablenotes}
            \small \item \textit{Note:} The table shows the covariates of the women in the month they give birth. Column (2) shows the mean for the treated group. Columns (3) shows it for the control group, and column (4) shows the standardised difference comparing the treated with the control group. The smaller the value, the more similar the two samples are \citep{Rosenbaum:1985}.
        \end{tablenotes}
        \end{threeparttable}
        \end{adjustbox}
\end{table}

\clearpage
\subsection{Results}
    
\begin{figure}[h!]
        \centering
        \begin{minipage}{14cm}
        \caption{Paternal leave effect: Antidepressant prescriptions}
        \label{fig:antidepri_diff_diff}
        \includegraphics[width=14cm]{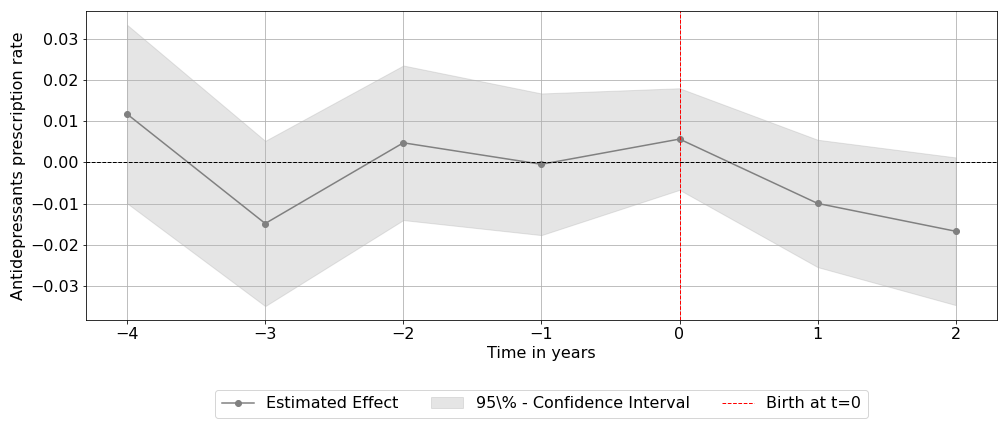}
        \footnotesize \textit{Note:} The figure shows the effect of the introduction of paternity leave in 2021 on the percentage of women having an antidepressant prescription in one year. The comparison group consists of women who had a child in the same months one year before. (01.01.2021 - 31.03.2021: $N = 2,500$, 01.10.2020 - 31.12.2020: $N = 2,233$, 01.01.2020 - 31.03.2020: $N = 2,173$, 01.10.2019 - 31.12.2019: $N = 2,086$)
        \end{minipage}
\end{figure}

Figure \ref{fig:antidepri_diff_diff} shows the effect of the introduction of paternity leave in 2021 on antidepressant prescriptions. The parallel trend assumption is not violated as we do not see any statistically significant effects in the years before childbirth. We do not find a statistically significant effect in the first two years after childbirth. However, there seems to be a trend towards a negative effect as the effect lies between zero and a decrease of 3 p.p. in the second year after childbirth. A problem with the analysis are the large confidence intervals. The large confidence intervals partly come from the combination of rare outcome variables and a small number of observations.\footnote{For example, only between 1\% and 2\% of women take antidepressants around childbirth. Hence, if we have a group of around 2000 observations, only 20 to 40 of those women take antidepressants.} Hence, the power is low, and we can also not claim that we have not found an effect. Additionally, as seen in the main part of the paper, the increase in the number of antidepressant prescriptions mostly happens after the second year after childbirth. Therefore, looking at a longer horizon would be valuable.

\subsection{Robustness Checks} \label{robustness_checks_paternity_leave}

\subsubsection{Decreasing Treatment Window to One Month}
\begin{figure}[h!]
        \centering
        \begin{minipage}{14cm}
        \caption{Paternal leave effect smaller treatment window: Antidepressant prescriptions}
        \label{fig:antidepri_durant_year_diff_diff_1_months}
        \includegraphics[width=14cm]{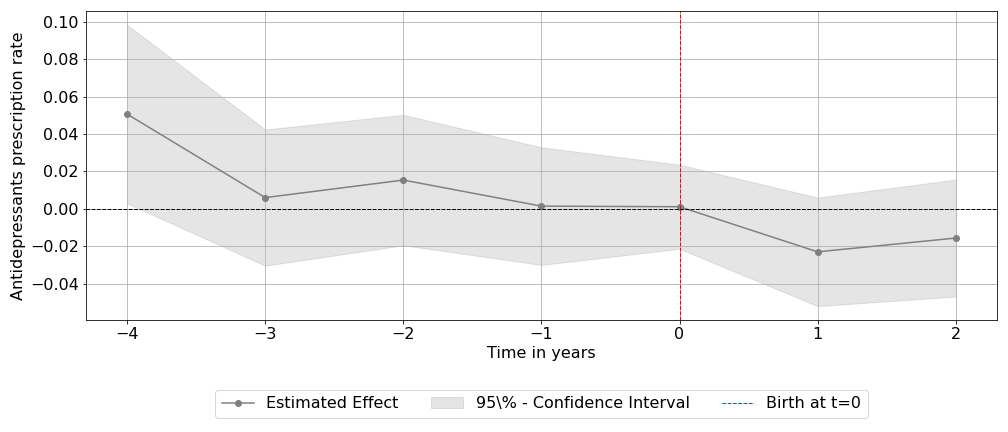}
        \footnotesize \textit{Note:} The figure shows the effect of the introduction of paternity leave in 2021 on the percentage of women having an antidepressant prescription in one year. The comparison group consists of women who had a child in the same months one year before. (01.01.2021 - 31.01.2021: $N = 862$, 01.12.2020 - 31.12.2020: $N = 2,768$, 01.01.2020 - 31.01.2020: $N = 748$, 01.12.2019 - 31.12.2019: $N = 685$)
        \end{minipage}
    \end{figure}

We decrease the treatment window to one month as a first robustness check. This means that the treatment group consists of women having a child between the 1st of January 2021 and the 31st of January 2021, and the control group consists of women having a child between the 1st of December 2020 and the 31st of December 2020. To eliminate the seasonalities, we compare them to the women having a child during the same time window one year before. Figure \ref{fig:antidepri_durant_year_diff_diff_1_months} does not show any statistically significant effects.

\subsubsection{Linear Specification}

As a second robustness check, we want to make sure that the results do not differ if we use a simple linear model that looks as follows:
\begin{align*}
 Y_{i} = \beta_0 + \beta_1 D_{i} + \beta_1 T_{i} + \beta_2 D_{i} T_{i} + \beta_3 X_{i} 
\end{align*}
with $D_i$ being the treatment indicator, $T_i$ being the time indicator and $X_i$ being the pre-treatment covariates. The pre-treatment covariates $X_{i}$ are only added to reduce the estimates' variance. The outcome variable of interest $Y_{i}$ is a binary variable stating if a woman has at least one antidepressant prescription in a year.

\begin{figure}[h!]
    \centering
    \begin{minipage}{14cm}
    \caption{Paternal leave effect linear specification: Antidepressant prescriptions}
    \label{fig:antidepri_diff_diff_linear}
    \includegraphics[width=14cm]{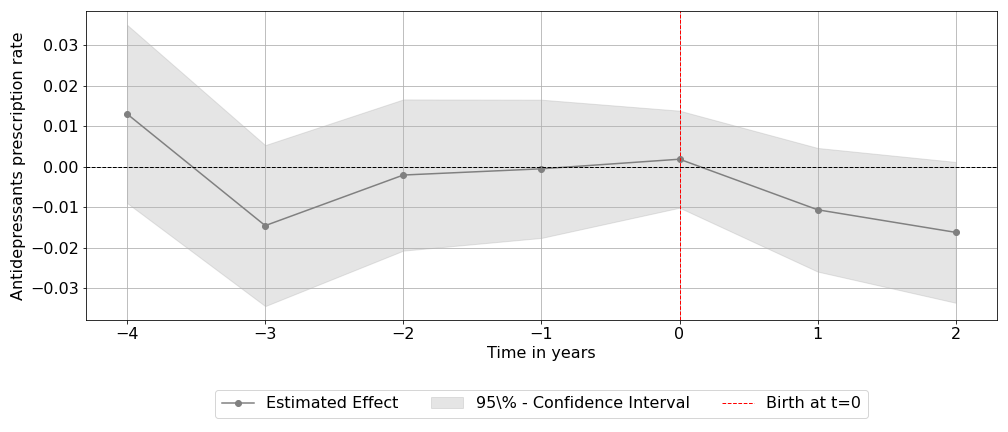}
    \footnotesize \textit{Note:} The figure shows the effect of the introduction of paternity leave in 2021 on the percentage of women having an antidepressant prescription in one year. The comparison group consists of women who had a child in the same months one year before. (01.01.2021 - 31.03.2021: $N = 2,500$, 01.10.2020 - 31.12.2020: $N = 2,233$, 01.01.2020 - 31.03.2020: $N = 2,173$, 01.10.2019 - 31.12.2019: $N = 2,086$)
    \end{minipage}
\end{figure}

A three-month treatment window, as in the main specification, is employed. Figure \ref{fig:antidepri_diff_diff_linear} confirms the robustness of the results to this specification, with the patterns remaining consistent. 

\newpage
\subsubsection{Placebo Reform}
\begin{figure}[h!]
    \centering
    \begin{minipage}{14cm}
    \caption{Paternal leave effect placebo reform three months: Antidepressant prescriptions}
    \label{fig:antidepri_diff_diff_plabeco_reform_3}
    \includegraphics[width=14cm]{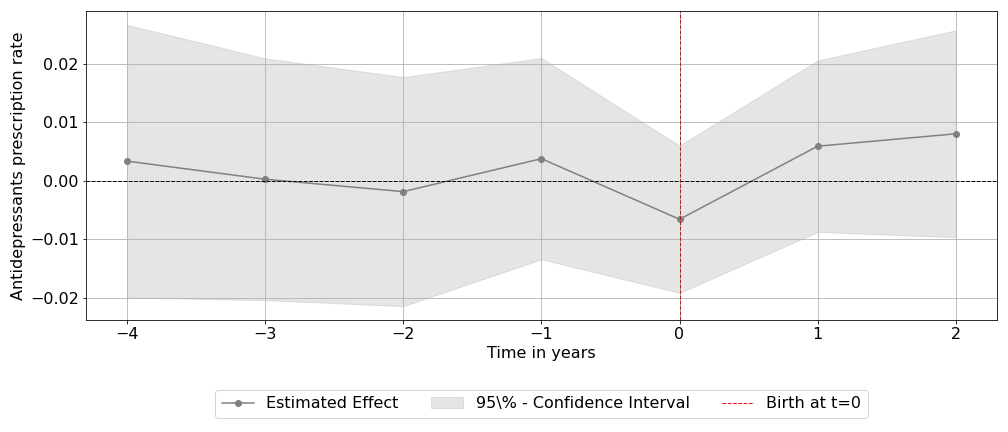}
    \footnotesize \textit{Note:} The figure shows the effect of the introduction of a placebo reform in October 2020 on the percentage of women having an antidepressant prescription in one year. The comparison group consists of women who had a child in the same months one year before. (01.01.2020 - 31.03.2020: $N = 2,371$, 01.10.2019 - 31.12.2019: $N = 2,263$, 01.01.2019 - 31.03.2019: $N = 2,372$, 01.10.2018 - 31.12.2018: $N = 2,182$)
    \end{minipage}
\end{figure}

As a last robustness check, we look at a placebo reform three and six months before the actual reform. For the placebo reform three months before, women giving birth between the 1st of October 2020 and the 31st of December 2020 are in the treatment group. Women giving birth between the 1st of July 2020 and the 31st of September 2020 are in the control group. The comparison group consists of women giving birth during the same months one year before. Figure \ref{fig:antidepri_diff_diff_plabeco_reform_3} shows that there is no statistically significant effect of the placebo reform on the percentage of women having an antidepressant prescription in the year after childbirth.

\begin{figure}[h!]
    \centering
    \begin{minipage}{14cm}
    \caption{Paternal leave effect placebo reform six months: Antidepressant prescriptions}
    \label{fig:antidepri_diff_diff_plabeco_reform_6}
    \includegraphics[width=14cm]{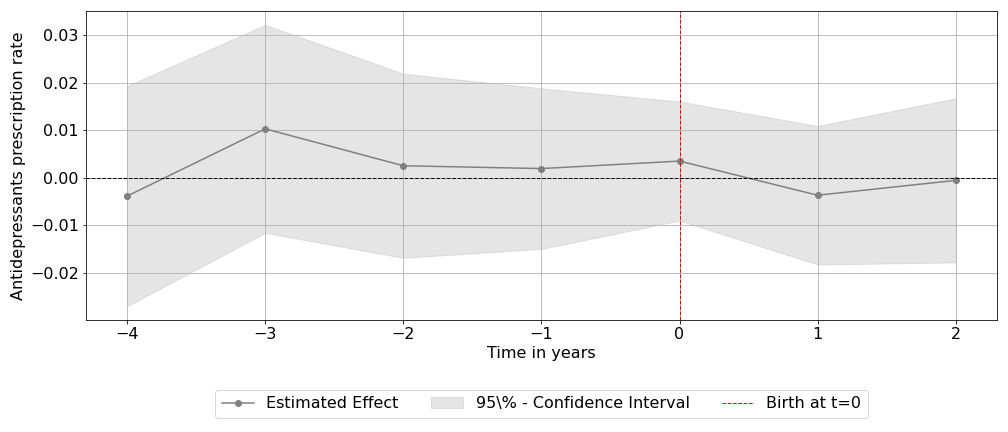}
    \footnotesize \textit{Note:} The figure shows the effect of the introduction of a placebo reform in July 2020 on the percentage of women having an antidepressant prescription in one year. The comparison group consists of women who had a child in the same months one year before. (01.07.2020 - 31.09.2020: $N = 2,3233$, 01.04.2019 - 30.06.2019: $N = 2,371$, 01.07.2019 - 31.09.2019: $N = 2,086$, 01.04.2018 - 30.06.2018: $N = 2,372$)
    \end{minipage}
\end{figure}

For the placebo reform six months before, women giving birth between the 1st of July 2020 and the 31st of September 2020 are in the treatment group, and women giving birth between the 1st of April 2020 and the 30th of June 2020 are in the control group. The comparison group consists of women giving birth during the same months one year before. Figure \ref{fig:antidepri_diff_diff_plabeco_reform_6} shows that there is no statistically significant effect of the placebo reform on the percentage of women having an antidepressant prescription in the year after childbirth. This strengthens the identification strategy.

\end{appendices}

\end{document}